\documentclass[letterpaper]{ar2e_mod}

\RequirePackage{latexsym}
\RequirePackage{amssymb}
\RequirePackage{natbib}
\input epsf.tex    

\def\ncl{{\cal N}_{\rm cluster}}
\def\eclump{\epsilon_{\rm clump}}

\def\phin{\phi_{\rm in}}
\def\mdin{\dot m_{\rm in}}

\def\micron{$\mu$}
\def\boms {B_{0,\,-6}}

\def\rdd {R_{\rm dd}}

\def\fm {f_{\rm mult}}

\def\geff{G_{\rm eff}}
\def\mds{\dot m_*}
\def\vkep{v_{\rm Kep}}
\def\rcore{R_{\rm core}}
\def\rbh{R_{\rm BH}}
\def\phibh{\phi_{\rm BH}}

\def\ecore {\epsilon_{\rm core}}

\def\tgs  {t_{g*}}
\def\epsff {\epsilon_{{\rm ff},S}}

\let\ga=\gtrsim
\let\la=\lesssim

\newcommand\K{{\;\rm K}}
\newcommand\g{{\;\rm g}}
\newcommand\Msun{{\;\rm\,M_\odot}}

\def\beq	{\begin{equation}}
\def\eeq	{\end{equation}}
\newcommand{\beqa}	{\begin{eqnarray}}
\newcommand{\eeqa}	{\end{eqnarray}}
\def\symbol#1{\ifmmode#1\else$#1$\fi}
\def\e{\symbol{^{-1}}}
\def\ee{\symbol{^{-2}}}
\def\eee{\symbol{^{-3}}}
\def\simlt{\lower.5ex\hbox{$\; \buildrel < \over \sim \;$}}
\def\simgt{\lower.5ex\hbox{$\; \buildrel > \over \sim \;$}}

\def\frac#1#2{{#1\over #2}}        
\def\avg#1{\langle #1\rangle}	   
\def\calb	{{\cal B}}
\def\calm	{{\cal M}}
\def\caln	{{\cal N}}
\def\calt	{{\cal T}}
\def\calw	{{\cal W}}

\def\vecv	{{\bf v}}

\def\krho   {{k_\rho}}
\def\avir   {\alpha_{\rm vir}}
\def\blos   {B_{\rm los}}
\def\bpos   {B_{\rm pos}}

\def\brot   {\beta_{\rm rot}}
\def\htwo   {H$_2$}

\def\vA	    {v_{\rm A}}
\def\cs	    {c_s}

\def\mkms   {{\rm km~s^{-1}}}
\def\kms    {{\mkms}}

\def\cm    {\;{\rm cm}}

\def\cmt    {{\rm cm}^{-3}}
\def\pc	    {\;{\rm pc}}

\def\msun   {{M_\odot}}
\def\mug    {\mu{\rm G}}

\def\yr     {{\rm yr}}

\def\gp     {{\gamma_{\rm p}}}

\def\mbe    {M_{\rm BE}}
\def\mcr    {M_{\rm cr}}

\def\mphi   {M_\Phi}

\def\muh    {\mu_{\rm H}}

\def\nh     {n_{\rm H}}

\def\pth    {P_{\rm th}}
\def\ptho    {P_{\rm th,\,0}}

\def\snt    {\sigma_{\rm nt}}
\def\sth    {\sigma_{\rm th}}

\def\tff    {{t_{\rm ff}}} 
\def\tgs    {{t_{g*}}}

\def\vcl    {{V_{\rm cl}}}

\usepackage[hyperindex,breaklinks]{hyperref}

\begin{document}

\jname{Annu. Rev. Astron. Astrophys.}
\jyear{2007}
\jvol{45}
\ARinfo{XXX}

\title{Theory of Star Formation\footnote{Posted with permission from the Annual Review of Astronomy and Astrophysics,
Volume 45, \copyright 2007 by Annual Reviews, http://www.annualreviews.org}}

\markboth{McKee \& Ostriker}{Theory of Star Formation}

\author{
Christopher F. McKee \affiliation{Departments of Physics and
    Astronomy, University of California, Berkeley, CA 94720; 
    cmckee@astro.berkeley.edu} 
Eve C. Ostriker \affiliation{Department
    of Astronomy, University of Maryland, College Park, MD 20742;
    ostriker@astro.umd.edu}}


\begin{abstract}

We review current understanding of star formation, outlining an
overall theoretical framework and the observations that motivate it.
A conception of star formation has emerged in which turbulence plays a
dual role, both creating overdensities to initiate gravitational
contraction or collapse, and countering the effects of gravity in
these overdense regions.  The key dynamical processes involved in star
formation -- turbulence, magnetic fields, and self-gravity -- are
highly nonlinear and multidimensional. Physical arguments are used to
identify and explain the features and scalings involved in star
formation, and results from numerical simulations are used to quantify
these effects.  We divide star formation into large-scale and
small-scale regimes and review each in turn.  Large scales range from
galaxies to giant molecular clouds (GMCs) and their substructures.
Important problems include how GMCs form and evolve, what determines
the star formation rate (SFR), and what determines the initial mass
function (IMF).  Small scales range from dense cores to the
protostellar systems they beget.  We discuss formation of both low-
and high-mass stars, including ongoing accretion.  The development of
winds and outflows is increasingly well understood, as are the
mechanisms governing angular momentum transport in disks.  Although
outstanding questions remain, the framework is now in place to build a
comprehensive theory of star formation that will be tested by the next
generation of telescopes.

\end{abstract}

\maketitle

\section{INTRODUCTION}

Stars are the ``atoms'' of the universe, and the problem of how stars
form is at the nexus of much of contemporary astrophysics.  By
transforming gas into stars, star formation determines the structure
and evolution of galaxies.  By tapping the nuclear energy in the gas
left over from the Big Bang, it determines the luminosity of galaxies
and, quite possibly, leads to the reionization of the universe.  Most
of the elements---including those that make up the world around
us---are formed in stars. Finally, the process of star formation is
inextricably tied up with the formation and early evolution of
planetary systems.

The problem of star formation can be divided into two broad
categories: 
``microphysics'' and 
``macrophysics''.  The microphysics of star
formation deals with how individual stars (or binaries) form. Do stars
of all masses acquire most of their mass via gravitational collapse of
a single dense ``core''?  How are the properties of a star or binary
determined by the properties of the medium out of which it forms? How
does the gas that goes into a protostar
lose its magnetic flux and angular momentum?  
How do massive stars form in the face
of intense radiation pressure?  What are the properties of the
protostellar disks, jets, and outflows associated with Young Stellar
Objects (YSOs), and what governs their dynamical evolution?

The macrophysics of star formation deals with the formation of systems
of stars, ranging from clusters to galaxies.  
How are giant molecular clouds (GMCs), the loci of most star
formation, themselves formed out of diffuse interstellar gas?
What processes determine the distribution of
physical conditions within star-forming regions, and why does star
formation occur in only a small fraction of the available gas? How is
the rate at which stars form determined by the properties of the natal
GMC
or, on a larger scale, of the interstellar medium in a
galaxy?  What determines the mass distribution of forming stars, the
Initial Mass Function (IMF)? Most stars form in clusters
\citep{2003ARA&A..41...57L}; 
how do stars form in such a dense
environment and in the presence of enormous radiative and mechanical
feedback from other YSOs?

Many of these questions, particularly those related to the
microphysics of star formation, were discussed in the classic review
of \citet{1987ARA&A..25...23S}. Much has changed since then.  Observers
have made enormous strides in characterizing star formation on all
scales and in determining the properties of the medium out of which
stars form. Aided by powerful computers, theorists have been able to
numerically model the complex physical and chemical processes associated with
star formation in three dimensions. Perhaps most important, a new
paradigm has emerged, in which large-scale, supersonic turbulence
governs the macrophysics of star formation.

This review focuses on the advances made in star formation since 1987,
with an emphasis on the role of turbulence. Recent relevant reviews
include those on 
the physics of star formation \citep{larson03}, and on the role of
supersonic turbulence in star formation 
\citep{2004RvMP...76..125M,2007prpl.conf...63B}.
The chapter by
Zinnecker \& Yorke in this volume of {\it ARAA} 
provides a different perspective on high-mass star formation, 
while that by Bergin \&
Tafalla gives a more detailed description of dense cores just prior to
star formation.
Because the topic is vast, we must necessarily exclude a number of
relevant topics from this review: primordial star formation
\citep[see][]{2004ARA&A..42...79B}, planet formation, astrochemistry,
the detailed physics of disks and outflows, radiative transfer, and
the properties of young stellar objects. 

In \S 2, we begin with an overview of basic physical processes and
scales involved in star formation, covering turbulence (\S 2.1), self
gravity (\S 2.2), and magnetic fields (\S 2.3). In \S 3, we
review macrophysics of star formation, focusing on: the physical state of
GMCs, clumps, and cores (\S 3.1), the formation, evolution, and
destruction of GMCs (\S 3.2), core mass functions and the IMF
(\S 3.3), and the large-scale rate of star formation (\S 3.4).
\S 4 reviews microphysics of star formation, covering low-mass star
formation (\S4.1), disks and winds (\S 4.2), and high-mass star
formation (\S 4.3).  We conclude in \S 5 with an overview of the
star formation process.


\section{BASIC PHYSICAL PROCESSES}

\subsection{Turbulence }
\label{turbulence}

As emphasized in \S 1, many of the advances in the
theory of star formation since the review of
\citet{1987ARA&A..25...23S} have been based on 
realistic evaluation and
incorporation of the effects of turbulence.  Turbulence is in fact
important in essentially all branches of astrophysics that involve
gas dynamics\footnote{ \citet{1949ApJ...110..329C}
presaged this development, in choosing the then-new theory of turbulence 
as the topic of his Henry Norris Russell prize lecture.}, 
and many communities have contributed to the recent
progress in understanding and characterizing turbulence in varying
regimes.  Here, we 
shall
concentrate on the parameter regimes of
turbulence applicable within the cold 
ISM,
and the
physical properties of these flows that appear particularly
influential for controlling star formation.  

Our discussion provides an overview only; pointers will be given to
excellent recent reviews that summarize the large and growing literature
on this subject.  General references include
\citet{1995tlnk.book.....F}, \citet{2003matu.book.....B}, and
\citet{2003LNP...614.....F}.  A much more extensive literature survey
and discussion of interstellar turbulence, including both diffuse-ISM
and dense-ISM regimes, is presented
by
\citet{2004ARA&A..42..211E} and \citet{2004ARA&A..42..275S}.
A recent review focusing on the detailed
physics of turbulent cascades in magnetized plasmas is 
\citet{2005astro.ph..7686S}.

\subsubsection{SPATIAL CORRELATIONS OF VELOCITY AND MAGNETIC FIELDS}
\label{turb_correl}
Turbulence is defined by the Oxford English Dictionary as a state of
``violent commotion, agitation, or disturbance,'' with a turbulent
fluid further defined as one ``in which the velocity at any point
fluctuates irregularly.''
Although turbulence is by definition an irregular
state of motion, a central concept is that order nevertheless persists
as scale-dependent spatial correlations among the flow variables.
These correlations can be measured in many ways; common mathematical
descriptions include autocorrelation functions, structure functions,
and power spectra.  

One of the most fundamental quantities, which is also one of the most
intuitive to understand, is the RMS velocity difference 
between two points separated by a distance $\bf r$.  With the velocity
structure function of order $p$ defined as $S_p({\bf r})\equiv \langle
|{\bf v(x)-v(x+r)}|^p\rangle $, this quantity is given as $\Delta
v({\bf r}) \equiv [S_2({\bf r})]^{1/2}$.  The autocorrelation function of the
velocity is related to the structure function: $A({\bf r})\equiv
\langle {\bf v(x)}\cdot {\bf v(x+r)} \rangle= 
\langle |{\bf v}|^2
\rangle - S_2({\bf r})/2$; note that the autocorrelation with zero lag is
$A(0)=\langle |{\bf v}|^2
\rangle$ since $S_2(0)=0$.  The power spectrum of velocity, $P({\bf
  k})\equiv |{\bf v(k)}|^2$, is the Fourier transform of the
autocorrelation function.  For zero mean velocity, the velocity
dispersion averaged over a volume ${\ell }^3$, $\sigma_v(\ell)^2$, is
equal to the power spectrum integrated with $k_{\rm min} =2\pi/{\ell}$.
If turbulence is isotropic and the system in which it is observed is 
spatially symmetric with each dimension $\approx \ell$, 
then the one-dimensional velocity dispersion
along a given line-of-sight (a direct observable) will be related to
the three-dimensional velocity dispersion by $\sigma=\sigma_v(\ell)/\sqrt{3}$.
Analogous structure functions, correlation functions, and power
spectra can also be defined for the
magnetic field, as well as other fluid variables including the density
(see \S\ref{densstruct}).  Delta-variance techniques provide similar
information, and are particularly useful for reducing edge effects
when making comparisons with observational data \citep{2001A&A...366..636B}.

For isotropic turbulence, $S_p$ and $A$ are functions only of 
$r=|{\bf  r}|$, and $P$ is a function only of $k=|{\bf k}|$.  The Fourier
amplitude 
$|{\bf v}({\bf k})|$
is then (on average) only
a function of $\ell=2\pi/k$, and
can be denoted by $v(\ell)$; to emphasize that these velocities 
are perturbations
about a background state, the amplitude of a given Fourier component is
often written as $\delta v(k)$ or $\delta v(\ell)$.
When there is a large dynamic range between
the scales associated with relevant physical parameters (see \S 
\ref{scales}), correlations often take on power-law forms.
If $P(k) \propto k^{-n}$ for an isotropic flow, then 
\begin{equation}
v(\ell) \propto \sigma_v(\ell)\propto \Delta v(\ell) \propto \ell^q 
\label{LWS}
\end{equation}
with $q =(n-3)/2$.  Sometimes indices $n'$ of one-dimensional
(angle-averaged), rather than three-dimensional, power spectra are
reported; these are related by $n'=n-2$.

The turbulent scaling relations reflect the basic physics governing
the flow.  The classical theory of \citet{kol41} applies to {\it
  incompressible} flow, i.e. one in which the velocities are negligible
compared to the 
thermal speed $\sth=(\pth/\rho)^{1/2}$ (where 
$\rho$ 
is
the density and $\pth$ is the thermal
pressure); $\sth$ is equal to the sound speed 
$\cs=(\gamma \pth/\rho)^{1/2}$ in an isothermal ($\gamma=1$) gas.  
In incompressible flows,
energy is dissipated and turbulent motions are damped only
for scales smaller than the Reynolds scale $\ell_\nu$ at which
the viscous terms in the hydrodynamic equations, $\sim \nu
v(\ell)/\ell^2$, exceed the nonlinear coupling terms between scales,
$\sim v(\ell)^2/\ell$; here $\nu$ is the kinematic viscosity.  At
scales large compared to $\ell_\nu$, and small compared to the
system as a whole, the rate of specific energy transfer $\dot
{\cal E}$ between scales is assumed to be conserved, and equal to the 
dissipation rate at the Reynolds scale.  From dimensional
analysis, $\dot {\cal E} \sim v(\ell)^3/\ell$, which implies
$n=11/3$ and
$q=1/3$ for the so-called ``inertial range'' in Kolmogorov
turbulence.  The Kolmogorov theory includes the exact result that 
$S_3(\ell)= -(4/5)\dot {\cal E} \ell$.

Because velocities $v(\ell)\sim \sigma_v(\ell)\sim \Delta v(\ell)$ in
molecular clouds are in general {\it not} small compared to 
$\cs$,
at
least for sufficently large $\ell$, one cannot expect the Kolmogorov
theory to apply.  In particular, some portion of the energy at a given
scale must be directly dissipated via shocks, rather than cascading
conservatively through intermediate scales until $\ell_\nu$ is
reached.  In the limit of zero pressure, the system would consist of a
network of (overlapping) shocks; this state is often referred to as
Burgers turbulence \citep{2001ntt..conf..341F}.  Since the power
spectrum corresponding to a velocity discontinuity in one dimension
has $P(k)\propto k^{-2}$, an isotropic system of shocks in three
dimensions would also yield power-law scalings for the velocity
correlations, with $n=4$ and $q=1/2$.  
Note that correlations can take
on a power-law form even if there is {\it not} a conservative inertial
cascade; a large range of spatial scales with consistent physics is
still required.

Turbulence in a magnetized system must differ from the unmagnetized
case because of the additional wave families and nonlinear couplings
involved, as well as the additional diffusive processes -- including
resistive and ion-neutral drift terms (see \S \ref{scales}).  When the
magnetic field $B$ is strong, in the sense that the Alfv\'en speed
$\vA\equiv B/\sqrt{4\pi \rho}$ satisfies $\vA\gg
v(\ell)$, a directionality is introduced such that the correlations of
the flow variables may depend differently on $\bf r_\parallel$, $\bf
r_\perp$, $\bf k_\parallel$, and $\bf k_\perp$, the displacement and
wavevector components parallel and perpendicular to 
$\hat {B}$.

For {\it incompressible} magnetohydrodynamic (MHD) 
turbulence, \citet{1995ApJ...438..763G}
introduced the idea of a critically-balanced anisotropic cascade, in
which the nonlinear mixing time perpendicular to the magnetic field
and the propagation time along the magnetic field remain comparable for
wavepackets at all scales, so that $\vA k_\parallel \sim
v(k_\perp,k_\parallel) k_\perp$.  Interactions between
oppositely-directed Alfv\'en wavepackets travelling along magnetic
fields cannot change their parallel wavenumbers 
$k_\parallel = {\bf k}\cdot \hat{\bf B}$, 
so that the energy transfers produced by these
collisions involve primarily $k_\perp$; i.e. the cascade is
through spatial scales $\ell_\perp= 2 \pi /k_\perp$ with
$v(\ell_\perp)^3/\ell_\perp\sim$ constant.  Combining critical balance
with a perpendicular cascade yields anisotropic power spectra (larger 
in the $k_\perp$ direction); at a given level of power, the theory
predicts $k_\parallel \propto k_\perp^{2/3}$.  Magnetic fields and
velocities are predicted to have the same power spectra.

Unfortunately, for the case of strong compressibility ($\cs \ll v$) and
moderate or strong magnetic fields 
($\cs \ll \vA \simlt v$),
which
generally applies within molecular clouds, there is as yet no simple
conceptual theory to characterize the energy transfer between scales
and to describe the spatial correlations in the velocity and magnetic
fields.  On global scales, the flow may be dominated by large-scale
(magnetized) shocks which directly transfer energy from macroscopic to
microscopic degrees of freedom.  Even if velocity differences are not
sufficient to induce (magnetized) shocks, for trans-sonic motions
compressibility implies strong coupling among all the MHD wave
families.  On the other hand, within a sufficiently small sub-volume
of a cloud (and away from shock interfaces), velocity differences may
be sufficiently subsonic that the incompressible MHD limit and the 
Alfv\'enic cascade approximately 
hold
locally.

Even without direct energy transfer from large to small scales in
shocks, a key property of turbulence not captured in classical models
is intermittency effects -- the strong (space-time) localization of
dissipation in vortex sheets or filaments, which can occur even with a
conservative energy cascade.  
(Shocks in compressible flows represent a different class of 
intermittent structures.)
Signatures of intermittency are
particularly evident in departures of high-order structure function
exponents from the value $p/3$, and in non-Gaussian tails of
velocity-increment probability distribution functions (PDFs) \citetext{e.g.
\citealp{1997AnRFM..29..435S,1996ApJ...463..623L}}.  Proposed methods to
account for intermittency in predicting correlation functions for
incompressible, unmagnetized turbulence have been discussed by
\citet{1994PhRvL..72..336S} and \citet{1994PhRvL..73..959D}.  
\citet{2002ApJ...569..841B} 
proposed 
an adaptation of this framework for the compressible MHD case, but 
omitted 
direct dissipation of
large-scale modes in shocks.
Research on formal turbulence theory is quite active
\citetext{see \citealp{2004ARA&A..42..211E} for a review of  
the recent theoretical literature relevant to the ISM}, 
although a comprehensive
framework remains elusive.

Large-scale numerical simulations afford a complementary theoretical
approach to model turbulence and to explore the spatial correlations
within flows.  Numerical experiments can be used to test formal
theoretical proposals, and to provide controlled, quantitative means
to interpret observations -- within the context of known physics --
when formal theories are either nonexistant or limited in detail.
In drawing on the results of numerical experiments, it is important to
ensure that the computational techniques employed adequately capture the
relevant dynamical processes.  For systems in which there are steep
gradients of velocities and densities, grid-based methods are more accurate
in following details of the evolving flow 
(such as development of instabilities)
than SPH methods, which have been shown to have difficulty capturing shocks
and other discontinuities \citep{2006astro.ph.10051A}.

Spatial correlations within turbulent flows have been evaluated using
numerical simulations in a variety of regimes.  Overall, results are
consistent with theoretical predictions in that power-law scalings in
the velocity and magnetic field power spectra (or structure functions)
are clear when there is sufficient numerical resolution to separate
driving and dissipative scales.  At resolutions of $512^3$ and above,
results with a variety of numerical methods show angle-averaged
power-law slopes $n=7/2-11/3$ (i.e. $3.5-3.67$) for incompressible 
(i.e. $v/\cs\ll 1$) MHD flows
\citep{2003PhRvE..67f6302M,2004PhRvE..70a6308H,2005PhRvL..95k4502M}
and $n=3.5-4.0$ for strongly compressible ($v/\cs\simgt 5$) flows both
with \citep{2003ApJ...590..858V,2007astro.ph..1795P} and without
\citep{2006ApJ...638L..25K} magnetic fields.  Consistent with
expectations, spectra are steeper for compressive velocity components
than for
 magnetic fields (and also
sheared velocity components if the
magnetic field is moderate or strong), 
and steeper
for more supersonic and/or more weakly magnetized models 
\citetext{see also \citealp{2002ApJ...573..678B,2004PhRvL..92s1102P}}.

When strong mean magnetic fields are present, there is clear
anisotropy in the power spectrum, generally consistent with the
scaling prediction of \citet{1995ApJ...438..763G}, for both
incompressible and compressible MHD turbulence
\citep{2000ApJ...539..273C,2001ApJ...554.1175M,2002ApJ...564..291C,
2003ApJ...590..858V,2003MNRAS.345..325C}.  

In order to identify the sources of turbulence in astronomical
systems, it is also important to determine the behavior of velocity
and magnetic field correlations on spatial scales larger than the
driving scale.  Numerical simulations, both for incompressible
\citep{2001ApJ...554.1175M,2004PhRvE..70a6308H} and compressible MHD
turbulence \citep{2003ApJ...590..858V}, show that the power spectra
below the driving wavenumber scale are nearly flat, $n\approx 0$; that is,
``inverse cascade'' effects are limited.  
For spatially-localized forcing (rather than forcing localized in
$k$-space), 
\citet{Nakamura2007} also found a break in the power spectrum, at
wavelength comparable to the momentum injection scale.
Thus, the
forcing scale for internally-driven turbulence in a system can be
inferred observationally from the peak or knee of the velocity correlation
function. 
If $v(\ell)$ continues to rise up to $\ell\sim L$,
the overall scale of a system, this implies that turbulence is either (i)
externally driven, (ii) imposed in the initial conditions when the system
is formed,
or (iii) driven internally to reach large scales.
Note that for systems forced at multiple scales, or both internally
and externally, breaks may be evident in the velocity correlation
function (or power spectrum).

\subsubsection{TURBULENT  DISSIPATION TIMESCALES}\label{turbdiss}

Recent numerical simulations under quite disparate physical regimes
have reached remarkably similar conclusions for the dissipation rates
of turbulence.  On dimensional grounds, the specific energy 
dissipation rate should
equal $\epsilon U^3/{\ell}_0$, where $E= U^2/2$ is the total
specific kinetic 
energy, ${\ell}_0$ is the spatial wavelength of the main
energy-containing scale (comparable to the driving scale for forced
turbulence; ${\ell}_0\le L$), 
and $\epsilon$ is a dimensionless coefficient.  For
incompressible turbulence, the largest-scale ($4096^3$ zones)
incompressible, unmagnetized, driven-turbulence simulations to date
\citep{2003PhFl...15L..21K} yield a dimensionless dissipation
coefficent $\epsilon=0.6$.  For driven incompressible MHD turbulence
(at $1024^3$ resolution), the measured dimensionless dissipation rate 
$\epsilon \equiv (1/2)(\dot E_{\rm turb}/E_{\rm turb})(\ell_0/U)$
also works out to be $\epsilon=0.6$ \citep{2004PhRvE..70a6308H}.
Quite comparable results also hold for strongly-compressible
($U/\cs=5$) turbulence at a range of magnetizations $\vA/\cs=0-10$;
\citet{1998ApJ...508L..99S} found that $\epsilon=0.6-0.7$ for
simulations at resolution up to $512^3$ zones.  
For decaying
compressible MHD turbulence, damping timescales are also comparable to
the flow crossing time ${\ell}_0/v({\ell}_0)$ on the
energy-containing scale
\citep{1998ApJ...508L..99S,1998PhRvL..80.2754M,1999ApJ...524..169M,1999ApJ...526..279P}.
Thus, although very different physical processes are involved in turbulence
dissipation under different circumstances, the overall damping rates
summed over all available channels (including shock, reconnection, and
shear structures) are nevertheless quite comparable.
Defining the turbulent dissipation timescale as
$t_{\rm diss}=E_{\rm turb}/|\dot E_{\rm turb}|$ and the flow crossing time over
the main energy-containing scale as $t_f={\ell}_0/U$,  
$t_{\rm diss}=t_f/(2\epsilon)$.  Since velocities in 
GMCs
increase up to
the largest scale, $\ell_0\rightarrow d$, the cloud diameter. 
Assuming that on average
$U=\sqrt{3} \sigma_{\rm los}$, the turbulent dissipation time based on
numerical results is
therefore given by
\begin{equation}
t_{\rm diss} \approx 0.5 \frac{d}{\sigma_{\rm los}}.
\end{equation} 
This result is in fact consistent with
the assumption of \citet{1956MNRAS.116..503M} that turbulence in GMCs
would decay within a crossing time.

The above results apply to homogenous, isotropic turbulence, but under
certain circumstances if special symmetries apply, turbulent damping
rates may be lower.  One such case is for {\it incompressible}
turbulence consisting of Alfv\'en waves all propagating in the same
direction along the magnetic field.  Note that for the
incompressibility condition $\nabla \cdot {\bf v}=0$ to apply,
turbulent amplitudes must be quite low ($v\ll \cs$).  Since Alfv\'en
waves are exact solutions of the incompressible MHD equations, no
nonlinear interactions, and hence no turbulent cascade, can develop
if only waves with a single propagation direction are present in this
case \citetext{see e.g.  \citealp{2004Ap&SS.292...17C} for a mathematical and
physical discussion}.  
A less extreme situation is to have an imbalance in the flux of
Alfv\'en waves propagating ``upwards'' and ``downwards'' along a given
magnetic field direction.  \citet{2001ApJ...554.1175M} show that in
decaying incompressible MHD 
turbulence, the power in both upward- and downward-propagating 
components decreases together until the lesser component is depleted.
\citet{2002ApJ...564..291C} 
quantify decay times of imbalanced incompressible 
turbulence, finding for example that if the initial imbalance is 
$\approx 50$\% or $\approx 70\%$, then the time to decay to half the initial
energy is increased by a factor $1.5$ or $2.3$,
respectively, compared to the case of no imbalance.

For even moderate-amplitude subsonic
velocities, however, Alfv\'en waves couple to other wave families and
the purely Alfv\'enic cascade is lost.  For 
strongly 
supersonic
motions, as are present in GMCs, the mode coupling is quite strong.
As a consequence, even a single circularly polarized Alfv\'en wave
cannot propagate without losses; a
parametric instability known as the ``decay instability''
\citep{1969npt..book.....R} develops in which three daughter waves (a
forward-propagating compressive wave and two oppositely propagating
Alfv\'en waves when 
$\beta\ll 1$) 
grow at the expense of the mother
wave.  The initial growth rate of the instability is 
$\gamma = (0.1-0.3)\ k \vA$ 
when $v(k)/\cs=1-3$ and $\beta \equiv 2 \cs^2/\vA^2=0.2$, and
larger for greater amplitudes and smaller $\beta$
\citep{1978ApJ...219..700G}.  The ultimate result is decay into
fully-developed turbulence
\citep{1994JGR....9913351G,2001A&A...367..705D}.  Thus, for conditions
that apply within GMCs, even if there were a localized source of purely
Alfv\'enic waves (i.e. initially 100\% imbalanced), the power would rapidly be 
converted to balanced, broad-spectrum turbulence with a short decay time.
The conclusion that turbulent damping times within GMCs are
expected to be comparable to flow crossing times has important
implications for understanding evolution in star-forming regions; these
will be discussed in \S\S \ref{dynam_state} and \ref{GMC_evol}.

\subsubsection{PHYSICAL SCALES IN TURBULENT FLOWS}\label{scales}

In classical incompressible turbulence, the only physical scales that
enter are the outer scale $\ell_0$ at which the medium is stirred, and the
inner ``Reynolds'' scale $\ell_\nu$ at which viscous dissipation occurs.
Assuming Kolmogorov scaling
$v(\ell)=v(\ell_0)(\ell/\ell_0)^{1/3}$ 
[for $v(\ell) \sim \sigma_v(\ell)\sim \Delta v(\ell)$],
the dissipation scale is 
\begin{equation}
\frac{\ell_\nu}{\ell_0}=\left[\frac{\nu}{\ell_0 v(\ell_0)} \right]^{3/4}\equiv
Re^{-3/4}.
\end{equation}
Here $Re\equiv v(\ell_0) \ell_0 /\nu$ is the overall
Reynolds number of the flow; if turbulence increases up to the largest
scales then $Re=U L/\nu$.  With $\nu\sim \cs \lambda_{\rm mfp}$ for 
$\lambda_{\rm mfp}$ the mean free path for particle collisions,
$\ell_\nu/\ell_0\sim (\lambda_{\rm mfp}/\ell_0)^{3/4}[v(\ell_0)/\cs]^{-3/4}$.  
In fact, the velocity-size scaling within
GMCs has power-law index $q$ closer to $1/2$ than $1/3$ on large
scales, because large-scale velocities are supersonic and therefore
the compressible-turbulence results apply.
Allowing for a transition from
$q=1/2$ to $q=1/3$ at an intermediate scale $\ell_s$ where
$v(\ell_s)=\cs$ (see below),  
$\ell_\nu = \ell_s^{1/4} \lambda_{\rm mfp}^{3/4}$.
Using typical GMC parameters so that $\lambda_{\rm mfp}\sim 10^{13}\cm$
and 
$\ell_s\sim0.03$ pc 
yields 
$\ell_\nu\sim 3\times 10^{-5}\pc$.  This is tiny compared to
the sizes, $\sim 0.1 \pc$,  of self-gravitating cores in which
individual stars form.

The length $\ell_s$ introduced above marks the scale
at which the 
RMS turbulent velocity is equal to the sound speed.
At larger scales, velocities are supersonic and compressions are strong;
at smaller scales, velocities are subsonic and compressions are weak.
Taking $v(\ell)=v(\ell_0) (\ell/\ell_0)^q$, the sonic scale is 
$\ell_s = \ell_0 [\cs/v(\ell_0)]^{1/q}$, or 
$\ell_s\approx \ell_0 [\cs/v(\ell_0)]^{2}$ 
when $q\approx 1/2$.
Density perturbations with
characteristic scales $\sim \ell_s$ will have order-unity amplitude in
an unmagnetized medium.  
In a magnetized medium, the amplitude of the
perturbation imposed by a flow of speed $v$ will depend on the
direction of the flow relative to the magnetic field.  Flows along the
magnetic field will be as for an unmagnetized medium, while flows
perpendicular to the magnetic field will create order-unity
density perturbations only if $v>(\cs^2+\vA^2)^{1/2}$.
Note that the thermal scale, 
at which the line-of-sight turbulent velocity
dispersion $\sigma_v/\sqrt{3}$ is equal to the one-dimensional thermal speed
$\sth$, is larger than $\ell_s$ by a factor $\approx 3$.

Another scale that is important for MHD turbulence in {\it fully-ionized} gas 
is the resistive scale; below this scale Ohmic diffusion would smooth
out strong bends in the magnetic field, or would allow folded field
lines to reconnect.  The resistive scale $\ell_\eta$ is estimated by
equating the diffusion term $\sim \eta B(\ell)/(4\pi \ell^2) $ to the flux-dragging
term $\sim v(\ell) B(\ell)/\ell$ in the
magnetic induction equation.  Defining the magnetic Reynolds number as
$Rm\equiv v(\ell_0) \ell_0 4\pi/\eta,$
and taking 
$v(\ell)\sim \cs (\ell/\ell_s)^{1/3}$ at small scales, this yields 
$\ell_\eta/\ell_\nu =(Re/Rm)^{3/4}$.
Since the ``magnetic Prandtl
number'' $Rm/Re$ is very large ($\sim 10^{6}$), this means that the
magnetic field could, 
for a highly-ionized medium, 
remain structured at quite small
scales \citetext{see \citealp{2002ApJ...566L..49C} for discussion of this in the
diffuse ISM}.

In fact, under the {\it weakly-ionized} conditions in star-forming regions, 
ambipolar diffusion (ion-neutral drift) becomes important
well before the resistive (or Ohmic diffusion) scale is reached.
Physically, the characteristic ambipolar diffusion scale $\ell_{AD}$
is the smallest scale for which the magnetic field (which is frozen to
the ions) is well-coupled to
the bulk of the gas, for a partially-ionized medium.  An estimate of
$\ell_{AD}$ is obtained by equating the ion-neutral drift speed, $\sim B_0
\delta B (m_i+m)/(4 \pi \rho_i \rho \alpha_{in} \ell)$, with the
turbulent velocity, $\delta v$. 
Here,
$\alpha_{in}=\langle \sigma_{in} |v_i -v_n|\rangle \approx 2 \times 10^{-9}
\;
{\rm cm}^3 {\rm s}^{-1}$ is the ion-neutral collision rate coefficient
\citep{1983ApJ...264..485D}, and $m_i, m$ and $\rho_i, \rho$ are
the ion and neutral mass and density.
The resulting ambipolar diffusion scale, assuming $m_i\gg m$ (for
either metal or molecular cations) and
$\delta v\approx \delta B/\sqrt{4\pi\rho}$, is
\begin{equation}
  \ell_{AD}= \frac{\vA}{n_i \alpha_{in}}
  \approx 0.05 \pc\left(\frac{\vA }{3\;\kms } \right) 
  \left(\frac{n_i }{10^{-3}\ \cmt } \right)^{-1}. 
\label{amb}
\end{equation}
Here, $\vA$ is the Alfv\'en speed associated with the large-scale
magnetic field $B_0$.  

The ambipolar diffusion scale (\ref{amb})
depends critically on the fractional ionization, which varies greatly
within star-forming regions.  Regions with moderate $A_V\simlt 5$ can
have relatively large ionization fraction due to UV photoionization,
whereas regions with large $A_V$ are ionized primarily by cosmic rays
(see \S\ref{ionize}).
For example, if the electrons are attached to PAHs in dense cores, 
the ion density is
$n_i\approx 10^{-3} (\nh/10^4
\cmt)^{1/2}(\zeta_{\rm CR}/3\times 10^{-17} {\rm s}^{-1})^{1/2}$
\citep{2005pcim.book.....T}
where $\zeta_{\rm CR}$ is the cosmic ray ionization rate per H atom..
Since $n_i\propto
n^{1/2}$, we can express equation (\ref{amb}) in terms of column density
and magnetic field strength as $\ell_{AD}/\ell = 0.09 (B/10\mu{\rm
G})/(N_{\rm H}/10^{21}\; {\rm cm}^{-2})$.

For spatial wavelengths
$\lambda=2\pi/k<\pi\ell_{AD}$, MHD waves are unable to propagate 
in the coupled neutral-ion fluid
at all, because the collision frequency of neutrals with ions, $n_i
\alpha_{in}$, is less than (half) the wave frequency $\omega=k \vA
$.  For $\lambda>\pi\ell_{AD}$, MHD waves are damped at a rate $\omega \pi
\ell_{AD}/\lambda$ \citep{1969ApJ...156..445K}.  Thus, at scales
$\ell\simlt \ell_{AD}$, 
the magnetic field will be essentially straight and uniform in
magnitude, 
and any further turbulent cascade
will be as for an unmagnetized medium.  The scale $\ell_{AD}$ is also
comparable to the thickness of the $C$-type shocks that are typical
under prevailing conditions within GMCs \citep{1993ARA&A..31..373D}.
Further discussion of the interaction between turbulence and ambipolar
diffusion is given by 
\citet{2002ApJ...567..962Z}, \citet{2002ApJ...570..210F},
and \citet{2004ApJ...603..165H}.

\subsubsection{DENSITY STRUCTURE IMPOSED BY
  TURBULENCE}\label{densstruct}

When turbulent velocities at a given scale are supersonic, they impose
density variations within the flow at that scale.  For star-forming
regions, in which turbulent velocities are increasingly supersonic for
scales $\simgt 0.1\pc$, the density becomes strongly structured over a
wide range of scales.  This density structure -- which is crucial to
the star-formation process -- can be characterized statistically
in a variety of ways.  

The simplest (one-point) statistic is the distribution of mass (or
volume) as a function of density, usually referred to as the density
PDF (probability density function).  For isothermal gas and supersonic
turbulence (either forced or decaying), a number of 3D numerical
simulations both with
\citep{2001ApJ...546..980O,2003LNP...614..252O,2004ApJ...605..800L}
and without \citep{1999intu.conf..218N,2000ApJ...535..869K} magnetic
fields have shown that the density PDF approaches a log-normal
distribution when self-gravity is unimportant.  This functional form
can be understood \citep{1994ApJ...423..681V,1998PhRvE..58.4501P} 
to arise as a
consequence of multiple, independent dynamical events that alter the
density according to 
$\rho/\bar\rho = \Pi_i (1+\delta_i)$ 
where $\delta_i$ is
$>0$ (or $<0$) for compressions (or rarefactions).  From the Central
Limit Theorem, $\log(\rho/\bar\rho)$ is the sum of independent random
variables, and should therefore approach a Gaussian distribution.
When the equation of state departs from a simple isothermal form, the
density PDF still follows a log-normal over a range of densities, but
aquires power-law tails either at high or low density depending on
whether the equation of state is softer or stiffer than isothermal
\citetext{\citealp{1998PhRvE..58.4501P,1998ApJ...504..835S}; see also e.g.
\citealp{2001ApJ...559L..41W}}.

For a log-normal distribution, the fraction of volume 
($V$) or mass ($M$)
as a function of $x\equiv \ln(\rho/\bar\rho)$ is given by $f(x)dx$ with
\begin{equation}
f_{V,M}= \frac{1}{\sqrt{2 \pi \sigma_x^2}}
\exp\left[\frac{-(x\pm|\mu_x|)^2}{2\sigma_x^2}\right]
\label{log-norm}
\end{equation}
where the mean and dispersion of the distributions are related by
$\mu_x=\sigma_x^2/2$, and the upper and lower signs correspond to
volume- and mass-weighting, respectively.  
For a log-normal
distribution, the mass-weighted
median density (half of the mass is at densities above
and below this value) is $\rho_{\rm med}= \bar\rho \exp(\mu_x)$,
whereas the mass-weighted mean density is 
$\langle\rho\rangle_M=\bar \rho \exp(2\mu_x)$.
Based on three-dimensional unmagnetized simulations,
\citet{1997ApJ...474..730P} propose that $\mu_x\approx 0.5 \ln(1+0.25{\cal
  M}^2)$.  
Other three-dimensional simulations with magnetic fields ($\beta=0.02-2$)
 have found $\mu_x\approx 0.5-1$ for ${\cal M}\approx 5-10$ 
\citep{2001ApJ...546..980O}.  These models confirm that the
mean density contrast generally grows as the turbulence level
increases,  but find no one-to-one relationship
between $\mu_x$ and $\cal M$ (or the
fast magnetosonic Mach number, ${\cal M}_F\equiv \sigma_v/(\cs^2+\vA^2)^{1/2}$).
The large scatter at large $\cal M$ is  because the flow is
dominated by a small number of large-amplitude 
modes (i.e. large ``cosmic variance''), some of which are compressive
and some of which are shear.
With magnetic fields, \citet{2001ApJ...546..980O} found that 
the lower envelope of the $\mu_x$ distribution 
increases with ${\cal M}_F$ according to 
$\mu_{x, \rm min}
= 0.2\ln(1+{\cal M}_F^2)+0.5$ for ${\cal M}_F=0.5-2.5$. 

Because the velocity field is spatially correlated, the density
distribution will also show spatial correlations over a range of scales.
Density correlations can be characterized in terms of the
autocorrelation function, the power spectrum, and structure functions
of various orders (cf \S \ref{turb_correl}); usually, analyses are
applied to $\delta \rho \equiv \rho -\bar \rho$.
Using delta-variance techniques, \citet{2000A&A...353..339M} show that
correlations in density decrease for wavelengths above the velocity driving
scale, and that there are relatively modest differences in the density
correlations between unmagnetized and magnetized models when all other
properties are controlled.

\citet{2005ApJ...630L..45K} have analyzed the dependence of the
spectral index on Mach number for three-dimensional turbulence forced
at large spatial scales, using isothermal, unmagnetized simulations at
resolution $512^3$.  For ${\cal M}\simlt 1$, the 
indices 
$n_\rho$ or $n_\rho'$
of the density
power spectrum $|\delta \rho(k)|^2$ 
are similar to those 
of the velocity
field in incompressible turbulence -- i.e. near $n=11/3$ or $n'=5/3$;
this is simply because $\delta \rho({\bf k})/\bar\rho \sim - \hat k
\cdot {\bf v}({\bf k})/\cs$ for low-amplitude quasi-sonic compressions
(note that even when ${\cal M}=1$, the Mach number for the
compressive component of the velocity field is $<1$).
As the Mach number
increases, the density power spectrum flattens, reaching 
$n_\rho'\approx 0.5$ for
${\cal M}=12$.  For comparison, a one-dimensional top hat --
corresponding to a large clump in 3D -- would have 
$n_\rho'=2$, 
whereas a 
one-dimensional
delta function -- corresponding to a thin sheet or filament in 3D -- 
would have 
$n_\rho'=0$.  
Note that for the density to take the form of multiple
delta functions, the velocity field
must generally be a
composite of step functions -- corresponding to shocks -- 
and has $n'=2$ for
the velocity power spectrum (as discussed above).
The low value of 
$n_\rho'$ 
at large Mach number 
implies the density structure becomes
dominated by curved sheets and filaments.  Curved sheets represent
stagnation regions (of the compressive velocity field) where shocked
gas from colliding flows settles, and filaments mark the intersections
of these curved sheets.

Other statistical descriptions of density structure include fractal
dimensions 
(e.g.,
\citealp{1996ApJ...471..816E,1998A&A...336..697S}),
multifractal spectra \citep{2001ApJ...551..712C}, and hierarchical
structure trees \citep{1992ApJ...393..172H}; see
\citet{2004ARA&A..42..211E} for a discussion.
The spatial correlation of density can also be characterized in terms
of clump mass functions.  Clump-finding techniques have been applied
to simulations of supersonic tubulent flows by a number of groups;
these results will be discussed and compared to observations in \S\ref{clumps}.

\subsubsection{OBSERVATIONS OF TURBULENCE}
\label{turbobs}

For observed astrophysical systems, the intrinsic properties of
turbulence cannot be directly obtained, due to line-of-sight
projection and the convolution of density and velocity in producing
observed emission.  A number of different techniques have been
developed, calibrated using simulations, and applied to observed data,
in order to deduce characteristics of the three-dimensional turbulent
flow from the available observations, which include spectral line data
cubes (from molecular transitions), continuum emission maps (from
dust), maps of extinction (using background stars), and maps of
polarization (in extinction and emission from dust).
\citet{2004ARA&A..42..211E} review the extensive literature on
observations of turbulence.  Here, we will mention just a few results.

The defining property of turbulent motion -- in contrast to, for
example, the purely random motions of gas particles in a
Maxwell-Boltzmann distribution or the highly systematic motions of
stars in a rotating system -- is the stochastic yet scale-dependent
behavior of flow correlations.  \citet{1981MNRAS.194..809L} was the
first to draw attention to the genuine ``turbulent'' nature of motions
internal to star-forming regions, as expressed by an empirical scaling
law of the form (\ref{LWS}) with $q=0.38$.  Using more homogeneous
data, \citet{1987ApJ...319..730S} obtained a ``linewidth-size''
scaling index $q\approx 0.5$ for GMCs as a whole.  
\citet{1988A&A...197..228P} pointed out that the linewidth-size
scaling $\sigma_v(\ell)\propto \ell^{1/2}$ observed in star-forming regions
is indeed what would be predicted for ``Burgers
turbulence,'' a more appropriate model than Kolmogorov turbulence
given the strongly supersonic conditions.

Many subsequent studies have been made of observed scaling behavior of
velocities, both for subsystems of a given star-forming region, and
for systems that are spatially disjoint.  A number of methods have
been developed for these investigations, including autocorrelation
analysis \citep{1994ApJ...429..645M} and delta-variance analysis
\citep{2006A&A...452..223O} applied to line centroids, the spectral
correlation function \citep{1999ApJ...524..887R}, velocity channel
analysis \citep{2004ApJ...616..943L,2006ApJ...653L.125P}, 
and principal component analysis
(PCA) \citep{2002ApJ...566..276B}.  
Overall,
analyses agree in finding power-law linewidth-size relations,
with similar coefficients and power-law expononents close to $q=0.5$.
The lack of features in velocity correlations at intermediate scales,
and more generally the 
secular increase in velocity dispersion up to sizes comparable to
the whole of a GMC, indicates that turbulence is driven on large
scales within or external to GMCs 
\citetext{e.g. \citealp{2002A&A...390..307O,2003ApJ...583..280B}}.

Interestingly, turbulence appears to have a ``universal'' character
within most of the molecular gas in
the Milky Way, in the sense that the same scaling laws with the
same coeffcients fit both entire GMCs and 
moderate-density substructures (observed via CO lines) 
within them.
Using PCA, \citet{2004ApJ...615L..45H} find a fit to the 
amplitudes of line-of-sight 
velocity components as a function of scale following
\begin{equation}
\delta v= 0.9  \left(\frac{L_{\rm pca} }{1\, \rm pc}\right)^{0.56\pm 0.02}~~~\kms
\label{GMC_lws}
\end{equation}
based on composite data of all PCA components from scales 
$L_{\rm pca}\sim 0.03-30$~pc in a sample of 27 
molecular clouds.  
Using data just {\it within} individual
clouds, \citet{2004ApJ...615L..45H} find a mean scaling exponent that 
is slightly lower, $q=0.49\pm 0.15$. 
Note that the lengths $L_{\rm pca}$ entering the relation
(\ref{GMC_lws}) are the 
characteristic scales of PCA eigenmodes, and may differ from size
scales defined in other ways.  
For example, the effective
GMC cloud diameters as measured by \citet{1987ApJ...319..730S} are
on average about four times 
the maximum $L_{\rm pca}$ found in each cloud.
Based on the scaling law (\ref{GMC_lws}) the sonic length will be similar, 
$\ell_s \sim 0.03\pc$ (allowing for varying
definitions of size),  in all GMCs.
We discuss this empirical result further in \S \ref{dynam_state};
note that strongly self-gravitating clumps with high densities and surface
densities depart from the relation given in equation 
(\ref{GMC_lws}).

For evaluating the density distribution, the most unbiased
measurements use dust extinction maps 
\citetext{see \citealp{2007prpl.conf....3L}
and references therein}.  
A promising new technique for observing the density distribution uses
scattered infrared light \citep{2006ApJ...636L.105F}, 
which can probe the structure of molecular clouds for visual extinctions of
$1-20$ magnitudes at very high spatial resolution \citep{2006ApJ...636L.101P}.
Consistent with the prediction of numerical
simulations \citep{ 2001ApJ...546..980O,2001ApJ...557..727V},
distributions of extinction follow log-normal functional forms to an
excellent approximation; distributions of integrated intensity
from molecular lines, on the other hand, 
are not log-normal \citep{2006AJ....131.2921R}, presumably due to
a combination of chemistry and/or optical depth effects.  
Column
density distributions of course cannot be directly inverted to obtain
volume density distributions.  Since the Fourier transform of the
column density, $N(k_x,k_y)$, is equal to $\delta\rho (k_x,k_y;k_z=0)$ (up to
an overall normalization; here $z$ is the line-of-sight 
direction), if
statistical isotropy holds 
then at
least the shape of the density power spectrum can be obtained from a
well-sampled map of column density.  Assuming isotropy,
integrated-intensity $^{12}$CO and $^{13}$CO line maps yield density
power 
spectra 
$|\delta \rho(k)|^2 \propto 
k^{-n_\rho}$ with $n_\rho
=2.5-2.8$
\citep{2001A&A...366..636B}, which is consistent with the large-$\cal
M$ results for density power spectra 
$n_\rho'=n_\rho
-2\approx 0.5$ obtained in
the simulations of \citet{2005ApJ...630L..45K}.  
In principle, features in the density
power spectrum should be evident
both at the sonic scale, $\ell_s$, and the Jeans scale
(these scales are comparable in star-forming 
regions -- \citealp{1995MNRAS.277..377P}). 
A first step toward identifying features in the density power spectrum, 
using velocity-integrated CO intensity, was 
taken by \citet{1997ApJ...488L.145B}.
It will be very interesting to investigate column density power spectra
based on high-resolution extinction maps in both self-gravitating GMCs
and unbound molecular clouds, evaluating the slopes and searching for
evidence of these breaks.

Other measures of density structure, including the fractal dimension
$D\approx 2.3$ empirically measured by \citet{1996ApJ...471..816E},
are in agreement with simulations of strongly compressible turbulence
\citep{2006ApJ...638L..25K}.
In addition, the typical range of density contrasts obtained for 3D
supersonic 
turbulence 
(see \S \ref{densstruct}) is consistent with the
compressions required 
to explain the low effective volume filling factors of gas 
deduced from CO observations of GMCs
\citep{1999ApJ...513..259O}.  
If 
$\langle\ln(n/\bar n)\rangle_M=\mu_x$
(eq. \ref{log-norm}),
then the mass-weighted mean
density is $\langle n/\bar n \rangle_M=
\langle (n/\bar n)^2 \rangle_V=\exp(2 \mu_x)$, so
that $\mu_x\approx 1.5$ yields a local density $n=10^3$ when
$\bar n=50$, in agreement with 
inferred filling factors $\sim 0.1$  
\citep{1987ApJ...312L..45B,1995ApJ...451..252W}.  
Observational estimates   
of the density and filling factor are often 
made under the assumption of a constant 
clump density, however, not the broad distribution of densities expected for
a log normal, which may introduce some differences.
In more detail, 
\citet{1999ApJ...525..318P} have shown
that the statistical properties of $^{13}$CO spectra seen in the
star-forming Perseus molecular cloud can be well reproduced by
synthetic non-LTE spectra created using simulation data cubes from
non-self-gravitating supersonic turbulence simulations.

\subsection{Self-Gravity}
\label{gravity}

The effects of self-gravity on a turbulent cloud can be analyzed
with the aid of the virial theorem, which in Lagrangian form (i.e.,
for a fixed mass) is 
\beq
\frac 12 \ddot I =2(\calt-\calt_s)+\calb+\calw
\label{eq:iddot}
\eeq
\citep{1953ApJ...118..116C,1956MNRAS.116..503M},
where $I=\int r^2 dm$ is proportional to the trace of
the inertia tensor of the cloud. \citetext{It is often assumed that
the sign of $\ddot I$ determines whether the cloud is expanding or contracting,
but in fact it determines the acceleration of the expansion or 
contraction---\citealp{2006MNRAS.372..443B}.} 
The term
\beq
\calt=\int_\vcl\left(\frac 32 \pth+\frac 12 \rho v^2\right)\;
	dV\equiv \frac 32 \bar \rho
{\sigma^2} \vcl
\label{eq:calt}
\eeq
is the total kinetic
energy in the cloud (thermal plus bulk), where 
$\sigma^2$
is the 1D mean square velocity 
(including both thermal $\sth^2$ and nonthermal [turbulent] 
$\snt^2\equiv \sigma_v^2/3$ terms)
in the cloud,
$\vcl$ is the volume of the cloud, and
$\calt_s=\oint \pth{\bf r}\cdot {\bf dS}$ 
is the surface kinetic term. The term
\beq
\calb=\frac{1}{8\pi}\int_\vcl B^2 dV + 
\frac{1}{4\pi}\oint {\bf r}\cdot ({\bf B }{\bf B}-\frac{1}{2} B^2 {\bf I})
\cdot {\bf dS}
\eeq 
is the net magnetic energy, including the effects of the
distortion of the field
outside the cloud. 
The volume and 
surface magnetic terms cancel for a completely
uniform magnetic field, since a uniform field exerts no force. 
Finally, 
\beq
\calw=-\int \rho {\bf r}\cdot \nabla \Psi\, dV
\eeq
is the gravity term, equal to the gravitational self-energy
$(1/2)\int \rho \Psi \, dV$
provided the acceleration due
to masses outside the system is negligible
\citetext{as is generally the case for 
dense clouds embedded in a diffuse
turbulent background---\citealp{2006astro.ph..7362D}}.

The virial theorem can also be written in Eulerian form, so that
it applies to a fixed volume \citep{1992ApJ...399..551M}; 
in that
case, the surface term for the kinetic energy includes the dynamic
pressure 
$\oint \rho\vecv\vecv\cdot {\bf dS}$, 
and the theorem itself includes a term
$
(1/2)(d/dt) \int(\rho \vecv r^2)\cdot {\bf dS}$ 
on the left-hand side. 
This form of the virial theorem
is particularly appropriate in a turbulent medium, in which the 
mass of a cloud is not necessarily fixed.
With this form of the equation, clouds that are actively forming or 
dispersing may have surface kinetic terms comparable to the volume 
kinetic terms.  

\citet{1999ApJ...515..286B} examined the
various terms in the Eulerian virial theorem  for
clumps and cores in their turbulent, self-gravitating MHD simulations,
and found that the 
time-dependent terms on the left-hand side are significant, and that the
surface terms are generally comparable to the
volume terms.  
\citet{2006astro.ph..7362D} confirmed this
and also found that only objects with large density contrasts are
virialized.
However, in contrast to the other terms in the virial theorem, the 
time-dependent
terms can change sign, so they become less important if the virial
theorem is averaged over time for an individual cloud (provided the
cloud lives more than a dynamical time) or if it is averaged over an
ensemble of clouds \citep{1999osps.conf...29M}.  In \S
\ref{dynam_state}, we discuss the application of the virial theorem
both to observed molecular clouds, clumps, and cores, and to
condensations identified within numerical simulations.

The virial parameter is defined to be proportional to the ratio of the
total kinetic energy to the gravitational energy
\citetext{\citealp{1992ApJ...395..140B}; cf. \citealp{1992A&A...257..715F}},
\beq
\avir\equiv\frac{5
{\sigma^2}R}{GM},
\label{alpha_vir_def}
\eeq
where 
the numerical coefficient is chosen so that
$\avir=1$ for a uniform, unmagnetized
gas sphere
in virial balance ($\calw=-2\calt$; but note that such a sphere is not in hydrostatic equilibrium).
This relation implies that the
mean pressure in a cloud 
varies as
the square of the 
mean surface density,
$\bar \Sigma$,
\beq
\bar P_{\rm tot}
=\phi_{\bar P}G\bar \Sigma^2,
\label{eq:pressure}
\eeq
where $\phi_{\bar P}\propto\avir$ is a numerical
factor of order unity for gravitationally bound objects 
\citep{2003ApJ...585..850M}. The total pressure includes
the magnetic pressure; fluctuating magnetic fields have an energy 
that is about 60\% of that of the turbulent 
kinetic energy \citep{1998ApJ...508L..99S} and contribute 
an effective pressure support 
that is about 30\% of the turbulent kinetic pressure support
\citep{2003ApJ...585..850M} (note that $\cal T-T_{\rm s}$ and $\cal B$
appear in equation \ref{eq:iddot} with coefficients 2 and 1, respectively).
Gravitationally bound
objects have $\avir\sim 1$, which 
(since $M~\sim\rho R^3$) defines gravitational
length, time, and mass scales, 
\beq
R_G=
\sigma
/(G\rho)^{1/2},~~~
t_G=1/(G\rho)^{1/2},~~~
M_G=
\sigma
^3/(G^3\rho)^{1/2}.
\label{eq:gravscales}
\eeq
These scales, derived
essentially from dimensional analysis, govern the structure 
and stability of self-gravitating clouds 
(the density 
$\rho$ can be chosen to be the central density, the mean density, or
the density at the surface, depending on the application).
The gravitational time scale is often expressed in terms of the free-fall time,
which is the time for a pressure-free, 
spherical
cloud to collapse to a point
due to its self-gravity,
\beq
t_{\rm ff}=\left(\frac{3\pi}{32G
\bar
\rho}\right)^{1/2}=1.37\times 10^6
\left(\frac{10^3~\mbox{cm\eee}}{
\bar
n
_{\rm H}}\right)^{1/2}
~~~\mbox{yr},
\label{eq:tff}
\eeq
where the numerical value is based on a He abundance of 10\% by number.

The simplest case of a self-gravitating cloud is a static
isothermal cloud with no magnetic field. For a given surface
pressure $P_{\rm th,\,0}=\rho_0\sth^2$, the critical
mass
$\mcr$
---i.e., the maximum mass for such a cloud 
to be in hydrostatic equilibrium (stable or unstable)---is the
Bonnor-Ebert mass \citep{1956MNRAS.116..351B, 1957ZA.....42..263E},
\beq
\mbe
\equiv
1.182\;\frac{\sth^4}{(G^3\ptho)^{1/2}}
=1.182\;\frac{\sth^3}{(G^3\rho_0)^{1/2}}.
\label{MBE_def}
\eeq
For conditions typical of dense clumps within low-mass star-forming regions,
this is of order a solar mass:
$\mbe=0.66 (T/10~\mbox{K})^{2}/
(\pth/3\times 10^5k_{\rm B}\;\mbox{cm\eee K})^{1/2}\;M_\odot$,
where $k_{\rm B}$ is Boltzmann's constant
and the pressure is normalized to the mean kinetic pressure in
a typical GMC (\S \ref{dynam_state})
(which is similar to the mean thermal pressure in dense clumps).
Note that the Bonnor-Ebert mass is very nearly equal
to the characteristic gravitational mass $M_G(\rho_0)$,
when evaluated with
the conditions at the surface of the cloud.
The radius of a Bonnor-Ebert sphere is $R_{\rm BE}=0.486 \sth/(G\rho_0)^{1/2}
=0.486R_G(\rho_0)$; this is comparable to the Jeans length (see below).

The importance of the magnetic field to cloud structure is
determined by the ratio of the mass to the magnetic critical mass $M_{\Phi}$,
which is defined by the condition that the magnetic energy equal the
gravitational energy, 
$\calb=|\calw|$, 
for a cold cloud in magnetostatic 
equilibrium:
\beq
M_{\Phi}\equiv c_\Phi\,\frac{\Phi}{G^{1/2}},
\label{eq:mphi}
\eeq
where $\Phi$ is the magnetic flux threading the cloud
(e.g., see the review by \citealp{1993prpl.conf..327M}).
Magnetic fields alone cannot prevent gravitational collapse
in {\it magnetically supercritical} clouds ($M>M_\Phi$), whereas
gravitational collapse is not possible in {\it magnetically subcritical}
clouds ($M<M_\Phi$); keep in mind, however, that
$M$ can change as the result of flows along the field,
and $M_\Phi$ can change due to ambipolar diffusion.
The numerical coefficient $c_\Phi$ depends on the internal distribution of density and
magnetic fields.
A cold cloud with a poloidal field and a
constant mass-to-flux ratio has
$c_\Phi=0.17$ \citep{1988ApJ...335..239T}, essentially
identical to the critical value of the mass-to-flux ratio for an 
infinite cold sheet,
$G^{1/2}(\Sigma/B)_{\rm cr}=1/(2\pi)\simeq 0.16$ \citep{1978PASJ...30..671N}.
For clouds with two other distributions of the mass-to-flux ratio,
\citet{1988ApJ...335..239T} found that the critical mass-to-flux ratio
for the central flux tube corresponds to $c_\Phi\simeq 0.17-0.18$.
For more complex field geometries, the magnetic flux does not determine the
mass that can be supported by magnetic stresses;
for example, if the field is poloidal, with half the
field pointing one way and half the other, so that the total flux is
zero, the 
mass that could be supported would initially be
$\mphi$, 
but it would go to zero as the
field reconnects. 
For a random field, arguments based on
\citet{1999ApJ...522..313M} suggest that the 
mass that can be supported by magnetic fields is
comparable to that in equation (\ref{eq:mphi}), but with $\Phi$
replaced by $\pi R^2 
\avg{B^2}^{1/2}
$.
Of course, when turbulent magnetic fields are present so are turbulent 
velocities, which lend their own support to the cloud (see below).

The magnetic critical mass can also
be expressed in terms of the mean density and magnetic field
in the cloud \citep{1976ApJ...210..326M},
\beq
\frac{M_B}{M}\equiv\left(\frac{M_\Phi}{M}\right)^3.
\label{eq:mb}
\eeq
For an ellipsoidal cloud of size $2Z$ along the axis
of symmetry and radius $R$ normal to the axis, this
becomes \citep{1992ApJ...395..140B}
\beq
M_B= 79 c_\Phi^3 \left(\frac{R}{Z}\right)^2 \frac{\bar v_A^3}{(G^3\bar\rho)^{1/2}}
=
1020
\left(\frac{R}{Z}\right)^2 \left(\frac{\bar B}{30\,\mu{\rm G}}\right)^3
\left(\frac{10^3~\mbox{cm\eee}}{\bar\nh}\right)^2~~~M_\odot,
\label{MB_eq}
\eeq
where the latter expression uses $c_\Phi=1/(2\pi)$.
Note that $M_B$ has the same form as the gravitational
mass $M_G$, with the velocity dispersion
$\sigma$ 
replaced
by the Alfv\'en velocity $\bar v_A\equiv \bar B/(4\pi\bar\rho)^{1/2}$.
Based on the idea that cores form from sheets that are supported by
kinetic pressure along the magnetic field and magnetic tension in the
(perpendicular) plane,
\citet{2004ApJ...601..930S} have introduced another mass scale,
$M_0
\equiv\pi^2\sigma^4/(G^{3/2}\bar B)$,
which yields values $\sim M_\odot$ when $\sigma \rightarrow \sth$
and $\bar B\rightarrow 30\,\mu$G.

Just as in the case of stellar structure, it is useful to
consider polytropic models of molecular clouds, in which
the pressure is a power-law function of the density,
\beq
P(r)=K\rho(r)^\gp,
\eeq
where $K$ is constant and $\gamma_p$ is often written
as $1+1/n$. Here, $P(r)$ and $\rho(r)$ represent the total pressure
and density averaged over the surface of a sphere of radius $r$.  
This approach is  based on the microturbulent approximation, in which
the turbulent pressure 
$\rho\snt^2$ is included in the total pressure
\citep{1951ProcRoySocA...210..18,1951ProcRoySocA...210..26};
this is equivalent to
assuming that the
random dynamical motions are isotropic.
For a given cloud at a given time, this is reasonable
for small-scale motions, but the approximation
becomes worse as the scale of the motion becomes
comparable to the scale on which the pressure is
being evaluated.  
However, just as in the case of the virial theorem, the microturbulent
approximation becomes better -- for objects that live more than
a dynamical time -- if a time average is taken.
Polytropes are spherical, so polytropic models apply only to objects with
well-defined centers; for such objects, an angular average is also necessary,
which improves the accuracy of the microturbulent approximation.
Star-forming clumps and cores often appear centrally concentrated and
are therefore suitable for modeling with a polytrope, whereas 
many GMCs do not appear to have well-defined centers and
are not very suitable for polytropic models.

For a polytrope, the velocity dispersion obeys
$\sigma^2=P/\rho\propto \rho^{\gp-1}$. If the mean density decreases
with increasing scale
(as it does for an object in hydrostatic equilibrium),
it follows that 
the velocity dispersion increases with scale for
$\gp<1$, which is consistent with observations of molecular clouds 
\citep{1988ApJ...334..761M}. 
[Because $n=1/(\gp-1)$ is negative in this case, 
such polytropes are often referred
to as ``negative-index polytropes.''] 
The stability of a polytrope depends on 
both $\gp$ and on its adiabatic index
$\gamma$, which describes the change in the pressure associated
with a given perturbation in density, $\delta\ln P=\gamma\delta\ln\rho$.
The value $\gamma=\frac 43$ is critical for spherical clouds: 
clouds with $\gamma>\frac 43$ are gravitationally stable for
arbitrarily large masses,
whereas those with $\gamma<\frac 43$ are unstable for sufficiently
large masses, or, at fixed mass, for sufficiently high ambient
pressures. Correspondingly,
the gravitational mass $M_G$ is independent
of density for $\gamma=\frac 43$. 
Polytropes with $\gp< \frac 65$ must be confined by an ambient
pressure \citep{1939isss.book.....C}, and their properties 
are determined by the pressure
of the ambient medium.
\citet{1999ApJ...522..313M} show that polytropes
with $0<\gp\leq 1$ have masses $\leq 1.182 M_G(\rho_0)$; 
the mean density and pressure
of these polytropes are $<3.8$ times the surface values.

As discussed in \S\ref{turbobs}, turbulent regions exhibit a
line width--size 
relation in which the velocity dispersion averaged
over a volume increases systematically with size scale, $\snt\propto
r^q$.   Observations often show $q\simeq 1/2 $,
the value expected for Burgers turbulence (see \S \ref{turb_correl}).
In general, this 
line width--size
relation reflects the statistical 
increase in velocity differences with separation
between two points, rather than the absolute increase in the local turbulent
velocity amplitude with distance from a common center.  
If the medium is gravitationally stratified, however, 
the central point has a physical significance, 
and it is not currently known whether 
in this situation $q$ varies significantly (locally or globally) 
from its value in a non-stratified medium.
Observations of individual low-mass cores indicate increasing 
linewidths away from the centers, with $q\simeq \frac 12$ on 
large scales \citep{1998ApJ...504..223G};
similar observations for star-forming clumps or high-mass cores, which are 
supersonically turbulent, are not yet available.
In polytropic models with 
$\snt\propto \rho^{(\gp-1)/2}\propto r^q$, 
the density follows 
a power-law in radius, $\rho\propto r^{-\krho}$, with $\krho=2q/(1-\gp)$.
In hydrostatic equilibrium, $\krho=2/(2-\gp)$ must hold, so that 
$q=(1-\gp)/(2-\gp)$; the value $q=1/2$ thus corresponds
to $\gp\rightarrow 0$. This has motivated the study of
equations of state for turbulent gas that include a pressure proportional
to the logarithm of the density, so-called ``logatropes''
\citep{1989ApJ...342..834L, 1996ApJ...457..718G}.
\citet{1996ApJ...469..194M} pointed out a difficulty with
previous logatropic models and developed a variant that
overcame this problem; however, 
their model
leads
to line widths that actually decrease near the edge of the cloud
\citep{2003ApJ...585..850M}. An alternative model for clouds 
in which
the inner regions
are
supported by thermal pressure and the
envelopes 
are
supported by turbulent pressure is the ``TNT''
(thermal/nonthermal) model,
in which the density is assumed to be given by
the sum of two power laws, one with
$\krho=2$, representing a singular isothermal sphere, and one with $\krho<2$,
representing a turbulent envelope \citep{1992ApJ...396..631M,1995ApJ...446..665C}.
A more rigorous formulation of this type of model is that of a composite polytrope, in which
the core and envelope of the cloud have different values of $\gp$
\citep{2000ApJ...528..734C}.

Cosmological simulations show that self-gravitating, 
pressureless
matter condenses into filamentary structures
\citep[e.g.,][]{2005Natur.435..629S}.  This reflects the nature of
evolution of cold, triaxial mass distributions under self-gravity
\citep{1965ApJ...142.1431L}: the first collapse is along the shortest axis, 
and the second collapse is along the (original) intermediate axis, 
resulting in a filament aligned along the (original) long axis.  
Molecular clouds often exhibit filamentary structure
as well 
\citep{1979ApJS...41...87S,1995ApJ...445L.161M,1998AJ....116..336N,
1999ApJ...512..250L}. This may reflect the 
effects of self-gravitational evolution, similar to cosmic structure formation.
However, it may also reflect the effects of strongly supersonic turbulence.  
Converging turbulent flows produce curved sheets of shocked gas at stagnation 
surfaces, and the loci of these sheet intersections are filaments.  
The morphology of cold, diffuse HI is similar to that in GMCs (e.g. 
\citealt{2003ApJ...586.1067H,2006astro.ph..8585M}), 
suggesting that at least some of the filamentary
structure in star-forming clouds originates with multi-scale supersonic 
turbulence; the filaments that are created by turbulent flows 
may also be (or become) self-gravitating.

Virial balance in filamentary clouds implies 
$GM/\ell
=Gm_\ell
\sim{\sigma^2}$,
where $\ell$ is a length along the filament
and $m_\ell$ is the mass per unit length.
\citet{2000MNRAS.311...85F} have shown that 
the 
virial gravity term for
a cylindrical cloud with 
an arbitrary density profile 
is $\calw_\ell=-Gm_\ell^2$, which, in contrast to the
spherical case, is unchanged by radial compression. They also showed
that the critical 
mass/length
is 
$m_{\ell,\,\rm cr}=2
\sigma^2/G$.  
Filaments
with $\gamma=\gp\geq 1$ are stable against compression, since the
ratio of kinetic energy to gravitational energy does not decrease
during compression. Isothermal filaments have a density
$\rho=\rho_c/[1+(r/r_0)^2]^2$, where $r_0=(2/\pi)^{1/2}R_G(\rho_c)$
\citetext{\citealp{1964ApJ...140.1056O}; the properties of filaments with $\gp\geq
1$ are not strongly affected by the ambient medium for $r\gg r_0$,
which is why their properties are determined by the central density
$\rho_c$.}
 However, observed filaments often have $\rho\propto 1/r^2$
rather than $1/r^4$
\citep{1999ApJ...512..250L}. \citet{2000MNRAS.311...85F} have shown
that isothermal filaments with helical 
magnetic
fields of the right magnitude
can give rise to such a density profile; alternatively,
\citet{2002PThPS.147...99N} have shown that negative index polytropes
with $\gp$ slightly less than unity have $\rho\propto 1/r^2$.

In general, masses
in excess of the critical mass are subject to fragmentation.
In an isothermal, uniform medium, the minimum wavelength for gravitational
fragmentation is the Jeans length,
\beq
\lambda_J=\left(\frac{\pi\sth^2}{G\rho_0}\right)^{1/2}=\pi^{1/2}R_G(\rho_0).
\label{LJeans}
\eeq
The corresponding Jeans mass is $M_J \equiv (4\pi/3)(\lambda_J/2)^3 \rho
=2.47 \mbe$,
where we have adopted the definition of \citet{1987gady.book.....B} (the Jeans
mass is elsewhere often defined as $\rho \lambda_J^3 =6 M_J/\pi$,
which is even larger than the Bonnor-Ebert mass).
For slabs and filaments, there is a fastest growing mode, which will
determine the spacing of fragments. For an isothermal slab
with
surface density $\Sigma$
and with $\rho_c\gg\rho_0$,
it is 
$\lambda_{\rm max,\, slab}\approx 2^{3/2}\lambda_J(\rho_c)
=4 \sigma_{\rm th}^2/(G\Sigma)$, 
where the Jeans length
is defined in terms of the midplane density $\rho_c$.
An isothermal filament with $\rho_c\gg\rho_0$ has 
$m_\ell \simeq m_{\ell,\,\rm cr}$ and 
$\lambda_{\rm max,\, fil}\approx 
1.25\times 2^{3/2}\lambda_J(\rho_c)$
\citep{1985MNRAS.214..379L}.
These estimates assume that the gas is optically thin; if it becomes opaque,
fragmentation stops. \citet{1976MNRAS.176..367L} show that fragmentation ceases for
masses $\la 0.004\, M_\odot$ (including He---see \citealt{2007prpl.conf..459W}), 
and that this is relatively insensitive
to parameters.

For a thin, rotating disk, 
rotation stabilizes self-gravitational contraction 
for wavelengths greater than 
the Toomre  length
$\lambda_T\equiv 4\pi^2G\Sigma/\kappa^2$, where $\kappa$ is the  
epicyclic
frequency \citep{1964ApJ...139.1217T}.  
In order for a rotating gas disk to fragment, the maximum instability  
scale imposed by angular momentum considerations must
exceed the minimum length for fragmentation set by thermal pressure.
The Toomre parameter 
$Q\equiv \kappa\sth/(\pi G\Sigma)=
(\lambda_{\rm max,\, slab}/\lambda_T)^{1/2}$, 
and must be $\la 0.7-1$
for gravitational fragmentation in an isothermal 
rotating disk, depending on the 
strength of magnetic fields \citep{1965MNRAS.130...97G,2002ApJ...581.1080K}.
Allowing for turbulence and for the additional gravity of a stellar
disk (for large-scale galactic instabilities), the critical $Q$ is
larger (see \S\ref{GMC_formation}).
Real gases are not strictly isothermal; 
\citet{2001ApJ...553..174G}
has shown that in a Keplerian disk, the cooling time $t_{\rm cool}$
and angular velocity $\Omega=\kappa$ must satisfy
the condition $t_{\rm cool}\la 3\Omega^{-1}$ for gravitational runaway
to occur
(the coefficient 3 is based on 2D simulations with 
$\gamma_{\rm eff}=2$,
allowing for dimensional reduction;
\citet{2005MNRAS.364L..56R} showed that this coefficient can change by
a factor of a few depending on the adopted $\gamma$).
Nonlinear instability develops when $Q$ is small even for 
adiabatic disks, but gravitational collapse of the condensations that 
form is ultimately halted if $\gamma$ is sufficiently large 
\citep{2001ApJ...559...70K}.

\subsection{Magnetic Fields}
\label{magnetic_fields}


The interstellar medium is strongly magnetized, whereas stars are weakly
magnetized. How the mass-to-flux ratio increases so dramatically during
star formation is one of the classic problems of star formation
\citep{1956MNRAS.116..503M}. We shall characterize this ratio by
the ratio of the mass to the magnetic critical mass
for poloidal fields,
$\mu_\Phi\equiv M/\mphi$ 
(eq. \ref{eq:mphi}, \S \ref{gravity}).
\citet{2005ApJ...624..773H}
found that the median field in the cold H~I phase of the ISM (the CNM) is
$|B_0|=6.0\pm 1.8\,\mug$, and that the CNM is organized into sheets
with column densities 
$2.6\times 10^{18} \;\mbox{cm\ee} \simlt N_H 
\simlt 2.6\times 10^{20}\;\mbox{cm\ee}$;
 the maximum column presumably reflects
the transition to molecular hydrogen. It follows that the CNM is
magnetically very subcritical, $\mu_\Phi<0.16$ 
(throughout this section, we evaluate $M_\Phi$ with
$c_\Phi=1/2\pi$, the value appropriate for sheets).
There are thus two parts to the magnetic flux problem:
How does the mass-to-flux ratio increase to $\mu_\Phi\ga 2$ so that gravitational
collapse can readily occur, and then how does it increase to
the very large values ($\sim 10^{5-8}$) characteristic of stars?

Astronomers have two primary methods to measure the strength of magnetic
fields in the dense ISM: the Zeeman effect, which measures the line-of-sight
component, $\blos$; and the Chandrasekhar-Fermi method \citep{1953ApJ...118..113C}, which
measures the component of the field in the plane of the sky, $\bpos$, by
comparing the fluctuations in the direction of 
${\bf B}_{\rm pos}$
with those in
the velocity field \citetext{see the reviews by \citealp{2005AIPC..784..129C} and
\citealp{HC05}; note that in the diffuse ISM, magnetic field strengths are
  also obtained by Faraday rotation and synchrotron observations,
  with results consistent with Zeeman observations.}
The morphology of the field, which is needed for the Chandrasekhar-Fermi method,
can be measured from dust polarization and from linear polarization of 
spectral lines \citep
{1981ApJ...243L..75G}.
The largest compilation of magnetic field strengths in molecular clouds
remains that of \citet{1999ApJ...520..706C},
although it must be noted that the
median temperature of the regions with detected fields is 40~K, significantly
greater than average.
 Inferring the intrinsic
field strength and column density from measurements of the 
line-of-sight components is somewhat
subtle \citep{2005ApJ...624..773H};
in particular, care must be exercised in evaluating the average value 
of ${B_\parallel/B}$
using logarithmic values, since $\avg{\log B_\parallel/B} < \log
\avg{B_\parallel/B}$. However,
it is straightforward to infer the
median values: $B_{\rm med}=2B_{\rm los,\,med}$ and, 
for sheets,  $N_{\rm med}=N_{\rm los,\,med}/2$.
Most of the structures Crutcher (1999) studied are relatively dense cores,
so it is plausible that they are not sheet-like; in that case,
the median value of $\mu_\Phi$ is 
$1.65\pm0.2$ (\citealp{HC05}; note that the values in Crutcher 1999 are
based on $c_\Phi=0.12$, whereas we are using $c_\Phi=1/2\pi$),
the cores
are supercritical and the magnetic field is unable to significantly
impede gravitational collapse. On the other hand, if the objects in
his sample are in fact sheet-like, then the median value of 
$\mu_\Phi$ is reduced to 
0.8,
and the typical core is about critical.
However, it should be noted that none of the cores have
observed line-of-sight fields strong enough to ensure that they are subcritical, and
for many cases only upper limits on the magnetic field strength are obtained.
Subsequent OH Zeeman observations by \citet{2001ApJ...554..916B} have
strengthened these conclusions, although these authors suggest that
observations with higher spatial resolution are needed to determine whether
the relatively low fields they infer (mostly upper limits) are in part
due to variations in the field  structure within the telescope beam.

\citet{1999ApJ...520..706C} also reached a number of other conclusions 
on the role of magnetic fields in cores and clumps
within molecular clouds from
his sample: the observed structures are in approximate virial equilibrium;
the kinetic and magnetic energies are in approximate 
equipartition,
as expected theoretically \citep{1995ApJ...439..779Z}; 
correspondingly, the Alfv\'en Mach number is
$\calm_A\simeq 1$; the observed motions are highly supersonic, with
$\calm_s\equiv \sqrt{3}\snt/c_s\simeq 5$; and, to within
the errors, $B\propto \rho^{1/2}$, which corresponds to
a constant Alfv\'en velocity 
\citetext{for sources with measured fields, as opposed to upper limits,
the average value is
$\vA\simeq 2$~km~s\e, as found previously by \citealp{1993prpl.conf..279H}}.
\citet{2000ApJ...540L.103B} showed that the dispersion of the Alfv\'en
Mach number is significantly less than that in the Alfv\'en velocity
in this sample. He argued that a constant value of $\calm_A$ is to 
be expected if  the clouds are strongly bound, so that the surface
pressure is negligible, and if $\mu_\Phi$ is about constant.
Adopting a median value $\calm_A=1.0$ from \citet{1999ApJ...520..706C}
gives a median value for the magnetic field of
\beq
B_{\rm med}\simeq 
30
\left(\frac{\nh}{10^3\;\mbox{cm\eee}}\right)^{1/2}
\left(\frac{\snt}{\mbox{1 km s\e}}\right)~~~\mug~~~~~~(\nh\ga 
2\times
10^3\;\mbox{cm\eee}).
\eeq
The value of the density in this relation is $N_{\rm H}/(4R/3)$, where $R$ is the
mean projected radius. Projection effects could cause the actual density
to differ from this, but the change is not large for triaxial clouds of the type
considered by \citet{2000ApJ...540L.103B}. 

As yet, observations of the mass-to-flux ratio on large scales, up to
that of GMCs, are not available. 
The definition of $\mu_\Phi$ implies
\beq
\bar B=\frac{G^{1/2}\Sigma}{\mu_\Phi c_\Phi}=3.80\left(\frac{N_{\rm H,\,21}}
	{\mu_\Phi}\right)~~~\mug
	=7.60\left(\frac{A_V}{\delta_{\rm gr}\mu_\Phi}\right)~~~\mug,
\label{eq:bcr}
\eeq
where the numerical evaluations are based on $c_\Phi=1/2\pi$,
the visual extinction is 
$A_V=N_{\rm H}\delta_{\rm gr}/(2\times 10^{21}\;\mbox{cm\ee})$, and
$\delta_{\rm gr}$ is the dust-to-gas ratio normalized to
the local interstellar value.
Typical Galactic GMCs have $N_{\rm H}=1.5\times 10^{22}$~cm\ee\  
(see \S\ref{dynam_state}), corresponding to
critical magnetic field strength (i.e., such that $\mu_\Phi=1$) of
$B_{\rm cr}=57\,\mu$G for the
large-scale mean field.  In regions with densities $n_{\rm H}\approx 2\times10^3\;\cmt$, 
the lowest for which molecular-line Zeeman observations are available, 
\citet{1999ApJ...520..706C} reports line-of-sight magnetic field
strengths of $\le 21\mu$G.  Allowing for an increase of up to a factor
of two for projection effects, and for the fact that the mean magnetic
field strength will not increase as the density is reduced by a
factor of $\sim 10$ to reach the volume-averaged value in GMCs, 
we infer
that GMCs are supercritical.
GMC magnetic fields are not too weak, 
however: 450~$\mu$m polarimetry of four GMCs shows that 
the orientation of the field appears to be preserved during the
formation of the
GMCs and that the energy in the field is comparable to the turbulent energy
\citep{2006ApJ...648..340L}.

Theoretical arguments are consistent with the empirical evidence
that GMCs as well as their sub-parts 
are supercritical with respect to their mean magnetic fields.
Models of self-gravitating, 
isothermal,
magnetized clouds show that large pressure
contrasts between the center of the cloud and the edge occur only
when the cloud is near its critical mass; furthermore, if the kinetic
energy is comparable to the magnetic energy, then large pressure contrasts
occur only for 
$M>M_\Phi$. Extending the earlier work of \citet{1976ApJ...207..141M},
\citet{1988ApJ...335..239T} found that for the cases they considered with
$8\pi P/B^2=1$, the central pressure significantly exceeds the surface pressure
only when $M$ is quite close to $M_{\rm cr}$, and that in these cases 
the cloud is magnetically supercritical, $M>M_\Phi$.
Using this work, \citet{1989ApJ...345..782M} showed that the critical mass is 
$M_{\rm cr}\simeq M_\Phi+\mbe$
for quiescent clouds;
he assumed that this relation applies to turbulent clouds as well, with
$\sth$ replaced by the total velocity dispersion $\sigma$, but the validity
of this assumption remains to be demonstrated.
Since, on large scales, the turbulent magnetic energy is likely 
comparable to or
larger than the mean magnetic energy, and the kinetic energy 
is at least as large as
the magnetic energy (and much greater than the thermal energy), 
then clouds with $\alpha_{\rm vir}\sim 1$ have
$M\ga 2 \mphi$.
\citet{1998ApJ...494..587N}
has given a similar, more precise argument that the 
smaller-scale and less-turbulent
cores that form stars
are also magnetically supercritical.
In both cases, the basic argument is that if gravity is strong enough to
overcome both kinetic energy (turbulent plus thermal)
and magnetic fields (turbulent and ordered) 
in order to form a bound object, then it is certainly strong
compared to the support from mean magnetic fields alone. 
Note that this
argument does not apply to objects that are not bound, but instead
are the result of colliding flows (\S \ref{GMC_formation}). 
It is of great importance to
determine observationally the relative importance of magnetic fields
and gravity in the large scale structure of molecular clouds.

Most detailed modeling of magnetic fields in non-turbulent clouds
is based on the assumption that the field is poloidal \citetext{e.g.,
\citealp{1987ppic.proc..453M}}; such fields always tend to support
clouds against gravity. On the other hand, the toroidal component of a
helical field exerts a confining force, and can lead to prolate clouds
\citep{1991ApJ...376..190T,2000MNRAS.311...85F}.  From a virial
analysis of several filamentary clouds, \citet{2000MNRAS.311...85F}
find that the self-gravity and the pressure of the ambient medium are
inadequate to account for the high mean pressures that are observed in
the clouds; they conclude that the data can be explained if these
clouds are confined by helical fields.  The principal uncertainty in
this analysis is that in most cases there is no direct measurement of
the ambient pressure.

There are two mechanisms for increasing the mass-to-flux ratio, 
flows along magnetic fields and ambipolar diffusion. 
In the part of a bound molecular cloud that is shielded from the interstellar radiation
field so that the ionization is due to cosmic rays, the
ambipolar diffusion time is about 10 times the free-fall time
in the absence of turbulence \citep{1987ppic.proc..453M}
and several times faster
than this
in the presence of turbulence \citep{2002ApJ...567..962Z,2002ApJ...570..210F,2005ApJ...631..411N}.
However, most of the mass of a GMC is ionized
primarily by FUV radiation from stars \citep{1989ApJ...345..782M}, and
in this gas the ambipolar diffusion time is much longer.
GMCs are very porous, and as a result an even larger fraction of
the volume of the cloud is likely to be ionized above the level
set by cosmic rays. It follows that flux-freezing is a good approximation
on large scales in molecular clouds.
\citet{1985prpl.conf..320M} introduced the concept of the ``accumulation length,''
$L_0$, the size of the region required to achieve a given
mass-to-flux ratio when flux-freezing applies, 
$\mu_\Phi=M/\mphi\propto n_0L_0/B_0$. In our notation this yields
\beq
L_0=\left(\frac{c_\Phi B_0}{\muh G^{1/2}n_0}\right)\mu_\Phi=
                       85\left(\frac{\mu_\Phi \boms}{n_0}\right)
                         ~~~\mbox{pc},
\eeq
where $\boms\equiv B_0/(1\;\mug)$. Using
$n_0\sim 1$~cm\eee\ for the mean density in the diffuse ISM (since GMC columns
are much greater than those of individual CNM clouds)
and $\boms\sim 6$ for the mean field in the solar vicinity 
(since the mean field should be similar to the CNM field---\citealp{2005ApJ...629..849P})
gives a large value for this length, $\sim 1$~kpc if GMCs have $\mu_\Phi\sim 2$.
In  fact, as we shall discuss in \S \ref{GMC_formation}, GMC
formation from large-scale self-gravitating galactic disk instabilities
indeed involves very large accumulation lengths, and yields
supercritical clouds.

As discussed above, 
current observations do not determine whether ambipolar diffusion is necessary
for the initiation of gravitational collapse. Theoretical simulations suggest
that in the absence of ambipolar diffusion, star formation is strongly
suppressed in magnetically subcritical regions, even if 
$\mu_\Phi$ is only slightly less than unity  \citep{2005ApJ...635.1126K}.
However, similar
simulations show that magnetic fields have a relatively small
effect in slowing the rate of star formation if the gas is supercritical
\citep{1999ApJ...513..259O,2001ApJ...547..280H,
2004ApJ...605..800L,2005ApJ...618..344V,2005ApJ...630L..49V,
2005ApJ...631..411N}.   The primary effect of magnetic fields may be
to shift the initial collapse to higher masses. 
Simulations with ambipolar diffusion in weakly ionized
plasmas are very challenging.
In the strong-coupling approximation, in which the ions are not treated
as a separate fluid but the field diffuses relative to the flow,
explicit MHD codes have time steps $\propto \Delta x^2$, which 
is prohibitive at high resolution \citep{1995ApJ...442..726M}. 
If the ions are treated as a separate fluid,
explicit codes must resolve Alfv\'en waves in the ions as well as Alfv\'en waves
in the coupled ion-neutral fluid. 
For ionizations $\la 10^{-6}$, the Alfv\'en velocity
in the ions can exceed $10^3$~km~s\e, leading to very small time steps.
A potential way around this problem is to increase the ion mass and decrease
the ion-neutral coupling constant so that the momentum exchange rate between
the ions and neutrals is unchanged \citep{2006ApJ...638..281O,lmk06}.

One regime in which ambipolar diffusion (or the lack of it) could
have a strong effect on the star formation rate is in the outer
layers of GMCs,
which
are dominated by FUV ionization. 
As remarked
above, these regions constitute the bulk of the mass of a GMC,
as much as $\sim 90\%$. 
FUV photoionization slows ambipolar diffusion, and
therefore star formation, when it dominates
cosmic-ray ionization, which occurs for
visual extinctions
 $A_V\la 4$~mag from the surface or
$\sim 8$~mag along a line of sight through the cloud
\citep{1989ApJ...345..782M}.
Suppression of star formation in the outer layers of GMCs has 
been confirmed in the L1630 region of Orion \citep{1997ApJ...488..277L} 
and in Taurus \citep{1998ApJ...502..296O}.
To the extent that ambipolar diffusion is essential for forming
molecular cores, the absence 
or near absence of sub-mm cores in 
the outer layers of Ophiuchus 
\citep{2004ApJ...611L..45J} and Perseus 
\citep{2006ApJ...638..293E,2005A&A...440..151H}
is qualitatively consistent
with this prediction. 
On the other hand, \citet{1993ApJ...412..233S}
find that there is a substantial distributed population
of young stars in L1641,
although this population is relatively old ($5-7$)~Myr.


\subsubsection{Ionization}
\label{ionize}


The chemistry of molecular clouds is a full subject in its own right. Here
we summarize several developments that affect the ionization, which governs the coupling
between the gas and the magnetic field. (1) Photodissociation regions (PDRs) are
regions of the ISM that are predominantly neutral and in which the chemistry
and heating are predominantly due to far-UV radiation 
\citetext{see the review by \citealp{1999RvMP...71..173H}}. 
Most of the non-stellar infrared
radiation and most of the millimeter and submillimeter CO emission in galaxies originates
in PDRs. In the typical interstellar radiation field,
photoionization dominates ionization by cosmic rays for
extinctions $A_V<4$~mag, which includes
most of the molecular gas in the Galaxy \citep{1989ApJ...345..782M}. 
(2) Polycyclic aromatic hydrocarbons
(PAHs), which contain a few percent of the carbon atoms, often dominate the
mid-IR spectrum of star-forming regions and galaxies. It is frequently assumed that
PAHs have a low abundance in molecular clouds due to accretion onto dust grains;
if this is not the case, they can dominate the ionization balance,
since electrons react with them very rapidly
\citep{1988ApJ...329..418L}. (3) H$_3^+$ is a critical ion in initiating ion-molecule
reactions in molecular clouds. For many years, the rate of dissociative recombination,
H$_3^++e\rightarrow$~\htwo\ +H or H+H+H, was uncertain, but careful laboratory
experiments have shown that the rate coefficient for this reaction is large:
a fit to the results of \citet{2003Natur.422..500M} gives
$\alpha_d($H$_3^+)=4.0\times 10^{-7}(T/10\;\mbox{K})^{-0.52}$~cm$^3$~s\e.
In order to maintain the observed abundance of H$_3^+$ in the face of
this high recombination rate, these authors inferred a very high cosmic-ray
ionization rate, 
$\zeta_{\rm 
CR
}=6\times 10^{-16}$~s\e\ 
per H atom (including secondary ionizations), in a diffuse molecular
cloud along the line of sight to $\zeta$~Per. Models involving two gas phases
give somewhat lower values of 
$\zeta_{\rm CR}$ 
\citep[e.g.,][]{2006PNAS..10312269D}, 
but the correct value
is now quite uncertain. In dense clouds, \citet{2006PNAS..10312269D} 
concludes that the ionization rate
is 
$\zeta_{\rm CR}\simeq 2.5-5\times 10^{-17}$~s\e. 
(4) Recent observations
have established that carbon-bearing molecules 
freeze out onto dust grains
at high densities ($\nh\sim 10^5$~cm\eee) in low-mass cores, with nitrogen-bearing molecules
freezing out
at higher densities \citep{2006astro.ph..2379D}. 
This affects the ionization,
since it removes abundant ions such as HCO$^+$ from the gas. 

Although the chemistry determining the ionization in molecular clouds
is complex, simple analytic estimates are possible. In the outer layers
of PDRs, carbon is photoionized so that $n_e\simeq n$(C). In 
regions ionized by cosmic rays, the degree of ionization is given by
\beq
x_e\equiv \frac{n_e}{\nh}\simeq\left(\frac{\zeta_{\rm CR
}}{\alpha \nh}\right)^{1/2}
\label{eq:xe}
\eeq
if the ionization is dominated by molecular ions (including PAHs),
where $\alpha$ is the relevant recombination rate in the chemistry that
determines the ionization fraction.
If PAHs are depleted, then
$\alpha\simeq 10^{-6}$~cm$^3$~s\e\ is the dissociative recombination rate for
heavy molecules provided
the density is high enough 
that
H$_3^+$ is destroyed primarily by reactions with such molecules;
for lower densities, where the ionization is dominated by H$_3^+$,
one has $\alpha=\alpha_d($H$_3^+$). 
If PAHs are sufficiently abundant that
most of the electrons are attached to PAHs, then 
$\alpha\simeq 3\times 10^{-7}$~cm$^3$~s\e\ \citep{2005pcim.book.....T}
and $n_e$ in equation (\ref{eq:xe}) includes the 
electrons attached to PAHs.
Metal ions can be readily included in the analytic theory 
\citep{1989ApJ...345..782M}, but they do not appear to be important
in dense cores \citep{2006Natur.442..425M}.
Equation (\ref{eq:xe}) is consistent with the results of 
\citet{2004ApJ...614..203P} 
at late times and at high densities for
$\alpha\simeq\alpha_d$(HCO$^+)$, 
which they took to be $2.5\times 10^{-6}$~cm$^3$~s\e.


\section{MACROPHYSICS OF STAR FORMATION}

\subsection{Physical State of GMCs, Clumps, and Cores}
\label{dynam_state}


The molecular gas out of which stars form is found in molecular clouds,
which occupy a small fraction of the volume of the ISM but, inside the
solar circle, comprise a significant fraction of the mass.  The
terminology for the structure of molecular clouds is not fixed; here
we follow the discussion in \citet{2000prpl.conf...97W}.  {\it Giant
  molecular clouds (GMCs)} have masses in excess of $10^4\;M_\odot$
and contain most of the molecular mass. Molecular clouds have a
hierarchical
structure that extends from the scale of the cloud
down to the thermal Jeans mass in the case of gravitationally bound
clouds,
and down to much smaller masses for unbound structures
\citep{1995ApJ...453..293L,1998A&A...331L..65H}.
Overdense regions (at a range of scales) within GMCs are termed {\it
 clumps}. {\it Star-forming clumps} are the massive clumps out of
which stellar clusters form, and they are generally gravitationally
bound. {\it Cores} are the regions out of which individual stars (or
small multiple systems like binaries) form, and are necessarily
gravitationally bound. As remarked above, this terminology is not
universal; e.g., \citet{2006astro.ph..3474W} use ``pre-stellar core'' to refer to a
core, and ``cluster-forming core'' to refer to a star-forming clump.

A molecular cloud is surrounded by a layer of atomic gas that shields
the molecules from the interstellar UV radiation field; in the solar
vicinity, this layer is observed to have a column density $N_{\rm
  H}\simeq 2\times 10^{20}$~cm\ee, corresponding to a visual
extinction $A_V=0.1$~mag \citep{1978ApJ...224..132B}.  A larger column
density, $N_{\rm H}\simeq 1.4\times 10^{21}$~cm\ee, is required for CO
to form \citep{1988ApJ...334..771V}. The layer of gas in which the
hydrogen is molecular but the carbon is atomic is difficult to
observe, and has been termed ``dark gas'' \citep{2005Sci...307.1292G}.

The mass of a molecular cloud is generally inferred from its
luminosity in the $J=1-0$ line of $^{12}$CO or $^{13}$CO.  Because
$^{12}$CO is optically thick, estimating the column density of \htwo\
molecules from the $^{12}$CO
line intensity $I_{\rm CO}$ (in units of K~km~s\e)
requires multiplication by an ``X-factor,'' an appropriate name since
it is not well understood theoretically; this is defined as $X\equiv
N(\mbox{\htwo})/I_{\rm CO}$. 
Various
methods have been used to infer the
value of $X$ in the 
Galaxy. In one method, observations
of $\gamma$-rays emitted by
cosmic rays interacting with the ISM give the total amount of
interstellar matter; the mass of molecular gas follows by subtracting
the neutral atomic hydrogen (H~I) contribution. With this technique,
\citet{1996A&A...308L..21S} infer $X=1.9\times
10^{20}$~cm\ee(K~km~s\e)$^{-1}$. 
In another method,
subtracting the
H~I-associated dust emission from the total observed dust emission in
the infrared gives a local value $X=1.8\times
10^{20}$~cm\ee(K~km~s\e)$^{-1}$ \citep{2001ApJ...547..792D}.  Note
that both
of these
methods account for all the molecular hydrogen gas, including the
``dark gas.'' Allowing for the atomic shielding layer around a
molecular cloud (but not the ``dark-gas'' layer),
\citet{1989ApJ...338..178E} predicted 
$X\propto (G_0/Z)^{3/8}/T_b$,
where $G_0$ is proportional to the intensity of FUV radiation that can
photodissociate \htwo, $Z$ is the metallicity, and $T_b$ is the
brightness temperature of the line.  \citet{1988ApJ...325..389M}
concluded that $T_b$ should be substantially reduced in regions of low
metallicity. Observing the $^{13}$CO line is advantageous in that it
is optically thin in all but the 
high-density cores.  Conversion from
$^{13}$CO intensity to column density involves an assumption of LTE
(using temperatures derived from $^{12}$CO), and a fixed \htwo\ to
$^{13}$CO abundance.  Because $^{13}$CO may be subthermally excited in
diffuse regions, however, column densities there will be underestimated.
Near-infrared extinction mapping \citep{2007prpl.conf....3L} offers the prospect of
obtaining more accurate masses, at least for nearby molecular clouds.

The observed mass distribution of GMCs is a power law with a
relatively sharp cutoff.  Let $d\caln_c(M)$ be the number of GMCs with
masses in the range $M$ to $M+dM$. Observations of GMCs inside the
solar circle (but excluding the Galactic Center) are consistent with
the mass distribution of the form \citep{1997ApJ...476..166W}
\begin{equation}
\frac{d\caln_c}{d\ln M}=\caln_{cu}\left(\frac{M_u}{M}\right)^{\alpha} ~~~~~(M\leq M_u),
\label{GMC_massfnct}
\end{equation}
with no GMCs above $M_u$. Here $\caln_{cu}/\alpha$ is equal to
the number of clouds eliminated from the distribution by the cutoff at $M_u$.
With $\caln_{cu}=63$, $\alpha=0.6$, and $M_u=6\times 10^6\;M_\odot$,
this cloud mass distribution accounts for all the molecular mass observed
inside the solar circle excluding the Galactic Center.
An independent analysis by 
\citet{2005PASP..117.1403R} finds a similar slope ($\alpha=0.5\pm 0.1$),
but a somewhat smaller maximum mass ($M_{u}=3\times 10^6\;M_\odot$,
although this does not include the several most massive clouds). 
These results are necessarily approximate due to the difficulties 
in identifying
clouds from position-velocity data in the inner Galaxy---in particular,
blending of clouds along the line of sight is likely to make the true value
of the slope steeper \citep{2005PASP..117.1403R}. 
However, the main implications are likely to be robust.
First, most of the mass in
GMCs is in large clouds---a significant fraction is in clouds with
$M>10^6\;M_\odot$, and $>80$\% is in clouds with $M>10^5\;M_\odot$ 
(see also \citealp{2006ApJ...641L.113S}).
And second, since $\caln_{cu}\gg 1$, the upper limit of the mass 
distribution, $M_u$, has a physical
significance \citep{1997ApJ...476..144M}. If there were no cutoff to the
distribution, one would expect about 100 GMCs more massive than
$6\times 10^6 \, M_\odot$ in the Galaxy, whereas there are none. 
This upper mass limit may be set
by the processes that form GMCs out of diffuse gas (see
\S\ref{GMC_formation}).
It should be noted that the GMCs are embedded in more massive HI 
``superclouds'' (sometimes encompassing multiple GMCs),
which also appear to be gravitationally bound \citep{1987ApJ...320..182E}.
In other Local Group galaxies, GMC mass distributions have similar
power laws to that in the Milky Way, with the exception of M33, 
which has $d\caln_c/d\ln M\propto M^{-1.5}$ 
\citep{2007prpl.conf...81B}.
In more distant galaxies, giant molecular associations 
(GMAs)
with masses up
to $\sim 10^7\Msun$ have been observed 
\citep{1988Natur.334..402V,1999ApJS..124..403S}.

\subsubsection{Dynamics of GMCs}

In a seminal paper, \citet{1981MNRAS.194..809L} summarized some of the
key dynamical features of GMCs in what are often referred to as
``Larson's laws.'' The first result is that GMCs obey a {\it line
  width--size relation}: GMCs are supersonically turbulent with
velocity dispersions that increase as a power of the size. For GMCs in
the first Galactic quadrant, almost all of which are inside the solar
circle, \citet{1987ApJ...319..730S} found 
\begin{equation}
\sigma=(0.72 \pm 0.07)R_{\rm pc}^{0.5\pm0.05} \;\kms,
\label{Sol_lws}
\end{equation}
where $R_{\rm pc}\equiv R/(1$~pc).
(Note that this value for $\sigma_{\rm pc}$ is based on a distance to
the Galactic Center of 10 kpc; we have not adjusted this value for a more
accurate distance since the change is within the errors.)
\citet{2004ApJ...615L..45H} find that this cloud-to-cloud relation
extends to the structure functions within individual GMCs as well (see
eq. \ref{GMC_lws} in 
\S \ref{turbobs}), and argue that this demonstrates the
universality of the turbulence in 
moderate-density gas in
molecular clouds (see below).

Larson's second law is that GMCs are gravitationally bound
($\avir\simeq 1$; see eq. \ref{alpha_vir_def}).  
(It should be noted that while molecular gas is generally bound in the
Galaxy, different physical conditions can lead to substantial amounts
of unbound molecular gas or bound atomic gas---\citealp{1993prpl.conf...97E}.)
\citet{1987ApJ...319..730S} determined the masses
of clouds in their sample using the virial theorem with $\avir=1.1$,
and then determined the $X$-factor. Including He and adjusting the
distance to the Galactic Center to 8.5 kpc from 10 kpc, their value of
the $X$-factor corresponding to a typical GMC with a mass of
$10^6\;M_\odot$ (see above) is $1.9\times
10^{20}$~cm\ee(K~km~s\e)$^{-1}$, the same as the value determined from
$\gamma$-ray observations; hence, the GMCs in their sample are
gravitationally bound on average. (Note that this argument is
approximate since the $\gamma$-ray value for the $X$-factor includes
the ``dark gas,'' whereas the value from the virial theorem includes
only part of this gas, depending on the morphology of the GMC.)
Observations of $^{13}$CO, which is optically thin, permit a direct
measurement of the mass, provided the abundance is known.  Such
observations of a sample of GMCs in the outer Galaxy, where blending
of different clouds along the line of sight is negligible, confirm
that molecular clouds with $M\ga 10^4\;M_\odot$ (i.e., GMCs) are bound
\citep{2001ApJ...551..852H}. 
Lower mass clouds become progressively
less bound \citep[see also][]{1990ApJ...348L...9M},
and unbound molecular clouds are for many purposes
equivalent to non-self-gravitating clumps within larger GMCs. 
Based on observations in $^{13}$CO and other species that are believed to
be optically thin, clump mass functions within GMCs
follow ${d\caln}/{d\ln M}\propto M^{-\alpha_{\rm clump}}$
with $\alpha_{\rm clump}=0.3-0.7$
\citep{1993prpl.conf..125B,1994ApJ...428..693W}.
The slope of the clump mass function is similar
to that for GMCs as a whole (see eq. \ref{GMC_massfnct}),
possibly because both are determined by turbulent processes within
larger, gravitationally-bound systems.

Larson's third law is that GMCs all have similar column densities. For the
\citet{1987ApJ...319..730S} sample, the mean column density is $\bar N_{\rm H}
=(1.5\pm 0.3)\times 10^{22} R_{\rm pc}^{0.0\pm 0.1}$~cm\ee; this corresponds
to an extinction $A_V=7.5$~mag with the local dust-to-gas ratio.  The 
corresponding mean surface density of GMCs is 
$\bar \Sigma = 170\Msun\pc^{-2}$.
However, GMCs in the outer Galaxy 
are observed to 
have smaller column densities
\citep{2001ApJ...551..852H},
in part because of the greater sensitivity of these observations.

As Larson pointed out, these three relations are not independent; any two of
them imply the third. Indeed, 
if we express the line width---size relation as 
$\sigma \equiv \sigma_{\rm pc} R_{\rm pc}^{1/2}$, then
\beq
\avir = \left(\frac{5}{\pi\;{\rm pc}}\right)\frac{\sigma_{\rm pc}^2}{G\Sigma}
=3.7\left(\frac{\sigma_{\rm pc}}{\mbox{1 km s\e}}\right)^2
\left(\frac{100\,M_\odot\;{\rm pc}^{-2}}{\Sigma}\right)
\label{eq:relate}
\eeq 
relates the three scaling laws.
Observations supporting a ``universal turbulence law"
(eq. \ref{GMC_lws}) in the Galaxy, and thus
small differences in 
$\sigma_{\rm pc}$ 
between inner- and outer- Galaxy GMCs
\citep{2004ApJ...615L..45H,2006ApJ...643..956H}, 
then imply that the value of $\Sigma$ is about the same for all 
GMCs with $\alpha_{\rm vir}\sim 1$,
regardless of Galactic location. 
Observational confirmation of this conclusion would be valuable.
Provided Larson's Laws apply,
the mean kinetic pressure 
within GMCs is
independent of mass and size, and is given by 
$\bar P_{\rm kin}=\bar\rho\sigma^2=
3\Sigma \sigma_{\rm pc}^2/(4\;{\rm pc})$. For inner-Galaxy GMCs, this is
$\bar P_{\rm kin}/k_{\rm B}\approx 3\times 10^5 \K \;\cmt$.  

These results can also be expressed in terms of the 
sonic length $\ell_s$ (see \S \ref{scales}), since
$\sigma_{\rm pc}=\cs (2/3\ell_{s,\,\rm pc})^{1/2}$ 
if 
$\sigma \approx \snt \gg 
\cs$,
and $\ell=2R$.
Gravitationally bound objects ($\avir\sim 1$)
that obey a line width--size relation with an exponent $\simeq 1/2$ 
necessarily have surface densities
$\Sigma=(10/3\pi\avir)
c_s^2/(G\ell_s)\sim c_s^2/(G\ell_s)$. 
The small observed variation in $\Sigma$  
for the set of inner-Galaxy GMCs 
is then equivalent to a small variation in $\ell_s$ in those clouds
(since $c_s$ is observed to be about constant).
In terms of $\ell_s$, the mean 
kinetic
pressure in GMCs is
$\bar P=\Sigma \cs^2/(2 \ell_s)$.



Do Larson's laws apply in other galaxies? 
\citet{2007prpl.conf...81B} summarize
observations of GMCs in galaxies in the Local Group, in which the
metallicity varies over the range $(0.1-1)$ solar. They find that the
GMCs in most of these galaxies would have luminous masses within a
factor two of their virial masses if $X=4\times
10^{20}$~cm\ee\ (K~km~s\e)$^{-1}$; alternatively, if $X$ has the same
value as in the Galaxy, then the GMCs are only marginally bound
$(\avir\simeq 2)$. They conclude that metallicity does not have a
significant effect on $X$ since the ratio of the virial mass to the CO
luminosity is constant in M33, despite a factor 10 variation in
metallicity. (Note, however, that Elmegreen 1989 argues that $X$
depends on the ratio of the metallicity to the intensity of the FUV
radiation field, which is not addressed by these results.)
Although there is insufficient dynamic range for clear evidence of a
relationship between linewidth and size based on current observations,
the data are consistent with $\sigma\propto R^{1/2}$
but with values of $\sigma_{\rm pc}$ (and therefore $\ell_s$) 
that vary from galaxy to galaxy.
\citet{2007prpl.conf...81B} 
also find that GMC surface densities have a relatively
small range within any given Local group galaxy, while varying from
$\sim 50\Msun\pc^{-2}$ for the SMC (L. Blitz, personal communication) 
to $> 100\Msun\pc^{-2}$ for M33.

There are currently two main conceptual frameworks for interpreting
the data on GMC properties.  One conception of GMCs is that they are
dynamic, transient
entities in which the
turbulence is driven by large-scale colliding gas flows that
create the cloud (e.g. 
\citealt{2005ApJ...633L.113H,2006ApJ...643..245V,2007prpl.conf...63B}).
This picture naturally explains why GMCs
are turbulent (at least in the initial stages), and why the line
width--size relation within clouds has an exponent of 1/2 -- simply
due to the scaling properties of supersonic turbulence.  
However,  it is less obvious
why $\avir\sim 1$ and why $\Sigma$ has a particular value, since
small-scale dense structures may form (and collapse) at stagnation
points in a high-velocity compressive flow before sufficient material
has collected to create a large-scale GMC.  
Indeed, based on simulations with a converging flow of $\sim 20\;\kms$
with no stellar feedback,
\citet{2007ApJ...657..870V} find that star formation occurs when 
the column density is $N_H\approx 10^{21}\cm^{-2}$, a factor of ten
below the mean observed value for GMCs.   They 
also find that $\avir$ remains near
unity after self-gravity becomes important,
although the kinetic energy 
is primarily due to the gravitational collapse of the cloud, not to
internal turbulence. 
\citet{1974ARA&A..12..279Z} argued many years ago that GMCs cannot
be in a state of global collapse without leading to an unrealistically high 
star formation rate.  
Proponents of the transient GMC picture counter by pointing out that most
of the gas in GMCs is unbound and 
never forms stars (e.g., \citealp{2005MNRAS.359..809C}); 
the global collapse 
is reversed by feedback from stars that form in the fraction of the 
gas that is overdense and bound. Indeed,
individual star formation proceeds more rapidly than global collapse in
essentially all
turbulent simulations
(see also \S \ref{GMC_evol}).  However, dominance of global collapse 
and expansion over large-scale random turbulent motions has not been confirmed
from observations.

In the second conceptual framework, GMCs are formed by large-scale
self-gravitating instabilities (see \S \ref{GMC_formation}), and the
turbulence they contain is 
due to a combination of inheritance from
the diffuse ISM, conversion of gravitational energy to turbulent
energy during contraction, and energy injection from newly formed
stars (\S \ref{GMC_evol}); the balance among these terms 
presumably shifts in time.  In the work of
\citet{1987A&A...171..225C,1989ApJ...338..178E,1988ApJ...334..761M,
  1999ApJ...522..313M,1999osps.conf...29M}, GMCs are treated as
quasi-equilibrium, self-gravitating objects, so that the virial
parameter is near-unity by definition. Whether or not equilibrium
holds, the virial parameter is 
initially
of order unity in scenarios involving
gravitational instability because GMCs separate from the diffuse ISM
as defined structures when they become gravitationally bound.  For a
quasi-equilibrium, the mean surface density is set by the pressure of
the ambient ISM (see \S \ref{gravity}), which in turn is just the
weight of the overlying ISM.  \citet{1989ApJ...338..178E} has given
explicit expressions for how the coefficient in the line width--size
relation and the surface density depend on the external pressure,
finding results that are comparable with observed values.  In
particular, the cloud surface density scales with the mean surrounding
surface density of the ISM.  
Even if GMCs are not equilibria, if they
are formed due to self-gravitating instabilities in spiral arms (e.g.
\citealt{2002ApJ...570..132K,2006ApJ...646..213K}) they must 
initially
have surface densities a
factor of a few above the mean arm gas density, consistent with
observations.  Provided that 
stellar feedback destroys clouds
within a few
(large-scale) dynamical times before gravitational collapse
accelerates, the mean surface density would never greatly exceed the value
at the time of formation.  Simple models 
of cloud evolution with stellar feedback
(e.g., \citealt{2006astro.ph..8471K}) suggest that the scenario of
slow evolution with $\avir=1-2$ is
self-consistent and yields realistic star formation efficiencies, but
more complete studies are needed.

The two approaches to interpreting GMC dynamics correspond to two
alternate views on GMC lifetimes.  \citet{2000ApJ...530..277E} 
argued that, over a wide range of scales,
star formation occurs in about $1-2$ dynamical crossing times of the
system, 
$t_{\rm cross}\equiv 2R/(\sqrt{3}\sigma)$.  
\citet{1999ApJ...527..285B} and \citet{2001ApJ...562..852H} 
focused on the particular case of star formation in Taurus,
and argued that it occurred in about one dynamical time.
The alternate view is that GMCs are gravitationally bound and live 
at least $2-3$, and possibly more, crossing times, 
$t_{\rm cross}\simeq 10M_6^{1/4}$~Myr,
where $M_6\equiv M/(10^6\;M_\odot)$ and a virial parameter $\avir\sim
1-2$ is assumed.
(Note that the crossing time in the nearby star-forming region in 
Taurus is $\sim 10^6$~yr, whereas in 
a large GMC it is $\sim 10^7$~yr.)
However, GMCs 
(as opposed to structures within them) cannot be too close to being
equilibria, since they do not appear to 
have a systematic, global
density stratification,
nor does the line width--size relation
within individual clouds differ from that in unstratified clouds.
(Note that logatropic clouds can account for
the observed relation  
$\delta v\propto \ell^{1/2}$ only if they are unbound---see
\S \ref{gravity}).
Estimates (empirical and theoretical) of GMC lifetimes will be  
discussed in \S \ref{GMC_evol}.
It should be borne in mind that the difference between the two 
scenarios is only a factor $\sim 2-3$
for GMC lifetimes, which makes it difficult to obtain an unambiguous
observational resolution purely based on timescales.
However, there are major physical distinctions between the limiting
cases of the scenarios that are under consideration -- e.g. collapse
triggered in colliding flows vs. a quasi-steady state supported by
internally-driven turbulence.  As complete
numerical simulations are developed to flesh out the current proposals,
it will be possible to distinguish among them using detailed kinematic 
observations.

\subsubsection{Clumps and Cores}
\label{section_clumps}

GMCs are highly clumped, so that a typical molecule is in 
a region with a density significantly greater than average.
\citet{1993ApJ...411..720L} finds that the 
typical density of molecular gas in the Galactic plane
is $\nh\simeq 3\times 10^3$~cm\eee; \citet{1993AIPC..278..311S} find
a somewhat higher value from a multi-transition study,
$\nh\simeq (4-12)\times 10^3$~cm\eee. However, the mean density in
GMCs is considerably less: since $M\propto \bar\nh R^3$ and
$\bar N_{\rm H}\propto \bar\nh R$, 
\beq
\bar\nh=
\frac{84}{M_6^{1/2}}\left(\frac{\bar N_{\rm H}}{1.5\times 10^{22}\;\mbox{cm\ee}}
  \right)^{3/2}~~~~~\mbox{cm\eee},
\eeq
where we have normalized the column density to the typical value
in the \citet{1987ApJ...319..730S} sample.
The effective filling factor of this molecular gas is then
\beq
f\equiv\frac{\bar\nh}{\nh}=\frac{0.028}{M_6^{1/2}}\left(\frac{
    3000\;\mbox{cm\eee}}{\nh}\right)\left(\frac{\bar N_{\rm H}}
  {1.5\times 10^{22}\;\mbox{cm\ee}}\right)^{3/2}.
\eeq
Note that since $f\leq 1$, clouds with $M \la 10^3\;M_\odot$
must have column densities less than the \citet{1987ApJ...319..730S} value
if their typical density is $\sim 3000$~cm\eee. The small filling factor of
molecular gas in GMCs is expected in turbulent clouds (\S \ref{densstruct}).
It should be noted that star-forming clumps are themselves clumpy,
and contain the cores that will evolve into stars.

The nature of the interclump medium is uncertain; it is not
even known if it is atomic or molecular \citep{2000prpl.conf...97W}. 
\citet{2006ApJ...647..404H} have suggested that the damping of
hydromagnetic waves incident from the ambient ISM could
maintain an interclump medium made up of warm H~I.

The physical conditions in clumps and cores have been thoroughly reviewed
by \citet{2006astro.ph..2379D} and \citet{2006astro.ph..3474W}, and we shall address only a few issues here.
First, how well are Larson's laws obeyed in clumps and cores?
Most $^{13}$CO clumps
are unbound, and therefore do not obey Larson's laws 
\citetext{e.g., \citealp{1987ApJ...323..170C}};  
the mass distribution
of such clumps can
extend in an unbroken power law from several tens of solar
masses down to Jupiter masses
\citep{1998A&A...331L..65H}.
On the other hand, \citet{1992ApJ...395..140B} 
found that most of the mass in the clouds 
is concentrated in the most massive clumps, and these appear to
be gravitationally bound. 
The virial parameter for the unbound clumps decreases with increasing
mass, in a manner similar to that observed for both the small molecular
clouds and clumps within GMCs in the outer Galaxy \citep{2001ApJ...551..852H}. 
They found that the velocity dispersion in the unbound clumps
is approximately independent of clump mass (or, since
the clump density is also about constant in each cloud, of
clump size): the unbound clumps do not obey a line width--size
relation. \citet{2001ApJ...551..852H} find the same result for clumps in the outer
Galaxy. By contrast, \citet{1992A&A...257..715F} found
that unbound clumps do obey a line-width size relation, 
albeit with considerable scatter.
\citet{1992ApJ...395..140B}
showed that the kinetic pressure in the unbound clumps in their study
is comparable to that in the host molecular cloud, which
is $P\simeq G\Sigma_{\rm MC}^2$ (eq. \ref{eq:pressure}).
Further evidence
on whether clumps or cores are bound is imprinted in their
shapes and density structure 
and
is discussed below.

It is difficult to determine from the data whether there is a line
width--size relation {\it within} individual clumps and cores.  For low-mass
cores, \citet{1998ApJ...504..207B} (see also
\citealt{1998ApJ...504..223G,2004A&A...416..191T}) 
found that the nonthermal linewidth decreases and then
reaches a minimum ``plateau'' level at a finite impact parameter $\sim
0.1\pc$ from
the center of the core.  Because of projection effects, however, it is
not possible to determine whether the observed turbulence pervades the
whole volume interior to that radial impact parameter, or whether
the turbulence is primarily in a shell
surrounding a more-quiescent core;
it is also possible that the non-thermal line width is due to
coherent oscillations of the cores \citep{2006astro.ph..2262K}.
A line width--size relation in the ensemble of
{\it different} gravitationally bound clumps and cores
 is expected only if
they have similar surface densities (eq. \ref{eq:relate}).
\citet{1999ApJS..125..161J} carried out a comprehensive study of cores
and star-forming clumps (in our terminology; ``dense cores'' in
theirs) observed in NH$_3$ and found that the objects with and without
associated IRAS sources each obeyed a nonthermal line width--size
relation with slopes of about 0.5 and 0.8, respectively.  When the
sample was divided into objects associated with or without clusters
(defined as having at least 30 embedded YSOs), the cluster sample had
a weak correlation between line width and size with a slope of only
about 0.2, whereas the non-cluster sample had a stronger correlation
with a slope of about 0.6.  However, as remarked above,
\citet{2001ApJ...551..852H} found no evidence for a line width--size
relation for small ($<10^4\;M_\odot$) molecular clouds in the outer
Galaxy; furthermore, they did not find evidence for a constant surface
density at any mass. \citet{1997ApJ...476..730P} observed a sample of
clumps that are forming high-mass stars, and did not find a line
width--size relation.

The lack of an observed linewidth-size relation in observed unbound
clumps within a given cloud is at first puzzling, since defined
volumes should sample from the overall structure function of the GMC,
which follows $\delta v \propto \ell^{0.5}$ (\S\ref{turbobs}).  Analysis of
turbulence simulations offers a resolution to this puzzle, suggesting
that many apparent clumps in moderate-density tracers such as
$^{13}$CO are not in fact single physical entities.  Observationally,
clumps are generally identified as connected overdense peaks in
position-velocity 
data cubes, with the line-of-sight velocity
acting as a surrogate for line-of-sight position.  Analysis of
simulations shows, however, that many ``position-velocity'' clumps in
fact consist of separate physical structures superimposed on the sky;
correspondingly, many physically-coherent structures have two or more
separate components when observed in line-of-sight velocity
\citep{2000ApJ...532..353P,2001ApJ...546..980O,2002ApJ...570..734B}.
Because low-contrast {\it apparent} clumps with any plane-of-sky size
may sample from the velocity field along the whole line of sight, the
linewidth varies only very weakly with size.  Since a fraction of the apparent
clumps sample velocities from a range of line-of-sight distances no larger than
their transverse extent, however, \citet{2001ApJ...546..980O} argued
that the lower envelope of the ``clump''
linewidth-size distribution should follow the scaling defined by
the true three-dimensional power spectrum; this is generally
consistent with observations \citep{1990ApJ...356..513S,1994ApJ...428..693W}.

What about Larson's third law? Since gravitationally bound clumps have
$P\simeq G\Sigma_{\rm bd\ clump}^2$, and the mean pressure in a bound
clump cannot be much greater than the ambient pressure
for a stable structure without an internal energy source,
it
follows that typically
$\Sigma_{\rm bd\ clump}$ 
is comparable to 
$\Sigma_{\rm GMC}$
\citep{1999osps.conf...29M}.
The high-mass, star-forming clumps studied by
\citet{1997ApJ...476..730P} violate this conclusion: they have
$\Sigma\sim 
4800 \, M_\odot~\mbox{pc\ee}\simeq 1$~g~cm\ee\ 
with considerable dispersion,
which is much greater than the typical GMC surface
density 
$\sim 
170 \,M_\odot\;$pc\ee. 
There are several
possible explanations for this, and it is important to determine
which is correct: Are these clumps 
just the innermost, densest parts of much larger clumps?
Do they have much higher pressures than their surroundings but are avoiding gravitational
collapse due to energy injection from star
formation? 
Or are they the result of a clump-clump collision that produced
unusually high pressures?

The density structure,
velocity structure, and shape 
of cores offer potential means for
determining whether they are dynamic objects, with short lifetimes, or
quasi-equilibrium,
gravitationally bound,
objects. The observation of the Bok globule B68 in
near-infrared absorption revealed an
angle-averaged
density profile consistent with
that of a Bonnor-Ebert sphere to high accuracy
\citep{2001Natur.409..159A}.  Since then, a number of other isolated
globules and cores have been studied with the same technique and fit to
Bonnor-Ebert profiles, showing that starless cases are usually close
to the critical limit, while cases with stars often match
supercritical profiles
\citep{2005ApJ...629..276T,2005AJ....130.2166K}.  Profiles of dense
cores have also been obtained using sub-mm dust emission (see 
\citealt{2006astro.ph..2379D}).  A recent study by
\citet{2005MNRAS.360.1506K} found that ``bright'' starless cores 
have density profiles consistent with supercritical 
Bonnor-Ebert spheres.  

Consistency of density profiles with the Bonnor-Ebert profile 
does not, however, necessarily imply that a core is bound.
Analysis of dense
concentrations that arise in turbulence simulations show that
Bonnor-Ebert profiles often provide a good fit to these structures
(provided they are averaged over angles),
even when they are transients rather than true bound cores 
\citep{2007prpl.conf...63B}. Even if a cloud with a Bonnor-Ebert
profile is bound, however, it need not be in equilibrium: 
\citet{2005ApJ...623..280M} and \citet{2005AJ....130.2166K} have shown
that density profiles of {\it collapsing} cores initiated from near-critical
equilibria in fact follow the shapes of {\it static} 
supercritical equilibria very closely.  
The reason for this is that 
initially
these cores collapse slowly,
so that they are approximately in 
equilibrium; at later times, they evolve via
``outside-in'' collapse 
to a state that is marginally Jeans unstable everywhere 
(\S \ref{lowmass}) so that
$\rho\propto r^{-2}$ except in a central core; 
highly supercritical Bonnor-Ebert spheres have profiles with the same
characteristic shapes.
Thus, not only the density structure but
also the level of the internal velocity
dispersion and detailed shape of the line profiles must be used in
order to distinguish between transient, truly equilibrium, and collapsing
objects \citep{2005ApJ...635.1151K}.

Core shapes also provide information on whether cores are transient or
are bound, quasi-equilibrium objects. In the absence of a magnetic field,
a quasi-equilibrium, bound cloud is approximately spherical. If the cloud
is threaded by a magnetic field that tends to support the cloud
against gravity, it will be oblate; if the field tends to compress the
cloud (as is possible for some helical fields--\citealt{2000MNRAS.311...85F}),
it will be prolate. If one assumes axisymmetry, 
the distribution of observed axis ratios implies dense cores are primarily 
prolate \citep{1996ApJ...471..822R}. 
However, this conclusion appears to be an artifact of the assumption of
axisymmetry:
using the method of analysis
for triaxial clouds developed by \citet{2000ApJ...540L.103B}, 
\citet{2001ApJ...551..387J} and \citet{2002ApJ...569..280J}
concluded that cores with
sizes $< 1$~pc are in fact oblate. \citet{2000ApJ...540L.103B} showed that if the 
magnetic field is aligned with
the minor axis, as in most quasi-equilibrium models, the projection of
the field on the plane of the sky will not generally be aligned with the projection
of the minor axis, and he argued that the limited polarization data available
are consistent with the theoretical expectation that
the field in the cloud is aligned with the minor axis of the cloud.
\citet{2003A&A...411..149K} showed that larger structures, extending up
to GMCs, are intermediate between oblate and prolate, and are clearly
distinct from the smaller objects. This is consistent with the analyses
of clump shapes in turbulence simulations by
\citet{2003ApJ...592..203G,2004ApJ...605..800L}, who found that the
majority of objects are triaxial.
The data on cores and small clumps are
thus consistent with (but do not prove) that they are bound, quasi-equilibrium
objects. Large clumps and GMCs appear to be farther from equilibrium.

A key feature of the cores that form individual 
low-mass
stars is that they have low nonthermal velocities, whether these cores are
found in isolation or clustered with other cores 
\citep{2006astro.ph..2379D,2006astro.ph..3474W}.
The mean one-dimensional velocity dispersion 
in starless cores based on 
the sample of \citet{1989ApJS...71...89B} is 
$0.11\;
\kms$, 
such that the
three-dimensional velocity dispersion is approximately sonic.  
This places constraints on theoretical models, and in
particular may constrain the nature of turbulent driving.
\citet{2005ApJ...620..786K} compared the results for cores identified
in two (unmagnetized) simulations with the same RMS Mach number $\approx 10$, one
driven on large scales and the other driven on small scales;
the time correlation of the driving force is short for both cases.
They found
that only with large-scale driving is the mean turbulence level within cores
approximately sonic; in the small-scale driving case the preponderance
of cores are supersonic.  
\citet{2005ApJ...620..786K} also found that the starless
cores in their large-scale-driving models 
are within a factor of a few of kinetic and gravitational
energy equipartition.  
In the 2D MHD simulations of \citet{2005ApJ...631..411N} that implement
driving by instantaneous injection of radial ``wind'' momentum when
(low-mass) stars are formed, the dense
cores that are identified also primarily have subsonic
internal motions.
Importantly, both types of models show that dense, quiescent cores can form in
a turbulent environment; the slow, diffusive formation of quiescent
cores central to 
the older picture of star formation does not seem to be required.

What happens to dense cores once they form?  Cores that have
sufficient internal turbulence compared to their self-gravity 
will redisperse within a crossing time.  Cores that reach low enough
turbulence and magnetization 
levels (allowing for dissipation) within a few local free-fall times
will collapse if $M>M_{\rm cr}$. \citet{2005ApJ...618..344V}
found that in globally supercritical 3D
simulations with driven turbulence, cores
that collapse do so
within 3--6 local free-fall times of their formation.
\citet{2005ApJ...631..411N} 
found via 2D simulations including ambipolar diffusion
that even when  
the mass in the simulation volume is 20\% less 
than critical,
supercritical cores can form; these then either collapse or redisperse
within several local
free-fall times.  In both cases (see also
\citealt{2005ApJ...635.1126K}), only magnetically 
supercritical cores collapse, as expected.
Quiescent cores that are stable against
gravitational collapse could in principle survive for a long time
\citep{1989ApJ...342..834L}, undergoing oscillations in response to
fluctuations in the ambient medium
\citep{2005ApJ...635.1151K,2006astro.ph..2262K}.  Because they are
only lightly bound, however, such ``failed cores'' can also be destroyed
relatively easily by the larger-scale, more powerful turbulence in the
surrounding GMC.  This process is clearly seen in numerical
simulations; \citet{2005ApJ...618..344V} and \citet{2005ApJ...631..411N} found
that the bound cores that subsequently disperse do so in 1 -- 6 $\times
t_{\rm ff}$.  
Quiescent, magnetically subcritical 
cores with thermal pressure 
$\rho_{\rm core}
\cs^2$
exceeding
the mean turbulent pressure 
$\bar \rho \sigma_{\rm nt}^2$
(so that the 
core would collapse in the absence of magnetic support)
cannot easily be destroyed,
however, and it is likely that they remain intact until they merge
with other cores to become supercritical.
Simulations have not yet
afforded sufficient statistics to determine the mean time to collapse
or dispersal as a function of core properties and cloud turbulence
level, or whether there is a threshold density above which ultimate
collapse is inevitable.

Observationally, core lifetimes can be estimated by
using chemical clocks or from 
statistical inference.
The formation of complex molecules takes $\sim 10^5$~yr
at typical core densities, but this ``clock'' can be
reset by events that bring fresh C and C$^+$ into the core, such as
turbulence or outflows \citep{2000prpl.conf...29L}.
A potentially more robust clock is provided by observations
of cold H~I in cores: 
\citet{2005ApJ...622..938G} 
infer ages of $10^{6.5-7}$ yr for five dark clouds from the low 
observed values of 
the H~I
/\htwo\ ratio.
These age estimates would be reduced if clumping
is significant
and hence the time-averaged molecule formation rate is
accelerated, but, as in the case of
complex molecules, they would be increased if
turbulent mixing were effective in bringing in fresh atomic hydrogen.
In simulations of molecule formation in
a turbulent (and therefore clumpy) medium, 
\citet{2006astro.ph..5121G} find that \htwo\ formation
is indeed accelerated when compared with the non-turbulent
case, although the atomic fractions they found
are substantially greater than those observed by \citet{2005ApJ...622..938G}.
If confirmed, these ages, which are considerably greater than
a free-fall time, would suggest that these dark clouds are
quasi-equilibrium structures.

Statistical studies of core lifetimes are based on comparing the number of
starless cores with the number of cores with embedded YSOs 
and the number of visible T Tauri stars. 
The ages of the cores (starless and with embedded YSOs) 
can then be inferred from the ages of the T Tauri population, 
provided that most of the observed starless
cores will eventually become stars. The results of several such
studies have been summarized by \citet{2006astro.ph..3474W}, who conclude that
lifetimes are typically $3-5 \,t_{\rm ff}$ for starless cores with densities
$n_{
\rm
H_2}=10^{3.5}-10^{5.5}\;\cmt$. 
This is not consistent with dynamical
collapse, nor is it consistent with a long period 
($>5 t_{\rm ff}$)
of ambipolar
diffusion. It is consistent with the ambipolar diffusion in observed
magnetic fields (\S \ref{magnetic_fields}), which are approximately
magnetically critical.  Of course, cores are created with a range of
properties, and observational statistics are subject to an evolutionary
selection effect:  cores that are born or become supercritical evolve 
rapidly into collapse, and are no longer counted among the
starless population.  Given a population with a range of intrinsic
lifetimes (but similar birth rates), the longest-lived objects will be 
the best represented.  The data rule out the possibility that most cores
are born very
subcritical and lose their magnetic flux slowly, over $\sim 10 t_{\rm ff}$.

The angular momentum of cores was initially regarded as a bottleneck
for star formation, but extensive theoretical analysis led to the
conclusion that magnetic fields would provide an effective braking
mechanism \citep[e.g.,][]{1985prpl.conf..320M,1987ppic.proc..453M}.
Observations have established that the angular momentum, or
equivalently, the rotational energy, of cores is indeed small
\citep[e.g.,][]
{1993ApJ...406..528G,1999ApJS..125..161J,2002ApJ...572..238C,2003A&A...405..639P}.
\citet{1993ApJ...406..528G} characterized the rotational energy by the
parameter 
\beq \brot\equiv \frac 13\left(\frac{v_{\rm
      rot}^2}{GM/R}\right), 
\label{eq:brot}
\eeq 
which is the ratio of the rotational
energy to the gravitational binding energy for a uniformly rotating,
constant density sphere; they found a median value $\brot\simeq 0.03$.
The specific angular momenta $j$ in this sample range from $6\times
10^{20}-4\times 10^{22}$ cm$^2$ s$^{-1}$, and increase with size
approximately as $j\propto R^{3/2}$. Interestingly enough,
\citet{2000ApJ...543..822B} showed that rotation arising from sampling
turbulent fluctuations with a Burgers power spectrum (and
normalization matched to observations) is adequate to account for the
observations;
in this picture, the role of magnetic braking on small scales is unclear.
(On the other hand, magnetic braking appears to be clearly significant in
regulating the spin of GMAs and hence GMCs --
\citealp{2003ApJ...599..258R,2003ApJ...599.1157K}.)
  Since $j\propto v R$ and $v\propto R^{1/2}$ for
large-scale turbulence, 
the observed $j\propto R^{3/2}$ relation is what would be
expected if core turbulence scales similarly.
\citet{2004ApJ...605..800L} indeed find agreement with this scaling
from cores identified in their simulations.
\citet{2003ApJ...592..203G} and \citet{2005A&A...435..611J} both find
that the mean specific angular momentum of cores in their models
(using grid-based MHD and 
[unmagnetized]
SPH, respectively) is given in terms of the
sound speed and large-scale Jeans length by 
$\sim 0.1 \cs \lambda_J$;
\citet{2005A&A...435..611J} show that if the mean density is adjusted
so that core masses match those in observed regions, then the mean
angular momentum distributions match as well.  
For cores that
collapse, \citet{2005A&A...435..611J} also found that the distributions of
$\brot$ are similar to those obtained by \citet{1993ApJ...406..528G}.

\subsection{Formation, Evolution, and Destruction of GMCs}

\subsubsection{CLOUD FORMATION}
\label{GMC_formation}



In principle, GMCs could form either by ``bottom-up'' or by
``top-down'' processes.  In bottom-up formation, successive inelastic
collisions of cold H~I clouds would gradually increase the mean cloud
size and mass until that of a (self-gravitating) GMC is reached 
\citep[e.g.][]{1965ApJ...142..568F,1979ApJ...229..567K}.  The difficulty
with this coagulation scenario, as was early recognized, is that it is
very slow; e.g. \citet{1979ApJ...229..567K} found that the time needed
for the peak of the mass distribution to exceed $10^5\msun$ is more
than $2\times 10^8$ yr.  
The binary collision time for spherical clouds of radius
$R_{\rm cl}$ and density $\rho_{\rm cl}$ is 
$t_{\rm collis}=(\sqrt{\pi}/3)(\rho_{\rm cl}/\bar\rho)(R_{\rm
  cl}/\sigma)$, where $\bar\rho$ is the density averaged over
large scales and $\sigma$ is the one-dimensional 
velocity dispersion over the mean intercloud separation
\citetext{see \citealt{1987gady.book.....B}, eq. 8-122; note that for
  considering agglomeration we neglect grazing collisions, choosing a
  maximum impact parameter $R_{\rm cl}$}.  
Expressed
in terms of the cloud ``gathering
scale'' 
$R_{\rm gath}\equiv [3M_{\rm cl}/(4\pi \bar\rho)]^{1/3}=190 \pc 
(M_{\rm cl,6}/\bar 
n_{\rm H})
^{1/3}$ 
in the
diffuse ISM, or in terms of 
the cloud surface density $\Sigma_{\rm cl}\equiv
M_{\rm cl}/(\pi R_{\rm cl}^2)$, the collision time is
\begin{equation}\label{collis_time}
t_{\rm collis}=\frac{\sqrt{\pi}}{3}
\left(\frac{\rho_{\rm  cl}}{\bar\rho}\right)^{2/3} \frac{R_{\rm gath}}{\sigma}
            =\frac{\sqrt{\pi}}{4}\frac{\Sigma_{\rm cl}}{\bar \rho \sigma}.
\end{equation}
The mean intercloud separation is comparable to $2 R_{\rm gath}$, which 
exceeds the atomic 
disk scale height $H\approx 150$ pc \citep{1995ApJ...448..138M}
for $M_{\rm cl,6}\equiv M_{\rm cl}/10^6 \Msun\linebreak[1]\simgt 0.04$. We can use 
equation (\ref{collis_time})
to estimate the collision time  if all the 
diffuse ISM gas were
apportioned into equal-mass clouds with equal surface density.  Using 
$\sigma \approx 7\; \kms$ for the 
nonthermal 
velocity dispersion in the diffuse ISM at large ($\simgt H$) 
scales \citep{2003ApJ...586.1067H}, 
$\Sigma_{\rm cl}\approx 170\;
\msun \pc^{-2}$
for the 
mean GMC column \citep{1987ApJ...319..730S}, and mean density 
$\bar n_{\rm H}=0.6\, \cmt$ 
typical of the diffuse ISM at the Solar circle  
\citep{1990ARA&A..28..215D}, this yields a collision 
timescale $>5\times 10^8$ yr.  
Gravitational focusing in principle 
decreases the cloud-cloud collision time, but in practice this does
not help in forming GMCs from atomic clouds since the reduction
factor for the collision time, 
$[1+ \pi G R_{\rm cl} \Sigma_{\rm cl}/\sigma^2]^{-1}$, is near unity
until the clouds are quite massive ($\simgt 10^5 \,\msun$).
Even if the background density were
arbitrarily (and unrealistically) 
enhanced by a factor 100 to approach $\rho_{cl}$, 
the total time of 40 Myr required to build clouds from 
$10^4\;\msun$ to $5\times 10^6\msun$ (by successive stages of collisions)
would still exceed the estimated
GMC lifetimes.  These lifetimes are set by
the time required to destroy clouds by a combination of
photodissociation and mechanical unbinding by expanding HII regions
(see \S \S \ref{GMC_evol}).
Thus, if coagulation were the only way to build GMCs, the process
would be truncated by destructive star formation 
before achieving the high GMC masses in which most molecular mass is
actually found.

Given the timescale problem and other difficulties of 
bottom-up GMC formation \citep[e.g.][]{1980ApJ...238..148B}, starting in the
1980's the focus shifted to top-down mechanisms involving
large-scale instabilities in the diffuse ISM 
(e.g., \citealp{1979ApJ...231..372E,1995mcsf.conf..149E}).
The two basic physical processes that could trigger growth of massive GMCs
involve (1) differential vertical buoyancy of varying-density regions
along magnetic field lines parallel to the midplane, or (2) differential 
in-plane self-gravity of regions with varying surface density.  The
first type of instability is generically termed a Parker instability 
\citep{1966ApJ...145..811P}.  The second type of instability is
generically a Jeans instability, although the simplest form of Jeans
instability involving just self-gravity and pressure 
cannot occur, due to galactic (sheared) rotation (see \S\ref{gravity}).  
If the background rotational shear is strong,
as in the interarm regions of grand design
spirals or in flocculent galaxies, 
there is no true instability but instead a process known as
swing amplification \citep{1965MNRAS.130...97G,1981seng.proc..111T};
the dimensionless shear rate must be $d\ln\Omega/d\ln R\la - 0.3$
for ``swing'' to occur \citep{2001ApJ...559...70K}.  If, on the other
hand, the mean background dimensionless shear rate is low (as in the inner
parts of galaxies where rotation is nearly solid-body, or as in spiral
arms), another type of
gravitational instability can develop provided magnetic fields are present 
to transfer angular momentum out of growing condensations
\citep{1987ApJ...312..626E,2001ApJ...559...70K}; this is referred to
as a magneto-Jeans instability (MJI).

The characteristic azimuthal spatial scale for Parker instabilities is 
$\lambda_\phi\approx 4\pi H$
\citep{1974A&A....33...55S}.  Growth rates are $\propto \vA/H$, which
tends to increase in spiral arms; thus these regions have
traditionally been considered most favorable for growth of Parker
modes \citep{1974A&A....33...73M}.  Numerical simulations have shown,
however, that Parker instability is not on its own able to create
structures resembling GMCs, because the instability is self-limiting
and saturates with only order-unity surface density enhancement 
\citep{1998ApJ...506L.139K,2000ApJ...545..353S,2001ApJ...557..464K,2002ApJ...581.1080K}.
Spiral arms are also the most favorable regions for self-gravitating 
instabilities \citep{1994ApJ...433...39E}, since the
characteristic (thin-disk) growth rate $\propto G \Sigma_{\rm gal}/\cs$
is highest there.  
(Here, $\Sigma_{\rm gal}$ is the mean gas
surface density averaged over large [$\sim$~kpc] 
scales in the plane of the disk.)
Since the spatial wavelengths of Parker and MJI
modes are similar, in principle growth of the former could help
trigger the latter within spiral arms 
\citep{1982ApJ...253..634E,1982ApJ...253..655E}.
In fact, it appears that turbulence excited in spiral shocks, together with 
vertical shear of the horizontal flow, may suppress growth of
large-scale Parker modes in arm regions \citep{2006ApJ...646..213K}.
Thus, while the Parker instability is important in removing excess
magnetic flux from the disk and in transporting cosmic rays 
\citep[e.g.][]{2000ApJ...543..235H}, it may be of limited
importance in the formation of GMCs.

Self-gravitating instabilities, unlike buoyancy instabilities, lead to
ever-increasing density contrast if other processes do not intervene.
The same is true for the swing amplifier if the (finite) growth is
sufficient to precipitate gravitational runaway.
The notion that there should be a threshold for
star formation depending 
on 
the Toomre parameter, 
\begin{equation}
Q \equiv \frac{\kappa\cs}{\pi G \Sigma_{\rm gal}}
= 1.4 \left(\frac{\cs}{7.0 {\rm\,km\,s^{-1}}} \right)
      \left(\frac{\kappa}{36\,{\rm km\,s^{-1}\,kpc^{-1}}}\right)
      \left(\frac{\Sigma_{\rm gal}}{13\,\Msun\pc^{-2}}\right)^{-1},
\label{Qdef}
\end{equation}
is based on the idea that star-forming clouds can form by
large-scale self-gravitating collective effects 
only if $Q$ is sufficiently low.  
Here, 
$\kappa^2 \equiv R^{-3}\partial( \Omega^2R^4)/\partial R $ is the
squared epicyclic frequency, and $\cs$ is the mean sound speed of the gas.
Numerical simulations have been used to determine
the nonlinear instability criterion,
finding that gravitationally bound clouds form provided 
$Q<Q_{\rm crit}\approx 1.5$ in 
model disks that allow for realistic vertical thickness, turbulent 
magnetic fields,
and a ``live'' stellar component 
\citetext{\citealt{2003ApJ...599.1157K,2005ApJ...626..823L,2006ApJ_submitted};
these models do not include global spiral structure in the gas 
imposed by variations in the stellar gravitational potential -- see below.}
These results agree with 
empirical findings for the mean value of $Q$ at the star
formation threshold radii in nearby galaxies
\citep[e.g.][]{1972ApJ...176L...9Q,1989ApJ...344..685K,2001ApJ...555..301M}.  
The masses of bound clouds formed via self-gravity
in galactic disk models where the background gas surface density is relatively
uniform are typically a few to ten times the
two-dimensional Jeans mass,
\begin{equation}\label{MJ2D}
M_{J,2D}\equiv \frac{\cs^4}{G^2\Sigma_{\rm gal}}
= 10^7 \Msun 
\left(\frac{\cs}{7\,\kms}\right)^4
\left(\frac{\Sigma_{\rm gal}}{13\Msun\pc^{-2}}\right)^{-1},
\label{MJ2Ddef}
\end{equation}
depending on the specific ingredients of the model 
\citep{2002ApJ...581.1080K,2003ApJ...599.1157K,2006ApJ_submitted}.

Observations of external galaxies with prominent spiral structure show
that most of the molecular gas is concentrated in the spiral arms
\citep[e.g.][]{2003ApJS..145..259H,2003ApJS..149..343E}, and within
the Milky Way the most massive clouds that contain most of the mass
and forming stars are strongly associated with spiral arms
\citep{1985ApJ...292L..19S,1989ApJ...339..919S,1998ApJ...502..265H,2006ApJ...641L.113S}.
The observed relationship between GMCs and spiral structure suggests that
molecular clouds are preferentially born in the high density gas that
makes up the arms; this is consistent with theoretical
expectations since growth rates for all
proposed mechanisms increase with the gas surface density. 
As noted above, collisional coagulation is expected to be 
too slow in spiral arms; gravitational instabilities are, however, faster
\citep[e.g.][]{1990ApJ...357..125E}.  Taking the ratio of the collision rate 
$t_{\rm collis}^{-1}$
to the characteristic growth rate of the MJI, $\pi G
\Sigma_{\rm gal}/\cs$, and setting $\sigma=\cs$ and 
$H\approx \cs^2/(\pi G \Sigma_{\rm gal})$ (since gas gravity dominates
stellar gravity in the arm), the result is 
$t_{\rm MJI}/t_{\rm collis}=2 \Sigma_{\rm gal}/(\sqrt{\pi} \Sigma_{\rm cl})$.  
Thus, the collision rate
is lower than the self-gravity contraction rate by roughly the surface
filling factor of clouds in the arm, $\sim 0.2-0.5$ if the arm surface
density is enhanced by a factor 3 -- 6 above the mean value.

To obtain realistic estimates of the masses and other properties of
clouds formed via gravitational instabilities, it is necessary to
include the effects of spiral structure in detailed numerical models.  Diffuse
gas entering a spiral arm will in general undergo a shock,
significantly increasing the background density
\citep[e.g.][]{1969ApJ...158..123R,1973ApJ...183..819S}; gas
self-gravity enhances the maximum compression factor and also tends to
symmetrize the gas density profile across the arm
\citep{1986ApJ...309..496L}.  In addition to strong variations in the
gas surface density, spiral structure induces corresponding local
variations in the gas flow velocity (both compression and streaming).  
Since the Jeans length is not small compared to the scale of
these variations, the background arm ``profile'' must be taken into
account in studying growth of self-gravitating condensations
\citep{1988ApJ...324...60B}.  \citet{2002ApJ...570..132K} performed
2D MHD simulations of this process, showing
that bound condensations develop both within the spiral arms
themselves, and also downstream from the arms in trailing gaseous
``spurs''.   Based on 3D
local MHD simulations \citep{2006ApJ...646..213K} and 2D ``thick
disk'' global simulations \citep{2006ApJ...647..997S}, bound gas
condensations formed in spiral arms (and arm spurs) have typical
masses $1-3\times 10^7\msun$.  This value is $\sim 10\times$ the
thin-disk 2D Jeans mass using the peak arm density in equation
(\ref{MJ2D}), or comparable to the value of the ``thick disk'' Jeans
mass $M_{J,\, \rm thick}=2\pi \cs^2 H/G$
which is obtained using the gravitational kernel 
$\Phi_k=-2\pi G \Sigma_k/(k + k^2H)$.

The masses of the bound structures formed via self-gravitating
instabilities are comparable to the upper end of the mass function of
GMCs in the Milky Way 
(see eq. \ref{GMC_massfnct}), allowing for H~I envelopes; they are
also comparable to the masses of 
GMAs 
that have been observed in external spiral galaxies 
\citep[e.g.][]{1988Natur.334..402V,
1999ApJ...522..165A,1999ApJ...513..720R}. 
In addition, the morphology and spacing of spiral-arm 
spurs predicted to form via self-gravity effects are consistent
with structures that have been observed at a variety of wavelengths
\citep[e.g.][]{1980ApJ...242..528E,2006astro.ph..6761L}.   This
spacing (of several times the Jeans length in the arm, 
$\cs^2/[G\Sigma_{\rm gal,\, arm}]$) 
is similar to that of giant HII regions arranged as 
``beads on a string'' along spiral arms in many grand design spirals 
\citep{1983MNRAS.203...31E}, and also similar to the spacings of giant
infrared clumps observed with {\it Spitzer} along the  spiral arms of 
interacting galaxies IC 2163 and NGC 2207 \citep{2006ApJ...642..158E}.
These giant IR clumps typically host a number of individual
(optically-observed) H~II regions associated with star clusters.
While current mm-wavelength observations have insufficient
resolution to identify sub-condensations within GMAs in external galaxies, 
it is expected that their internal turbulence 
would create density substructure, just as the internal turbulence
of GMCs fragments them into clumps.  
Since the mean density within GMCs is comparable to the typical density of cold
clouds in the atomic medium, the pre-existing cloudy structure of the diffuse
ISM would contribute to, but not dominate, the internal
structure within GMCs.
In this ``top-down'' picture, the
more massive, self-gravitating, substructures within GMAs 
(or analogous atomic ``superclouds'')
would then 
become gravitationally bound GMCs.

Although large-scale self-gravitating instabilities appear necessary for
forming massive GMCs, and many low-mass GMCs may form via
fragmentation of massive GMCs or GMAs, it remains possible that a
proportion of the low-mass GMCs form through other mechanisms.
Several recent studies have explored the possibility of GMC
assembly via colliding supersonic flows 
\citep[e.g][]{1995MNRAS.275..313C,1995ApJ...441..702V,1999ApJ...527..285B,2005ApJ...633L.113H,
2007ApJ...657..870V}; in this scenario 
the post-shock gas in the stagnation region (which  
in fact becomes turbulent) represents the nascent GMC.  For diffuse
ISM gas at 
mean density $\bar \rho$ with relative (converging) velocity
$v_{\rm rel}$, a total column of shocked gas $\Sigma_{\rm cl}$ builds up over
time 
\begin{equation}\label{accum-time}
t_{\rm accum}=\frac{\Sigma_{\rm cl}}{\bar\rho v_{\rm rel}}= 1.6\times 10^7\yr
\left(\frac{N_{\rm H}}{10^{21}\cm^{-2}}\right)
\left(\frac{\nh}{1\;\cmt}\right)^{-1} 
\left(\frac{v_{\rm rel}}{20\;\kms }\right)^{-1}; 
\end{equation}
note that, modulo order-unity coefficients, this time is
the same as the result in equation (\ref{collis_time}), with the
velocity dispersion of the cloud distribution replaced by the relative
velocity of the converging flow.  Correlated flows
can only be maintained up to the flow timescale over the
largest spatial scale of the turbulence, $\sim 2H\sim 300\;\pc$.  With 
$v_{\rm rel}$ equal to the RMS relative velocity $\sqrt{6}\sigma$ 
for a Gaussian with 1D
velocity dispersion $\sigma\approx 7\;\kms$, this is $\approx 2\times
10^7\,\yr$, yielding a column $\approx 10^{21}\cm^{-2}$ for $\bar n
\approx 1\; \cmt$
(note that the shock velocity is about $v_{\rm rel}/2$).
If the interstellar magnetic field does not
limit the compression of the shocked gas 
(an artificial assumption, requiring flow along field lines only),
the post-shock gas would have high
enough density for significant amounts of
H$_2$ to form within the overall accumulation time,
and the shielding from the diffuse UV is sufficient for
CO to begin to form
\citep{2001ApJ...562..852H,2004ApJ...612..921B}.  However, 
it should be noted that for a shock velocity of 10~km~s\e,
corresponding to a relative velocity of 20~km~s\e,
less than half the C is in CO after $10^8$~yr according to
the 1D calculations of \citet{2004ApJ...612..921B}.  
Turbulence-induced clumping can accelerate molecule formation rates 
\citep{2000ApJ...530..277E,2006astro.ph..5121G}, 
alleviating the timescale problem.  
The molecule formation rate is proportional to the mass-weighted mean density 
$\langle n\rangle_M$, which
is larger than the volume-weighted mean density 
$\bar n\equiv\langle n\rangle_V$ in a turbulent flow 
(see \S\ref{densstruct}). Since $\langle n\rangle_M/\bar n\sim 10$ for
typical turbulence levels in GMCs, this reduces the typical 
molecule formation time \citep{2005pcim.book.....T},  
$\sim 2\times 10^9\, \yr\, (T/10\K)^{-1/2}/\langle n\rangle_M$, 
to 1-2 Myr.
Even so, 
the discussion above shows that the maximum
column density produced in the 
colliding-flow scenario is $\sim 10^{21}$~cm\ee, which is
lower by an order of magnitude than the mean value of the column
of molecular gas
in Milky Way GMCs; thus, this process can account for at most
a small fraction of the
molecular gas mass in GMCs. 
Because gravitational instabilities
are suppressed by the high $Q$ values in interarm regions, on the
other hand, the turbulent
accumulation mechanism may 
be more important
there.  Potentially, this may
account for the observed difference (see above) between arm and interarm GMC
masses in the Milky Way, 
as well as for the very low surface densities $\Sigma_{\rm cl}$ 
observed for many of the
outer-Galaxy molecular clouds \citep{2001ApJ...551..852H}.

Finally, we note that the dynamical considerations for
gravitationally-bound cloud formation apply whether the diffuse gas is
primarily atomic, as is the case in the Solar neighborhood and the
outer portions of galaxies more generally, or whether the diffuse gas
is primarily molecular, as is true in the inner portions of many galaxies.
The time and length scales involved depend on the effective pressure
in the diffuse gas, which includes thermal as well as
turbulent and magnetic terms.  If the diffuse gas is primarily
molecular (or cold atomic) by mass, then the mean turbulent and Alfv\'en
speeds will exceed the thermal speed in the dense gas. 
The thermal sound speed $\cs$ in equations 
(\ref{Qdef}) and (\ref{MJ2Ddef}) must then be replaced by an
appropriately-defined $c_{\rm eff}$ that incorporates the effects of turbulent
kinetic and magnetic pressures (the form of $c_{\rm eff}$ 
would depend on the detailed multiphase structure of the gas).
Similarly, 
the characteristic timescale for self-gravitating cloud formation becomes
\begin{equation}
t_{J,2D}=\frac{c_{\rm eff}}{ G \Sigma_{\rm gal}} 
=3\times 10^7 {\rm yr} \left(\frac{c_{\rm eff} }{10\, \kms  }  \right) 
\left(\frac{\Sigma_{\rm gal} } {100 \Msun \pc^{-2} }  \right)^{-1}.
\label{tJ2D}
\end{equation}
Turbulent velocity dispersions
and magnetic field strengths are observed to be similar in the cold
and warm diffuse gas in the Solar neighborhood
\citep{2003ApJ...586.1067H,2005ApJ...624..773H}, and both observations
and simulations \citep{2005ApJ...629..849P} show that magnetic
pressure is generally a factor two larger than the thermal pressure.
 
The transition from having primarily atomic to primarily molecular gas
typically occurs where 
the total gas surface density
$\Sigma_{\rm gal}\approx 12\Msun\pc^{-2}$
\citep{2002ApJ...569..157W,2004ApJ...612L..29B} and where the mean
midplane pressure is inferred to lie in the range $P/k=10^4-10^5 \K \,\cmt$ 
\citep{2006ApJ...650..933B}.  
This transition
occurs due to a combination of increased self-shielding (hence a lower H$_2$
dissociation rate) as 
$\Sigma_{\rm gal}$ 
increases, and increased density
(hence increased H$_2$
formation rate) as both 
$\Sigma_{\rm gal}$ 
and the stellar surface density increase toward the
center of a galaxy \citep{1993ApJ...411..170E,2004ApJ...612L..29B}.

\subsubsection{CLOUD EVOLUTION AND DESTRUCTION}
\label{GMC_evol}


GMCs are born in spiral 
arms downstream from the 
large-scale 
shock fronts, 
and begin life in a very
turbulent state. As they contract under the influence of gravity, this
turbulence decays, although the rate of decay is slowed by
compression. 
\citet{1956MNRAS.116..503M} conjectured that turbulence would decay in
about a crossing time, and for that reason rejected turbulence as a
mechanism for supporting clouds against gravitational collapse
(see also \citealp{2006ApJ...646.1043M}, who argue for magnetic support).
Indeed, in simulations that do not include energy injection, 
the contraction 
eventually evolves into free-fall collapse
\citetext{cf. \citealp{2007ApJ...657..870V}, 
who simulated the formation of
molecular clouds
in colliding flows, as discussed above}, 
which is generally not observed.
Furthermore, as discussed in 
\S \ref{dynam_state},
turbulence in molecular
clouds is observed to be ubiquitous, suggesting that  there is some
mechanism acting to inject turbulent energy into clouds.  
How important is energy injection to the structure and evolution of molecular clouds?

Broadly speaking, there are two modes of energy injection, external
and internal.
External mechanisms tap the turbulence in the diffuse ISM, and because
these modes are large scale, external driving would tend to yield a power
spectrum that rises all the way to the largest scale in the GMC (see
\S \ref{turb_correl}).  In terms of total amplitudes, however, 
external driving may have limited practical importance.  
For example, while cloud-cloud collisions 
could drive turbulence, they may be too rare to be
important (except possibly within the denser portions of spiral arms; see
above).   Magnetorotational instability (MRI) in the
Galactic disk \citep{1999ApJ...511..660S} connects GMCs to the diffuse
ISM by threaded field lines and could therefore help drive GMC
turbulence, but it is difficult for this to be effective since GMCs
live for only a small fraction of a rotation period (see below), and
since MRI is modest compared to other kinetic turbulence drivers in inner
galaxies \citep{2005ApJ...629..849P}.  Supernovae are the dominant
energy source in most of the diffuse interstellar medium
\citep{2004RvMP...76..125M}, with instabilities in spiral shocks
making a significant contribution in spiral arm regions
\citep{2006ApJ...649L..13K}.  It is difficult, however, for these (or
other) processes to transmit energy from the diffuse ISM into
molecular clouds, which are much denser \citep{1989A&A...216..207Y}.
In fact, the density contrast between molecular clouds and the ambient
medium means that energy tends to be reflected from clouds rather than
being transmitted into them 
\citetext{e.g.,
  \citealp{1999ApJ...527..266E} and \citealp{2002ASPC..285...13H}
showed this for externally generated
  Alfv\'en waves}.  Thus, ``external'' energy is primarily limited to
the turbulence GMCs inherit from their formation stages.

Internal energy injection is due to feedback from protostars 
and newly formed stars.
This mechanism is observed to be important, but it remains to be shown
that it can account for the ubiquity of turbulence 
given that
star formation is intermittent in both space and time.
For example, \citet{1988ApJ...334L..51M} find that 1/4 of a sample of
inner-Galaxy GMCs show no evidence for the presence of O stars.
\citet{1980ApJ...238..158N} were the first to analyze
the importance
of energy injection by stellar outflows. At the time of their work,
bipolar outflows from protostars were unknown, and they focused on
winds from T Tauri stars.  They suggested that these winds would blow
cavities that would govern the structure of the clouds.
\citet{1983ApJ...264..508F} and \citet{1983ApJ...273..243F} considered
rotationally driven winds from protostars as well as T Tauri winds.
They estimated the star formation rate required to keep swept-up
shells colliding at a rate high enough to keep the cloud turbulent,
and showed that this was roughly consistent with estimates of the
Galactic star formation rate. The effect of stellar energy injection
on the surrounding molecular cloud can be described in terms of an
energy equation for the cloud 
\citep{1989ApJ...345..782M,1999osps.conf...29M,2006astro.ph..8471K},
$de/dt={\cal G}-{\cal L}, $
where $e$ is the energy per unit mass 
(including gravitational energy)
and $\cal G$ and $\cal L$
represent energy gains and losses, respectively.  
The injection of energy into a cloud by stars is often accompanied by mass loss
from the cloud.
Under the simplifying assumption that mass lost from the cloud does not change
the energy per unit mass, 
the specific energy losses are due to the decay of MHD
turbulence, 
which occurs at a rate (see \S\ref{turbdiss})
\beq
{\cal L}=2.5\;\frac{\sigma^3}{R} \left(\frac{2R}{\lambda}\right).
\eeq 
For a cloud of fixed size, the largest driving scale is $\lambda\simeq 2R$;
for a cloud  undergoing global expansion and contraction, the largest
scale is $\lambda\simeq 4R$.
Next, consider the energy gains.  Since
the shocks associated with the energy injection are radiative, most of
the energy injected by outflows is radiated away. The outflow energy
available to drive turbulence in the cloud is about 
$(\sqrt{3}/2)p_w\sigma$, 
where $p_w=m_wv_w$ is the momentum, $m_w$ the mass, and
$v_w$ the velocity of the (proto)stellar outflow; as a result, 
${\cal
  G}=(\sqrt{3}/2) (\dot M_*/M)(p_w\sigma/m_* )$, 
where $\dot M_*$ is the star
formation rate and $M$ is the cloud (or clump) mass.
Balancing the energy gains and losses, ${\cal G}={\cal L}$, then gives the star
formation rate necessary to maintain the turbulent motions. 
\citet{1980ApJ...238..158N}, \citet{1983ApJ...264..508F}, and
\citet{1989ApJ...345..782M}
all estimated outflow momenta $p_w/m_*\simeq
50-70$~km~s\e\ for a typical stellar mass $m_*=0.5 M_\odot$
and found that the energy injection rate was sufficient to support
star-forming clouds against collapse. 
\citet{2006ApJ...640L.187L} and \citet{Nakamura2007}
have carried out 3D MHD simulations
of a forming star cluster, and found that the rate at which energy 
is input from protostellar winds maintains the surrounding gas  
in approximate virial equilibrium.
\citet{2007astro.ph..1022M}  has given a 
general discussion of turbulence driven by protostellar
outflows. In particular, he has shown on the basis of dimensional analysis
that if outflows occur at a rate 
per unit volume
$\cal S$ and inject an 
average momentum $\cal I$
into a medium of density $\rho_0$, then on scales small compared to that
at which the outflows overlap, the line width---size relation
is $\sigma\propto ({\cal S I}r/\rho_0)^{1/2}$, where the coefficient 
is of order unity.

There are both observational and theoretical caveats to this picture.
Observationally, \citet{2000prpl.conf..867R} cite $p_w/m_*\simeq 0.3
v_K$, where $v_K$ is the Keplerian velocity at the stellar surface;
for $v_K\simeq 200$~km~s\e, this is $p_w/m_*\simeq 60\;\kms$, in
agreement with the values used in these models of energy
injection and in agreement with both X-wind and disk models of
protostellar outflows. However, some recent observations are in
striking disagreement with these estimates: based on observations of
CO outflows by \citet{1996A&A...311..858B},
\citet{2005AJ....129.2308W} infer a
much lower 
average outflow momentum
$p_w=1.2M_\odot$~km~s\e\ (it should be noted, however, that they infer
a much larger momentum, $p_w=18 \;M_\odot$~km~s\e, in a protostellar
jet). \citet{2005ApJ...632..941Q} discovered a number of CO cavities
in NGC 1333, which they identified as fossil outflows with $p_w\sim
1\; M_\odot$~km~s\e; this estimate of the momentum has been confirmed in 
numerical simulations of fossil outflows \citep{2006ApJ...653..416C}.
Like Walawender et al., they concluded that these
outflows could support a region of active star formation like NGC 1333
against gravity, but outflows could not support the larger Perseus
cloud in which it is embedded. 
Clearly, more observational work is needed 
to determine the rate of protostellar energy injection.

The theoretical caveat is that protostellar outflows are unlikely to 
be effective on the scale of GMCs.
Turbulent driving yields a flat power
spectrum at scales larger than the input scale of turbulence,
regardless of magnetization
(see
\S\ref{turb_correl}). Because the scale of protostellar jets and outflows is
small compared to that of a GMC, driving by low-mass stars is
inconsistent with observed turbulence spectra that continue rising up
to scales of tens of pc (see \S\ref{turbobs});
this effect is partly ameliorated by the clustering of stars,
since protostellar outflows from a cluster will extend to larger 
scales than those from individual stars.
Protostellar outflows remain a viable energy source for clumps and cores within
GMCs, however.

Massive stars can inject more momentum into GMCs through H~II regions
than do the much more numerous low-mass stars through their outflows
\citep{2002ApJ...566..302M};
in addition, H~II regions can inject energy on the scale of the GMC, overcoming
the limitation of protostellar outflows in this regard.
 The dominant destruction mechanism for
GMCs is via photoevaporation by blister H~II regions (see below). The
momentum given to a GMC by the loss of a mass $\delta M$ in a blister
H~II region is about $2 c_{\rm II}\delta M$, where $c_{\rm II}\simeq
10$~km~s\e\ is the sound speed in the H~II region
(the factor 2 represents the sum of the thermal and ram pressures in 
a blister H~II region).
If the overall
star formation efficiency is 
$\epsilon_{\rm GMC}$, 
then a cloud of initial mass
$M$ will form a mass of stars 
$\epsilon_{\rm GMC} M$ 
and will lose a mass
$(1-\epsilon_{\rm GMC})M$ 
via photoevaporation.
The ratio of the momentum given
to the cloud by H~II regions to that given by protostellar winds is
then 
\begin{equation}
 \frac{p_{\rm H\,II}}{p_w}=\frac{(1-\epsilon_{\rm GMC})}{\epsilon_{\rm GMC}}
\left(\frac{2c_{\rm II}}{ 60\;\mbox{km~s\e}}\right)
\simeq\frac{(1-\epsilon_{\rm GMC})}{3\epsilon_{\rm GMC}}, 
\end{equation} 
where we used the value
$p_w/m_*\simeq 60$~km~s\e\ from \citet{2000prpl.conf..867R}; if
protostellar momentum injection is less efficient than this, then H~II
regions are correspondingly more important. 
(Note that if the cloud is disrupted before it is photoevaporated, as can happen
to smaller GMCs---see below---both the mass loss and the star formation will
be reduced, but $p_{\rm H\,II}/p_w$ will be relatively unaffected.)
For GMCs, which have star
formation efficiencies 
$\epsilon_{\rm GMC}\sim$ 
a few percent, H~II regions
dominate the energy injection by an order of magnitude. Large OB
associations do not occur in small molecular clouds, so
photoevaporation is less important in such clouds;
\citet{2002ApJ...566..302M} estimates that protostellar outflows
dominate the energy injection for clouds with masses $\la 4\times
10^4\;M_\odot$.

The same H~II regions that inject energy to support GMCs also destroy
them.  \citet{1980ApJ...238..148B} estimated that H~II regions would
inject enough energy to unbind a GMC in about $10^7$~yr. Using a
simple analytic model for the evolution of blister H~II regions
developed by \citet{1979MNRAS.186...59W} and confirmed in numerical
simulations by \citet{1989A&A...216..207Y},
\citet{1997ApJ...476..166W} concluded that an expanding H~II region
would sweep up more mass than it ionized, so that a very large H~II
region would disrupt the cloud before ionizing it. The condition for
cloud disruption---that the cloud be engulfed by the H~II region---is
difficult to meet for massive GMCs ($\ga 10^6\;M_\odot$), but
relatively common for small GMCs. Indeed, the Orion molecular cloud
may well have been disrupted by previous generations of star
formation. 
If one assumes that the specific star formation rate is independent of GMC
mass, 
then the observed rate of star formation in the Galaxy implies
that GMCs with masses $\sim 10^6\;M_\odot$ live for about 30~Myr.
\citet{2002ApJ...566..302M} assumed that the star formation rate in a
cloud is self-regulating and found that the specific star formation
rate increases with cloud mass; as result, he obtained smaller
lifetimes for large GMCs,
$\sim 20$~Myr. \citet{2006astro.ph..8471K} obtained a similar value
with a more complete model for GMC evolution (see below).

Observational estimates of GMC lifetimes are difficult, although
chemical clocks can be used to measure the lifetime of clumps
and cores within them (see \S
\ref{section_clumps}).
For the LMC and
M33, 
\citet
{2007prpl.conf...81B}
discuss empirical measures of GMC lifetimes based
on their spatial correlation with H~II regions and young clusters
and associations.
About 25\% of the GMCs show no evidence for high-mass star formation.
This can be interpreted as a delay in the onset of star formation
due either to the effects of magnetic fields 
\citep{1989ApJ...345..782M,2004ApJ...616..283T}, 
or to the time required
to create cluster-forming clumps in a turbulent cloud 
\citetext{from simulations 
this time is 
$\sim 1-3\, t_{\rm ff}$;
e.g.
\citealp{2001ApJ...546..980O,
2001ApJ...547..280H,2005ApJ...635.1126K,2005ApJ...618..344V}};
alternatively, if the star formation rate undergoes significant
fluctuations (e.g., \citealp{2006astro.ph..8471K}), it could simply
represent a lull in the star formation rate.
In the LMC, about 60\% 
of star clusters with ages $<10$~Myr are within
40 pc of a GMC, while older clusters have no significant spatial
correlation with GMCs; 
\citet{2007prpl.conf...81B} therefore infer that GMCs are destroyed
within about 6 Myr of cluster formation. 
There
are also 
slightly more than twice as many clouds harboring small
H~II regions as those containing large H~II regions and
clusters;
they infer a lifetime of 14 Myr for this stage.
These statistics imply that
the typical time interval between when an H~II region turns on and when
the cloud is destroyed
is $\sim 20$~Myr.  
Including the GMCs without high-mass stars, they infer a total GMC lifetime
of 27 Myr.
There are several caveats to this analysis, however.
First, the GMCs are identified only by their CO emission; since
CO is more readily destroyed in the low-metallicity environment of the LMC
than in the Galaxy \citep{1988ApJ...334..771V}, 
some GMCs could have been missed,
both in the early stages of evolution when the density is low and
in the late stage when the UV flux is high.
Second, the mean mass of the clouds increases in going from the
starless sample and the small-H~II sample to the cluster sample;
in fact, the two largest clouds are in the cluster sample,
and overall just 10-15\% of massive ($M>10^{5.5}\Msun$) 
GMCs in the LMC and M33 lack (high-mass) star formation.
The naive interpretation of this secular trend is that massive GMCs
(which contain most of the molecular mass) have a more rapid onset of
star formation than do low-mass clouds, although this is not a unique
explanation. 
Next, the
associations are unbound and dissolve rapidly; clusters dissolve
more slowly, 
but one still cannot assume that their relative numbers accurately
track the relative lifetimes.
It is not clear how these latter two issues affect the analysis.
Finally, large associations could displace the clouds by the
rocket effect rather than
destroy them 
(a process termed ``cloud shuffling'' by \citealp{1979ApJ...231..372E}). 
Nonetheless, observations of extragalactic GMCs offer
the most promising avenue for getting an observational handle on
GMC evolution.

How do GMCs and the star-forming clumps within them
evolve in the presence of stellar and protostellar feedback?
In the quasistatic approximation, a molecular cloud will be in approximate
virial equilibrium. Under the additional assumption that the 
rate-limiting step in
star formation is ambipolar diffusion, \citet{1989ApJ...345..782M} found 
that GMCs contract under
the influence of their self-gravity until the contraction is halted by stellar
feedback; the equilibrium is stable and the column density 
is $A_V\simeq 4-8$~mag, comparable to the observed value.
Dropping the assumption of quasi-static evolution and using the time-dependent
virial theorem for a spherical, homologous cloud, \citet{1999sf99.proc..353M} found
that a star-forming clump undergoes several bursts
of star formation.
\citet{2006astro.ph..8471K} extended this work by
using a theory for the star formation rate that is
consistent with the Kennicutt-Schmidt law 
\citetext{\citealp{2005ApJ...630..250K} --- see \S \ref{large-scale_SFR}}
and the full time-dependent virial and energy equations to solve for
the evolution of GMCs that are large enough ($\ga 10^5\, M_\odot$)
that H~II
regions dominate the energy injection \citep{2002ApJ...566..302M}. 
Their
time-dependent integrations assume spherical, homologous evolution for
each GMC, and 
follow the formation and
dynamical effects
of many individual H~II regions, with stars selected from
the IMF. Blister H~II regions act to destroy the clouds, but they also provide
a confining pressure due to the recoil associated with the mass loss. The
prinicipal results of the 
time-dependent models
are (1) clouds live for 
a few crossing times,
$\sim 30$~Myr for clouds with $M\ga 10^6\;M_\odot$,
in agreement with observational estimates; (2) clouds are close to equilibrium,
with virial parameters $\avir\simeq 1-2$; (3) the column density of the clouds is
$N_{\rm H}\simeq 10^{22}$~cm\ee, in agreement with observation; (4)
large clouds are destroyed by photoevaporation, but small clouds
($M\la 2\times 10^5\;M_\odot$) are disrupted before half their mass is
photoionized; and (5) GMCs convert $\sim 5-10\%$ of their mass into stars
before they are destroyed. H~II regions are unable to
support GMCs with columns significantly greater than
$10^{22}$~cm\ee. \citet{2006astro.ph..8471K} conjecture that such clouds can occur in
regions in which the mean density is not much less than the density in
the GMCs, so that external driving is more efficient; such conditions
could occur in starbursts.
Testing and extension of these cloud evolution/destruction models via
full 3D numerical simulations has not yet been
attempted, but development and verification of the necessary computational
codes is well underway
\citep{2006ApJ...647..397M,2006astro.ph..5501M,2006astro.ph..6539K}.

\subsection{Core Mass Functions and the IMF}
\label{clumps}

\subsubsection{OBSERVATIONS OF THE STELLAR IMF AND THE CMF}
\label{obs_imf_cmf}

How is the distribution of stellar masses, or initial mass function
(IMF), established?  This is one of most basic questions a complete
theory of star formation must answer, but also one of the most
difficult.  Current evidence suggests that the IMF is quite similar 
in many different locations 
throughout 
the Milky Way, with the possible exception of star
clusters formed very near the Galactic Center
\citetext{\citealp{1998ASPC..142..201S} presents evidence 
for significant variations in the IMF,
but \citealp{1999ApJ...515..323E} argues that much, if not all, 
of this is consistent with 
the expected statistical variations}.  
The standard IMF of
\citet{2001MNRAS.322..231K} is a three-part power-law with breaks at
$0.08 \Msun$ and $0.5\Msun$; i.e.  $d\caln_*/d\ln m_*\propto m_*^{-\alpha}$ with
$\alpha=1.3$ for $0.5<m_*/\Msun<50$, $\alpha=0.3$ for
$0.08<m_*/\Msun<0.5$, and $\alpha=-0.7$ for $0.01<m_*/\Msun<0.08$.  The
slope of the 
IMF
at $m_*\simgt \Msun$ was
originally identified by \citet{1955ApJ...121..161S}, who found
$\alpha=1.35$. Up to $\sim 1\Msun$, a log-normal functional form
$d\caln_*/d\ln(m_*) \propto \exp[ -(\ln m_* -\ln m_c)^2/(2\sigma^2)]$ provides
a smooth fit for the 
observed mass distribution
\citep{1979ApJS...41..513M}, with 
\citet{2005imf..conf...41C}
finding
that
$m_c\approx 0.2 \Msun$ and $\sigma \approx 0.55$ applies both for 
individual stars in the disk and in
young clusters;
the system IMF (i.e., counting binaries as a single systems) has
$m_c=0.25 M_\odot$.
Thus, the main properties of the IMF
that any theory must explain are (i) the ``Salpeter'' power-law slope at
high mass, (ii) the break and turnover slightly below $\sim 1 \Msun$,
(iii) the upper limit on stellar masses 
$\sim 150 M_\odot$ 
\citep{2000ApJ...539..342E,2005Natur.434..192F,2005ApJ...620L..43O},
and (iv) 
the universality of these features over a wide range of 
star-forming environments, apparently independent of the mean
density, turbulence level, magnetic field strength, and to large
extent also metallicity.
Theory predicts that there should be a lower limit on (sub)stellar masses
\citep{1976MNRAS.176..367L}, but this has not been confirmed observationally.

Important additional information has been provided by recent mm and submm
continuum surveys covering both cluster regions and larger
areas in star-forming systems
(e.g. \citealt{1998A&A...336..150M,1998ApJ...508L..91T,
  2000ApJ...545..327J,2001ApJ...559..307J,
2001A&A...372L..41M,2004Sci...303.1167B,
2005ApJ...625..891R, 
2006ApJ...644..990R, 2006A&A...447..609S,2006ApJ...638..293E,
2007MNRAS.374.1413N}).
Within continuum maps, high-density concentrations representing
(starless) cores have been identified in sufficient numbers (and with
sufficent resolution) that core mass functions analogous to the IMF
can be defined. Similar core mass functions (CMFs) may be derived using
extinction data from well-sampled maps \citep{2007prpl.conf....3L}, 
and molecular
line maps in high-density tracers \citep{2002ApJ...575..950O}.
Studies of the CMF using extinction maps are only just beginning, but they
promise to be very important given the lower systematic errors that
are possible with this method. An
excellent recent summary of the statistical properties of observed 
cores is given by \citet{2006astro.ph..3474W}.

The CMFs derived from many independent studies and methods are in good 
agreement with each other, and are remarkably similar in functional
form to the stellar IMF. 
In particular, regardless of the total mass and size of the
star-forming cloud, 
and regardless of whether cores are well
separated or highly clustered, the high-end CMF (above $1\Msun$) is
consistent with a power law.  Applying a uniform analysis to data from
11 high- and low-mass star-forming regions,
\citet{2006astro.ph..7095R} find 
$\alpha=0.8-2.1$, with the mean
value $\alpha=1.4$.
Observed CMFs for
relatively nearby clouds 
in the references cited above
also
show a peak and turnover at low mass in the range 
$\sim 0.2-1 \Msun$.
For
distant clouds, the peak core mass is larger, but lack of resolution and
hence low-mass incompleteness affects these results.  Observed
well-resolved CMF distributions are thus very similar to the stellar
IMF, but shifted to higher mass by a factor of 
a few.
For CMFs derived from mm and submm observations, this
factor involves some uncertainty associated with conversion from
dust emissivity to total mass.  The CMF derived from extinction in the
Pipe nebula,
which is not subject to this uncertainty,
is shifted 
to higher mass
by a factor of 3 with respect to the standard stellar
IMF \citep{2007A&A...462L..17A}.

The mirroring of the ``universal'' IMF by the (possibly also
universal) CMF suggests that the stellar mass distribution is imposed
early in the star-forming process.  The final mass of a star appears
to be controlled by the available reservoir of the core
from which it forms, rather than, for example, being defined by a
termination of accretion due to internal stellar processes.  The shift
of the observed CMF relative to the IMF nevertheless implies that
stellar feedback and other processes in the collapse or post-collapse
stage affect stellar masses.  Magnetized protostellar disk winds are
believed to reduce the stellar mass compared to core mass by a factor
of a few (see discussion in \S 
\ref{section_outflows}).
  In particular, 
\citet{2000ApJ...545..364M} predicted that the efficiency of a single
star formation event in an individual core is
$\ecore\simeq 0.25-0.7$,
depending on
the degree of flattening,
which is comparable to the values implied by the observations cited above.  
Because the efficiency is not sensitive to the
parameters involved, this implies a similar shift from CMF to IMF at
all masses.  Given the uncertainty in the
CMF normalization, the inefficiency of single star formation may
account for essentially the whole CMF $\rightarrow$ IMF shift. 
Some further fragmentation of presently-observed massive cores during their
collapse may also occur, but provided that the majority of the mass
goes into a single object, this will leave the high-mass end of the 
CMF relatively unchanged.  Since the CMF is already dominated by
low-mass cores (by mass as well as by number), the addition of
low-mass stars formed as fragments from collapsing high-mass cores would 
negligibly affect the low-mass end of the IMF.  

Molecular line observations of low-mass cores,
whether found in isolation (as in Taurus) or in close proximity to
other cores in a dense, cluster-forming clump (as in $\rho$ Oph), show that
these cores have very low nonthermal internal velocities
\citep{andre06}.  
Since weak internal turbulence implies that little
density substructure is present within these cores, they are unlikely
to undergo subsequent fragmentation during collapse, except to form
binaries.  The low-mass portion of the CMF should therefore
be conserved in mapping to the IMF, modulo mass removal by outflows.
Although cores in the high-mass end of the CMF are turbulent and thus 
in principle subject to further fragmentation, the agreement between 
the CMF and IMF 
suggests that this is not a dominant effect.

The environments of observed prestellar cores provide further clues to the
processes involved in their formation.  Most stars form in
clusters 
(\citealp
{2003ARA&A..41...57L};
see \S \ref{clusters})
 and correspondingly, most (starless)
molecular cores are part of larger cluster-forming dense clumps.
These cluster-forming clumps\footnote{Referred to as
``cluster-forming cores'' by \citet{2006astro.ph..3474W}}, 
as observed for example in 
Ophiuchus,
Serpens, Perseus, and Orion, have supersonic internal turbulent
linewidths (even though the individual cores within them are subthermal).
Compared to isolated cores, the cores in clusters tend to
be more compact in overall size and have higher densities and column densities;
they are also lower in mass
\citep{2006astro.ph..3474W}.  The column densities of 
cluster-forming clumps are generally quite large, and in particular
they 
exceed the mean column densities of the GMCs in
which they are formed.  In Perseus, where 80\% of the mm cores lie in
groups and 50\% are in clusters \citep{2006ApJ...638..293E}, 50\%  of
the total cloud mass is at $A_V<4$ and 80\% is at $A_V<6$, 
whereas 90\% of the mass in prestellar cores is in larger structures
that have $A_V>6$, and 50\% is at $A_V>8$ \citep{2006ApJ...646.1009K}.  
Similarly in Ophiuchus, the prestellar cores are found in high-column density
regions ($A_V>15$ for $>90\%$ of the core mass), while most of the
cloud's mass has much lower column densities (70\% is at $A_V<7$) 
\citep{2004ApJ...611L..45J}.  The prestellar cores themselves represent only a
tiny fraction of the total cloud mass: 5\% in Perseus
\citep{2006ApJ...638..293E}, and 3\% in Ophichus \citep{2004ApJ...611L..45J}; 
this is comparable to the net observed star formation efficiency over the lifetime of a GMC (see \S \ref{large-scale_SFR}).  
On the largest scales, GMCs generally consist of collections of
filaments, and both the clusters of cores and most of the individual
isolated cores are embedded in these filaments.  The structure
formation that produces cores, and eventually stars, is therefore 
clearly a hierarchical process.

\subsubsection{THEORETICAL PROPOSALS AND NUMERICAL RESULTS}

Many theories have been proposed that aim to explain the IMF or some
aspect of it, and more recently to explain the CMF as well (see
\citealt{2001ASPC..243..255E}  and 
\citealt{2006astro.ph..3447B} for recent reviews).
 While none of the
proposals to date have won general acceptance, several have introduced
elements that are likely to be important in the eventual theory that
is developed.  
Numerical simulations have been valuable in
demonstrating that the general characteristics of observed CMFs arise
naturally in turbulent, self-gravitating flows,
and they have also
been useful in testing certain specific hypotheses.  However, many
features that are seen in the simulations are not yet understood in a
fundamental sense,
and limited numerical resolution
may affect some existing results.

A variety of different numerical models have been used in 
computational studies of 
the mass distributions of bound and unbound condensations
in turbulent systems.
Most models have adopted an isothermal equation of state:  using SPH 
techniques,
\citet{2001ApJ...549..386K}, \citet{2003MNRAS.343..413B}, 
\citet{2006MNRAS.368.1296B}, and \citet{2001ApJ...556..837K}
analyzed decaying-turbulence models with
a variety of power spectra, and \citet{2001ApJ...556..837K} and 
\citet{2006ApJ...637..384B} analyzed 
driven turbulence
models.
Using grid-based codes in the unmagnetized case, \citet{2006ApJ...637..384B} 
and \citet{2007astro.ph..1795P}
analyzed driven-turbulence models.
Using grid-based codes and including magnetic fields,
\citet{2003ApJ...592..203G} analyzed
decaying-turbulence models, and 
\citet{1997ApJ...474..292V},
\citet{2002ApJ...570..734B},
\citet{2004ApJ...605..800L} and 
\citet{2007astro.ph..1795P}
analyzed driven-turbulence models.  \citet{2004MNRAS.353..769T}
analyzed decaying-turbulence unmagnetized models from a grid-based
code.

Other simulations have investigated the effects of non-isothermal
equations of state. 
\citet{2003ApJ...592..975L}  analyzed
driven-turbulence SPH simulations that used a range of polytropic indices.
\citet{2003MNRAS.339..577B} and \citet{2005MNRAS.356.1201B} analyzed
the results of SPH decaying-turbulence 
simulations with a switch from
isothermal to $T\propto \rho^{0.4}$ at density 
$10^{-13}\g\, \cmt$,
to represent the transition from optically-thin to -thick conditions.
\citet{2005A&A...435..611J} and \citet{2006MNRAS.368.1296B} 
investigated the result of switching from  
a weakly cooling to weakly heating barotropic equation of state at
a range of densities $n\sim 10^4 -10^7\cmt$, using driven- and 
decaying-turbulence SPH simulations, respectively.

The distributions obtained by applying clump-finding techniques to
simulation ``data cubes'' share many characteristics,
generally showing clump mass functions dominated by the low-mass end and 
tails at high mass that are (marginally) consistent with a power laws
having indices similar to the Salpeter value. 
In detail, however, the mass functions 
depend on the adopted clump-finding algorithm and
on parameters such as density threshold levels and smoothing scales,
as well as on physical properties including the Mach number and the
history of a system.

In many simulations that include self-gravity, 
the high-end slope tends to become
shallower over time, as massive objects grow larger.  
This change in slope may not represent realistic evolution, if massive 
condensations in fact
should undergo fragmentation that the simulations do not follow.
Failure to capture fragmentation during collapse could affect results
from either
grid-based or SPH simulations.  Fragmentation is seeded by turbulence,
which imposes fluctuations in the density \citep{1973PASJ...25....1S}.  
These fluctuations grow as a condensation collapses, and in principle
could ultimately result in fragmentation if they become locally Jeans unstable
\citep{1962ApJ...136..594H,1973dses.conf.....C}.
Fluctuations at smaller and smaller 
mass scales
would grow to be highly nonlinear if collapse were to proceed
unchecked, so in order to capture this numerically, 
turbulence at scales  below the sonic scale  
would have to be resolved {\it  before} collapse commences in a given region.  
However, it is likely that in reality
fragmentation of collapsing high mass condensations
is prevented by real
physical effects, rather than numerical effects:  accretion onto stars
formed early in the collapse process heats the surrounding gas
significantly, which helps limit further fragmentation 
\citep{2006ApJ...641L..45K}, and for condensations
in cluster-forming regions,
outflows from nearby stars inject small-scale 
turbulence that may provide support
sufficient to prevent rapid localized collapse 
(\citealt{2006ApJ...641L.121T,Nakamura2007}; see also
\citealt{2001ApJ...556..837K}).
Until physical processes enter
to limit further 
breakup of massive condensations during their evolution, self-similarity
implies they would fragment due to the same 
turbulent processes that produced the massive
condensations in the first place.  This is presumably why the IMF in clusters
is the same as that in distributed star formation.
Observed dense clumps
break up into individual small cores when imaged 
at high resolution,
suggesting that much of the fragmentation is in place prior to
collapse.
Indeed, the fact that the turnover of the IMF is similar to the 
Jeans or Bonnor-Ebert mass evaluated at the mean turbulent kinetic pressure 
within GMCs (see \S \ref{dynam_state}),
$P_{\rm kin}/k_{\rm B}\sim 3\times 10^5$~K~$\cmt$, 
suggests that the ambient pressure that
sets the ``typical'' star's mass is
not significantly increased above this level by collapse prior to 
fragmentation.
However, note that the correspondence between the turnover in the IMF 
and the Bonnor-Ebert mass evaluated at the mean turbulent
pressure appears to break down in
star clusters like the Orion Nebula Cluster \citep{1998ApJ...492..540H} 
and globular clusters \citep{2000ApJ...534..870P},
which are believed to have formed at substantially greater pressures 
\citep{2003ApJ...585..850M};
the reason for this is not clear.

A common numerical ``shortcut'' to studying cluster formation is to
focus on just a single cluster, rather than the whole hierarchical
system; this allows the collapse and fragmentation to be better
resolved.  Models of this kind initiate a simulation at high 
density with comparable internal turbulent and gravitational energy.
However, this approach misses an aspect of the real situation which
may be quite important: self-gravitating massive condensations 
develop out of
non-self-gravitating gas in which pertubations have {\it
  already} been imposed by turbulence.  In simulations where the
initial kinetic energy does not exceed the gravitational energy,
collapse occurs before the turbulence is able to imprint a realistic
density structure on the system, such that the subsequent
fragmentation may also be unrealistic.  In particular, this may lead
to massive fragments continuing to grow over time as they capture
low-turbulence unstructured gas from their surroundings (``competitive 
accretion''
---see \S \ref{bondi}).
To obtain a reliable measure
of the high-end CMF from numerical models, it will be necessary to
perform simulations that include large scales as well as cluster
scales, and adequately resolve massive 
condensations both prior to and during 
collapse.  In addition, physical processes representing the
feedback from star formation must be properly included in order to impose
realistic limits on fragmentation, coalescence, and accretion 
after collapse begins. 

Another feature of numerical simulations 
that
is at least
qualitatively in accord with observations is the presence of a
resolved peak and turnover in the CMF.  Exactly how the location of this peak
depends on model parameters, however, is not well yet determined.  
In some simulations, the CMF peak is found to be at masses comparable to
the initial Jeans mass of the system
(these are primarily low-Mach-number simulations),
while in other simulations the
peak is at much lower mass 
(these are primarily at high Mach number).
The turbulent power spectrum can also
affect the position of the CMF peak, and in some simulations the peak
is seen to move to larger mass over time.  In fact, the position of
the peak for an isothermal simulation with a fixed turbulence scaling
law must 
be a function of two dimensionless
parameters, the ratio of the total mass in the system to the initial
Jeans mass, and the turbulent Mach number on the largest scale.  For
magnetized simulations, an additional parameter is the ratio of mass to
the magnetic critical mass.
Although limited dependence on parameters has been explored,
a comprehensive and controlled study has not yet been performed.
Note that the mass-weighted density
in a turbulent system increases as the turbulent Mach number increases
(see \S \ref{densstruct}), so that the Jeans mass at the ``typical''
(mass-weighted)
cloud density decreases as the turbulence level increases, for a given
mean (volume-weighted) density and Jeans mass. This probably accounts for why 
the peak of the CMF was found to be far below the {\it mean} 
Jeans mass in studies with high $\cal M$, 
and close to the mean Jeans mass in 
studies with lower $\cal M$.

A recurrent theme in star formation theory is that the characteristic
mass -- defined by the peak of the IMF -- is the Jeans mass at some
preferred density.  An upper limit on the preferred density, and hence
a lower limit on the fragment mass, is the value at which which the optical
depth is unity over a Jeans length; this yields a minimum fragment
mass $\approx 0.007\Msun$ \citep{1976MNRAS.176..367L}.  More recently,
\citet{2005MNRAS.359..211L} has argued that the thermal coupling of
gas to dust at densities above 
$\nh=n_c \approx 10^6\;
\cmt$ 
results
in a shift from weakly-decreasing to weakly-increasing temperature as
a function of density ($T\propto \rho^{-0.27}$ changes to to $T\propto
\rho^{0.07}$ at $T_{\rm min}\sim 5\K$), and that the Jeans mass 
$\sim 0.3\Msun$ at this inflection point sets the preferred mass scale
in the IMF.  Part of Larson's argument is that if structure is
filamentary, then the filaments will contract 
radially while $\gamma<1$;
fragmentation 
into protostellar cores would occur when
the filament's central density reaches $n_c$
and $\gamma$ exceeds unity.
This argument does not take
into account, however, that the mass per
unit length of a filament may be determined primarily by the
turbulence which originally creates it.
In this case, the density $n_c$ defines a Jeans length 
(see eq. \ref{LJeans} in \S \ref{gravity}), so that the
mass scale that emerges would be set by this (fixed)
length scale $\sim \lambda_J(n_c)$ times 
the (variable) filament mass per unit length.
The simulations of \citet{2005A&A...435..611J}, which vary
the density $n_c$ at which the temperature minimum occurs,
provide qualitative support for Larson's proposal in that the peak of the
CMF moves to lower mass as $n_c$ increases.  The scaling of peak mass
with $n_c$ in the simulations is not consistent with the predicted 
$m_{\rm peak}\propto n_c^{-0.95}$ scaling, however. 
In addition, these models did
not test dependence on other parameters that may be important, 
such as the Mach number of the turbulence or the total mass of the system.

A recent comprehensive proposal to explain the CMF and IMF 
has been advanced by
\citet{2002ApJ...576..870P, 2004ApJ...617..559P} (hereafter ``PN'').  
They argue that because the strength of any given compression (in a shock) is
related to its corresponding (pre-shock) spatial scale $\ell$, a
power-law turbulence spectrum $|v(\ell)|\propto \ell^q$ will result in
a distribution of clump masses  
that
itself follows a power law.  In
particular, they propose that the clump mass function 
produced by turbulence in a magnetized medium will obey
$
d\caln(m)/d\ln m \propto m^{-3/(3-2q)}.
$
They further
propose that at a given mass $m$, the fraction 
of clumps
created by turbulence
that collapse is obtained by integrating the
density PDF  down to the density at 
which that mass would be Jeans unstable, i.e.
$\rho_{\rm min}= \pi^5 \sth^6/(36 m^2 G^3)$.
With this prescription,
at high mass
the limit of the integral $\rho_{\rm min}\rightarrow 0$ 
and PN find $\alpha=3/(3-2q)\approx 1.4$.
The position of the CMF peak
would depend on the properties of the density PDF; for a log-normal 
PDF ($f_M$; see eq. \ref{log-norm}
in \S \ref{densstruct})
with $\mu_x=0.5-2$, the peak mass would be 
between $0.8-0.1$ times the Jeans mass at the mean (volume-weighted)
density in the cloud. 

The proposal of PN is attractive in its overall thrust, 
and analysis of numerical simulations \citep{2007astro.ph..1795P} 
shows promising consistency with some of
the model predictions, such as a steepening of the CMF (larger $\alpha$)
with steeper velocity power spectrum (larger $q$).  
The PN proposal, however,
also suffers from missing links in its theoretical underpinnings:
(1)
The effective
value of $v_A$ is defined by PN
such that the typical compression $\rho'/\rho$ in a shock moving at $v$
is a factor $~v/v_A$
(in fact, compression factors depend on the magnetic field {\it direction}
as well as strength).
This effective $v_A$ is assumed to be independent of 
scale, and 
for numerical comparisons with data they adopt a value
small compared to the typical value in a GMC of $\sim 2$~km~s\e.
(2) The argument used to obtain 
$\alpha=3/(3-2q)$ for turbulent clumps
assumes that each pre-shock volume
$\ell^3$ maps to a number of post-shock volumes of mass $m$ 
that is independent of $\ell$; 
i.e.
$\ell^3\caln(m)/L^3=\mbox{const.}$ 
While this is plausible, 
other arguments can be made that draw on
the scale-free nature of turbulence, yet yield different results.
\citet{1996ApJ...458..739F} and
\citet{1996ApJ...471..816E} 
have argued that in non-self gravitating
turbulence one obtains $m\caln(m)=\mbox{const}$.
This scaling corresponds to converting a constant fraction of the mass
or volume behind every shock into clumps,
$\ell'^3 \rho'\caln(m)/(L^3 \rho)=\mbox{const.}$, where
$\ell'=\ell\vA/v=\ell \rho/\rho'$ is the post-shock scale.
One might also propose that the filling factor of
post-shock clumps 
within the whole volume should be constant, i.e. 
$\ell'^3\caln(m)/L^3=\mbox{const}$.  This 
leads to $\caln(m)\propto  m^{-(3-3q)/(3-2q)}$,
or $\alpha=0.75$ for $q=1/2$.
While the assumption
$\ell^3\caln(m)/L^3=\mbox{const.}$ in the PN formulation 
yields results that are in agreement with measured CMFs, 
a physical argument is needed to explain why this is the correct choice
among several plausible alternatives.
In particular, since PN's argument for the slope $\alpha=3/(3-2q)$
involves only turbulence, why does this value of $\alpha$ disagree with the
distinctly-shallower 
empirical mass spectrum ($\alpha \sim 0.5$; see \S\ref{dynam_state}) 
of non-self-gravitating clumps in GMCs?
(3) The argument PN use to obtain a formula for the mass function does not
appear to take account of substructure within clumps at a given mass
scale 
although the presence of substructure is implicit in their picture.
In particular, they assume that any region that is unstable by
the {\it thermal} Jeans criterion will collapse.
An implicit requirement for this is that at each density, a contiguous volume
containing a mass in excess of the Jeans mass is present.
More generally, 
since 
hierarchical density structures
are clearly important in nature (most
cores and stars are clustered), any fundamental theory should
identify how this this comes about.  
Given these difficulties, it appears premature to
accept the PN proposal in its current form, although 
it is
promising as a basis for future development.

\subsection{The Large-Scale Rate of Star Formation}
\label{large-scale_SFR}



Much of this review focuses on the detailed physical processes of star
formation at and below GMC scales.  To understand the structure of a
given galaxy, however, or the evolution of a population of galaxies
over cosmological timescales, often only a very gross characterization
of the star formation processes -- such as the overall star formation
rate (SFR) and the resulting IMF -- is adequate.  Many empirical
studies of disk galaxies characterize the SFR in terms of the number
of stars formed per unit time per unit area $\dot \Sigma_*$; this is
usually reported using either averages over the whole of a galaxy
within some outer radius $R$, or using azimuthal averages over an
annulus of width $dR$ to give $\dot \Sigma_*(R)$.  Both of these
methods average over regions that may have widely varying SFRs, and
the results must be carefully interpreted as strong nonlinearities are
involved.  Fortunately, with the data becoming available from
large-scale galactic mapping surveys (e.g.  SONG and SINGS;
\citealp{2003ApJS..145..259H,2003PASP..115..928K}), it will soon be
possible to characterize SFRs on scales large compared to individual
GMCs but small enough to separately measure, e.g., SFRs for 
arm and interarm regions.

More fundamental than $\dot \Sigma_*$ is the star formation or gas consumption
timescale.
This is defined by
$\tgs\equiv \Sigma_g/\dot \Sigma_*=M_g/\dot M_*$, where
$\Sigma_g$ is the gas surface density; the second equality assumes
that the same area average is used for the total gas mass $M_g$ and
star formation rate $\dot M_*$. The resulting 
timescale depends on the gas tracer(s) chosen, which determines
the range of gas densities included in $\Sigma_g$.  For a chemical
species tracing gas in a class of structures 
denoted by $S$ that have 
mean internal gas density 
$\langle \rho\rangle_V = \rho_S$,
and total mass $M_S$, a convenient fiducial time for comparison to 
$t_{S*} \equiv M_S/\dot M_*$ is the free-fall time obtained by
using $\rho_S$ in equation (\ref{eq:tff}).  The star formation or 
gas consumption rate is then 
\begin{equation}
\dot M_* \equiv \epsilon_{{\rm ff}, S}\frac{M_S}{t_{{\rm ff},S}}, 
\label{eq:epsff}
\end{equation}
where the efficiency over a free-fall time is
$\epsilon_{{\rm ff},S}= t_{{\rm ff},S}/t_{S*}=
\Delta M_*(t_{{\rm ff},S})/M_S$
(see \citealp{2005ApJ...630..250K} and
\citealp{2006astro.ph..6277K}; note that they 
denote this quantity by SFR$_{{\rm ff},S}$).  
Note that the star formation efficiency 
$\epsff$
over the {\it free-fall
time} at the mean local density of structures $S$
differs from the star formation efficiency 
$\epsilon_S$
over the {\it mean lifetime} of individual
structures in class $S$, 
which is discussed below.
The definition in equation (\ref{eq:epsff}) is
particularly useful for describing star formation on local scales
within GMCs, in which different molecular transitions trace a
relatively limited range of densities, and in which densities can be
obtained for individual
structures that are spatially resolved and have  mass measurements 
from 
molecular line,
dust continuum or extinction observations.   
Since most of the molecular gas, and essentially all star formation,
is found
within GMCs, the 
SFR in a region with local surface density in GMCs,
$\Sigma_{\rm GMC}$, is given by
$\dot \Sigma_* = \epsilon_{\rm ff,GMC} \Sigma_{\rm GMC}/t_{\rm ff,
  GMC}$.
Here, $t_{\rm ff, GMC}$ is calculated using the free-fall time within
GMCs in a given region.  
Typical mean densities within GMCs are $n_H\sim 100\; \cmt$, implying 
$t_{\rm ff, GMC}\sim 4$ Myr, but
this may vary 
due to the effects of spiral arms, for example.
In clumps or cores within GMCs, $\rho_S$ can be larger by orders of
magnitude compared to $\rho_{\rm GMC}$, yielding a corresponding
decrease in $t_{\rm ff}$.

The values of $\epsilon_{{\rm ff},S}$ are generally low ($\simlt 0.01$) 
over a wide
range of density tracers (see below), and vary only weakly with $\rho_S$.  
\citet{2006astro.ph..6277K} point out that this suggests that turbulence is
limiting star formation, although magnetic regulation is also possible
(but probably not on GMC scales, since they appear to be magnetically
supercritical -- \S\ref{magnetic_fields}).
In the turbulence-regulation picture,
the low
overall efficiency of star formation on GMC scales (over their own
free-fall times) is dictated by the low fraction of gas that
concentrates into
structures that are sufficiently dense to collapse before being
redispersed by turbulence. 
The weak variation of $\epsff$ with density follows naturally if
the density obeys a log-normal distribution, which is consistent both
with numerical simulations of supersonic turbulent flows and with
observations of extinction statistics
(\S\S\ref{densstruct},\ref{turbobs}).  For a log-normal distribution
defined by equation (\ref{log-norm}), let 
$M_S$ be the mass with densities in the range $\delta x_S$
surrounding $x_S\equiv \ln(\rho_S/\bar\rho)$, where
$\bar\rho \equiv \langle \rho \rangle_V$ is obtained by dividing total
GMC mass by total GMC volume. Then 
since the star formation rate $\dot M_*$ is independent of tracer,
\begin{equation}
\frac{\epsilon_{{\rm ff}, S}}{\epsilon_{\rm ff,GMC}}
= \frac{\sqrt{4\pi \mu_x}  }{\delta x_S  } 
\exp\left[\frac{(x_S-2\mu_x)^2}{4\mu_x} - \frac 34 \mu_x    \right],
\label{eps_ratio}
\end{equation}
where
$\mu_x \equiv \langle
\ln\rho/\bar\rho\rangle_M$ is the (mass-weighted) mean within a GMC. 
With $\mu_x\sim 1.5$ [corresponding to mass-weighted mean density 
$\avg{\rho/\bar\rho}_M=\exp(2\mu_x)\sim 20$, typical of GMCs] and
$\delta x_S\sim 1$, $\epsilon_{{\rm ff},S}/\epsilon_{\rm
  ff,GMC}$ is between 1.4 and 3.4 for $1<\rho_S/\avg{\rho}_M<10$;
larger values of $\mu_x$ keep $\epsilon_{{\rm ff},S}
\sim \epsilon_{\rm GMC,ff}$ 
over a larger range of densities.
Approximate constancy of $\epsilon_{{\rm ff}, S}$ over a range of densities
implies the approximate relation 
$M_S\propto t_{\rm ff}\propto \rho_S^{-1/2}$ from equation (\ref{eq:epsff}).
Physically, this is because
the equilibrium fraction of mass in a GMC in structures 
at densities
significantly above $\bar\rho$
is equal to the ratio
of the destruction time to the formation time of those structures,
$t_{{\rm dest},S}/t_{{\rm form},S}$. 
In a turbulent medium,
the destruction time is of order the dynamical time of the structure,
which decreases with increasing density
and decreasing size, 
whereas the formation time is of order the dynamical time of the GMC
for all structures, since the large-scale flow dominates.
Note, however, that
this discussion does not apply to regions in which
self-gravity is strong but turbulence is weak, as occurs in 
low-mass
pre-stellar cores.  In such cores, $\epsilon_{\rm ff}$ rises by an order of magnitude or more to
$\sim 0.1$. 
Quiescent cores have individual lifetimes of a
few $t_{\rm ff}$ (see 
\S \ref{section_clumps}), 
and net efficiency 
of star formation in each core $\sim 1/3$ 
(see \S\ref{obs_imf_cmf} and \S\ref{section_outflows}).    
These structures have evolved to have 
internal densities (and hence self gravity) 
high enough that they can resist destruction
by the ambient turbulence.  
In regions such as forming clusters, where self-gravity causes strong
departures from the overall log-normal density distribution in GMCs and
high gravity is offset by locally-driven turbulence, the relation
(\ref{eps_ratio}) would also not be expected to apply.

Even within a given density regime, there may be significant
cloud-to-cloud variations in local conditions such that 
$t_{S*}$ need not be a universal quantity even for structures
observed in a given tracer.  
Indeed, \citet{1988ApJ...334L..51M} showed
that for Milky Way GMCs with virial masses (traced in CO)
$M_{\rm CO}=10^4-5\times 10^6\;\Msun$ and infrared luminosities
$L_{\rm IR}\propto\dot M_*$, the ratio $t_{\rm GMC,*}\propto M_{\rm CO}/L_{\rm IR}$
varies over two orders of magnitude and is not correlated with
$M_{\rm CO}$.  
\citet{1997ApJ...476..166W} came to a similar conclusion from their analysis
of OB associations and GMCs in the Galaxy. With a total GMC mass
$\simeq 10^9 M_\odot$ in the Galaxy and a star formation rate of several $M_\odot$~yr\e,
the mean value of $t_{\rm GMC,*}\approx 3 \times 10^8$~\yr,
which translates to 
$\epsilon_{\rm ff,GMC} \sim 0.01$ if $\bar n_H\sim100\, \cmt$ in GMCs.
For dense gas clumps in GMCs, however, it appears
that conditions are more uniform, such that there is less scatter
in 
$t_{S*}$
 for dense gas tracers.  In particular,
\citet{2005ApJ...635L.173W} show that the ratio $L_{\rm HCN}/L_{\rm
IR}\propto M_{\rm dense\, clumps}/\dot M_*$ measured in Milky Way star-forming 
regions agrees with the same values measured in high-redshift
galaxies \citep{2004ApJ...606..271G}, for which there is only one
order of magnitude 
scatter.  \citet{2005ApJ...635L.173W} estimate a 
corresponding star formation timescale of 
$t_{\rm HCN,*}=8\times 10^7\;$yr.  
If the typical density of HCN-emitting gas 
is $\sim 10^5\, \cmt$, 
the corresponding efficiency per free-fall time is  
$\epsilon_{\rm ff, HCN} \sim 0.002$.  \citet{2006astro.ph..6277K}
apply slightly different factors to convert total HCN and IR
luminosities to gas masses and star formation rates, and obtain 
$\epsilon_{\rm ff, HCN} \sim 0.006$.  These values of $\epsilon_{\rm
  ff}$ are small compared to those for individual cores ($\sim 0.1$), which in
clustered regions (where most stars form) 
have densities $\sim 10^7 \cmt$ \citep{2006astro.ph..3474W} that are
large compared to the densities traced by HCN.

In spite of the large scatter in $t_{S*}$ from one local region to
another (in various density tracers),
empirical studies have shown that when averaged over large scales, 
$t_{g*}$ is correlated with the global properties of gas in a galaxy.
The early studies of \citet{1959ApJ...129..243S,1963ApJ...137..758S}
sought to characterize the star formation rate
as a power law (with index $>1$) in the mean gas 
density (both volume and surface density); 
this would 
then translate to $t_{g*}$ (or $t_{\rm ff}/\epsilon_{\rm ff}$) 
that varies as a negative power of gas density.
More recently, following \citet{1989ApJ...344..685K}, 
a number of empirical studies of disk galaxies 
have identified and explored 
``Kennicutt-Schmidt'' (or KS) laws
of the form 
$\dot \Sigma_* \propto \Sigma_g^{p+1}$, for which 
$t_{g*}\propto \Sigma_g^{-p}$.  
The original study of Kennicutt
investigated  correlations of 
$\dot \Sigma_*(R)$ (based on $H\alpha$)
with the total $\Sigma_g(R)$ 
(including both atomic and molecular gas);
he found an index $p=0.3$ for  
$\Sigma_g(R)$ above a threshold level.
\citet{1998ApJ...498..541K} studied correlations of global averages of
$\dot \Sigma_*$ with $\Sigma_g$ (again combining atomic and molecular
gas).  For the whole sample including normal galaxies, the centers of 
normal galaxies, and starbursts, the fitted index is $p=0.4$; the
index is slightly larger for just normal spirals. 
Recent years have seen a number of other studies of the $\Sigma_g$ --
$\dot\Sigma_*$ relationship based on annular averages in galaxies, 
using H$\alpha$, radio continuum, or far-IR to
measure star formation, and using either the total gas surface density or just
the molecular gas contribution from CO observations
\citep{2002ApJ...569..157W,2002A&A...385..412M,2003MNRAS.346.1215B,2004ApJ...602..723H,2005PASJ...57..733K,2007A&A...461..143S}.  
Most of these studies have found $p$ in the range $0.3-0.4$, although
larger values of $p$  have been obtained in some analyses that include
both atomic and molecular gas.
For dense gas as traced by HCN,
\citet{2004ApJ...606..271G} found a linear relationship between the
integrated star formation rate and the total mass of dense gas,
i.e. $p=0$, based on a sample including both normal galaxies and
luminous/ultraluminous IR galaxies.  For the same sample, the fitted
SFR-gas mass 
index is $p=0.7$ for less-dense molecular gas observed in CO lines.
All of these fits involve (at least) an order of magnitude
scatter about the mean relation.  Taken together, these results imply
that the amount of dense gas available for star formation increases
nonlinearly with the global amount of lower-density gas, but that the
star formation rate in this dense gas is independent of global
galactic properties.

A second approach to characterizing global SFRs 
introduces the global timescale associated with the galaxy, 
the orbital period $t_{\rm orb}=2\pi/\Omega$.
For grand design spirals, the SFR is expected to be proportional
to the rate at which gas passes through spiral arms,
since GMCs are expected (and
observed) to form rapidly in the high-surface-density gas behind the
spiral shock (e.g.
\citealt{1969ApJ...158..123R,2002ApJ...570..132K,2006ApJ...646..213K,2006ApJ...647..997S}).
\citet{1973IAUS...52..257S} appears to have been the first
to propose this idea, and showed that it is roughly consistent
with observations of star formation in the Galaxy. 
\citet{1986ApJ...311L..41W} proposed that GMCs, and hence stars,
are the result of atomic 
cloud-cloud collisions at a rate 
$\propto \Sigma_{\rm HI}^2(\Omega-\Omega_p)$,
where $\Omega_p$ is the pattern speed.
More generally,
\citet{1989ApJ...339..700W} suggested that the star formation rate
should scale as
 $\dot \Sigma_*\propto \Sigma_g \Omega$. This has been
 confirmed by \citet{1998ApJ...498..541K}; the resulting two forms for the KS law are
 \beq
 \dot\Sigma_*=0.017 \Sigma_g\Omega\simeq 
 (2.5\pm0.7)\times 10^{-4}\left(
\frac{\Sigma_g}{ 1\;M_\odot\;\mbox{pc}^{-2}}\right)^{1.4\pm 0.15}\;
M_\odot\;\mbox{yr\e\ kpc\ee}.
\label{eq:ks}
\eeq
 The fact that there are two forms of the star formation law implies
 that there is a correlation between $\Sigma_g$ and $\Omega$;
\citet{2005ApJ...630..250K} found $\Omega\propto \Sigma_g^{0.49}$
for a sample comprised of both normal and starburst galaxies 
\citep{1998ApJ...498..541K,1998ApJ...507..615D}. 
The reason for this correlation is not known at present,
but may be related to an overall tendency
for velocity dispersions to increase at large 
surface
densities
(see below).  
The corresponding gas consumption time is
$t_{g*}/t_{\rm orb}\approx 10$ with $t_{g*}$ evaluated for
the entire galaxy and $t_{\rm orb}$ evaluated at the outer edge
of the star formation. Subsequent observations have found
$t_{\rm mol,*}/t_{\rm orb}
\sim 10-100$ when considering the molecular gas alone
 \citep{2002ApJ...569..157W,2002A&A...385..412M}.

Since most star formation is observed to take place within bound GMCs, it
is useful to introduce $f_{\rm GMC}\equiv \Sigma_{\rm GMC}/\Sigma_g$, 
i.e. the fraction of gas that is found in GMCs. The
surface densities must be averaged over a region containing a large
number of GMCs, since the specific star formation rate has very large fluctuations;
the average can be over a local patch of a galaxy, an azimuthal ring, or an entire
galaxy.   Equation (\ref{eq:epsff})
implies 
\begin{equation}
\dot \Sigma_* = \epsilon_{\rm ff,GMC} 
\frac{  \Sigma_g  f_{\rm GMC}}
{ t_{\rm ff, GMC}}.
\label{SFR-ff}
\end{equation}
This form of the star formation law
is particularly useful if most of the gas is in GMCs, $f_{\rm GMC}\simeq 1$.
Since the gas density in the midplane
$\rho_g\propto \Omega^2/GQ^2$ 
in terms of the Toomre $Q$ parameter,
and since
$t_{\rm ff,GMC}^{-1}\propto \rho_{\rm GMC}^{1/2}\propto \rho_g^{1/2}$, it follows that
$\dot\Sigma_*
\propto \epsilon_{\rm ff,GMC}\Sigma_g f_{\rm GMC}/(Q t_{\rm orb})$
(\citealp
{2005ApJ...630..250K};
see below), 
which is similar to the 
``orbital time'' 
form of the KS law.

An alternative expression for the SFR follows by noting that if GMCs
form from diffuse gas at a rate $M_{\rm diffuse}/t_{\rm diffuse}$
and are destroyed by star formation at a rate $M_{\rm GMC}/t_{\rm
  GMC}$, then $f_{\rm GMC}=t_{\rm GMC}/t_{\rm lc}$, where  
$t_{\rm lc}\equiv t_{\rm diffuse} + t_{\rm GMC}$ is the life-cycle time for gas in the galaxy.
The mean efficiency of star formation in any
GMC {\it over its lifetime} $t_{\rm GMC}$ is
$\epsilon_{\rm GMC}= \epsilon_{\rm ff,GMC}(t_{\rm GMC}/t_{\rm ff,GMC})$;
in the Milky Way, the observed average value of $\epsilon_{\rm GMC}$
is about~0.05 
(e.g., \citealp{1997ApJ...476..166W}),
corresponding to $t_{\rm GMC}/t_{\rm ff,GMC}\approx 5$.
The star formation rate is then
\begin{equation}
\dot \Sigma_* = \epsilon_{\rm GMC}\frac{ \Sigma_g}{t_{\rm lc}}
\label{SFR-lc}
\end{equation}
This equation leads to a simple interpretation of the
empirical result $\dot \Sigma_* \propto \Sigma_g \Omega$ for
grand design spirals.
In this situation
the mean lifecycle time of the gas should be equal to the timescale between
successive encounters with spiral arms, 
$t_{\rm lc}= (2\pi/m)(\Omega -\Omega_p)^{-1}$ for an $m$-armed spiral.
If $\epsilon_{\rm GMC}$ varies only modestly with radius,
then well inside corotation
(which is most of the star-forming disk) the overall star formation
rate should obey 
$\dot \Sigma_* \propto \Sigma_g \Omega$.  
More generally, consider an arbitrary disk galaxy
in which the gas is primarily diffuse, so that the GMC formation time
$t_{\rm diffuse}$ is  much greater than the GMC destruction time (or lifetime) 
$t_{\rm GMC}$, and as a result 
the life-cycle time $t_{\rm lc}\simeq t_{\rm diffuse}$. 
The characteristic timescale $t_{\rm diffuse}$
for formation of self-gravitating structures in a disk with surface
density $\Sigma_g$ is the two-dimensional Jeans time 
$t_{J,2D}=c_{\rm eff}/(G \Sigma_g)$ (eq. \ref{tJ2D}).
(Actual GMC formation
timescales differ from $t_{J,2D}$ due to rotation and
disk-thickness effects -- see references in \S \ref{GMC_formation}).
Using the definition of the 
Toomre $Q$ parameter (eq. \ref{Qdef}), 
 $t_{\rm orb}= (t_{J,2D}/Q)(2
\kappa/\Omega)$, so that from equation (\ref{SFR-lc}), 
$\dot\Sigma_* \sim 3\epsilon_{\rm GMC} \Sigma_g (Q t_{\rm orb})^{-1}$.  
Thus, both the diffuse-dominated and GMC-dominated cases yield
$\dot \Sigma_* \propto \Sigma_g/(Q t_{\rm orb})$, assuming that the
efficiency factors are comparable in different regions of a galaxy and
from one galaxy to another.  Since star formation tends to deplete the
gas in any region until $Q$ is near the critical value $Q_{\rm crit}\simeq 1.5$
(theory: \citealp{1972ApJ...176L...9Q};
observation: \citealp{2001ApJ...555..301M,2002ApJ...569..157W,2002A&A...385..412M,2003MNRAS.346.1215B}), this yields the ``orbital time'' form of
the KS law  
(including the normalization)
when $\epsilon_{\rm GMC}\approx 0.05$.  
GMC formation on a timescale $\sim t_{J,2D}$, implying a star
formation rate 
$\dot \Sigma_* \propto \Sigma_g^2/c_{\rm eff}$ if 
$\epsilon_{\rm GMC}$ is weakly-varying,
also yields the other form of
the KS law if the effective velocity dispersion increases with surface
density according to $c_{\rm eff} \propto \Sigma_g^{1-p}$.
A significant increase in the gas velocity dispersion at small radii,
where $\Sigma_g$ is generally larger,
has been noted in several galaxies
\citep{1993ApJ...418..687K,1996ApJ...471..173S,2002A&A...388....7W,
2004A&A...413..505L,2007A&A...461..143S},
although there is no quantitative theory for this increase.

A quantitative theory for the galactic star formation rate must determine the
star formation efficiency (e.g., $\epsilon_{\rm ff,GMC}$
or $\epsilon_{\rm GMC}$) as well as the corresponding overall rate.
(An exception is the theory of \citealp{1997ApJ...481..703S}, who
notes that the
porosity of hot gas in a galaxy is determined by the SFR;
the SFR can thus be expressed in terms of the porosity, but
this remains uncertain.) \citet{2000ApJ...536..173T}  proposed that 
the overall star formation
rate is determined by cloud-cloud collisions, but he set the efficiency
based on comparison with observations. 
\citet{2002ApJ...577..206E,2003Ap&SS.284..819E} suggested that the
star formation rate per unit volume is $\dot\rho_*\simeq
\epsilon_{\rm core} f_{\rm core}(G\rho_{\rm core})^{1/2}\rho$,
where $f_{\rm core}$ is the fraction of gas in dense cores;
the value of this was set by comparison with observation.
Simulations by \citet{2003ApJ...590L...1K}, by
\citet{2005ApJ...620L..19L,2005ApJ...626..823L,2006ApJ...639..879L},
and by \citet{2006ApJ...641..878T}
show that the fraction of high-density
gas scales as $\Sigma_g^{1.4}$, but the definition of
``high-density'' is arbitrary 
and the dependence of the SFR on this definition is not known.

A theory for the star formation efficiency per free-fall time has been
given by \citet{2005ApJ...630..250K}; 
for galaxies in which $f_{\rm GMC}\simeq 1$, 
this is proposed as a complete theory of the KS law.
The first three assumptions underlying the theory have been discussed above:
they assume that star formation occurs primarily in GMCs, so that the
star formation rate is described by equation (\ref{SFR-ff}),
that the density PDF in GMCs is log normal,
as inferred from simulations of turbulence in gas that
is approximately isothermal (\S \ref{densstruct}),
and that the IMF has the standard form. The final assumption
is that low-mass stars form in all gas dense enough
that the sonic length 
in the surrounding turbulent gas
exceeds the Jeans length ($\lambda_J < \ell_s$) 
\citep{1995MNRAS.277..377P},
with an efficiency $\epsilon_{\rm core}\sim 1/2$ 
from 
the theoretical  estimate of
\citet{2000ApJ...545..364M}.
The condition that $\lambda_J < \ell_s$ 
ensures that critical Bonnor-Ebert spheres
are not torn apart by turbulence; the corresponding critical
density implies that the thermal pressure
in the cores matches the turbulent pressure in the environment,
$\rho_{\rm core}/\bar\rho\simeq \calm^2$. 
An important part of this last assumption is that the regions 
that are dense enough
to satisfy $\lambda_J < \ell_s$ have masses large enough to collapse.
The normalization for the star formation rate is based on the simulations
of \citet{2003ApJ...585L.131V}.
These assumptions
lead to a star formation efficiency 
$\epsilon_{\rm ff, GMC}\simeq 0.017 \avir^{-0.68}(\calm/100)^{-0.32}$.
Since the virial parameter in GMCs is of order unity and the Mach numbers
are somewhat smaller than 100 in regions where they have been 
observed (and $\simlt$ a few 100 even in unresolved starbursts), 
this corresponds to a typical    
$\epsilon_{\rm ff, GMC}\sim 0.02$.
For galaxies in which the gas is not fully molecular, Krumholz and
McKee adopt the
phenomenological result for $f_{\rm GMC}$ obtained by
\citet{2006ApJ...650..933B}.
They show that the resulting star formation
rate agrees well with Kennicutt's observed relations (eq. \ref{eq:ks}).

Finally, we remark that controversy continues to surround the question
of what physical process defines the observed outer-disk
thresholds $R_{\rm th}$  for active star formation.  \citet{1989ApJ...344..685K} and 
\citet{2001ApJ...555..301M} argue that disk thresholds are set by
gravitational stability considerations in shearing, rotating disks,
and find a mean value of $Q\approx 1.4$ when they adopt a constant
value $c_{\rm eff}= 6\;\kms$ for their sample.  Numerical simulations
of isothermal gas disks, including both disk thickness effects and the
gravity from an active stellar disk, quantitatively support this
conclusion \citep{2005ApJ...620L..19L,2006ApJ_submitted}.  
On the other hand, \citet{2004ApJ...609..667S}, building on the suggestion
of \citet{1994ApJ...435L.121E}, argues that star
formation thresholds are defined by the condition that the pressure is high
enough that a 
cold component of the atomic ISM can exist.  
This transition point depends on the
UV intensity and metallicity 
(e.g., \citealp{2003ApJ...587..278W}), 
but typically corresponds to threshold
surface density $\sim 3-10 \;M_\odot\pc^{-2}$.  
Schaye essentially reverses the argument for a $Q$ threshold: he argues
that when a significant fraction of
the ISM becomes cold, $Q$ drops significantly and 
gravitational instability ensues.
An advantage of this proposal is that it can naturally account for 
the 
isolated patches of star formation that occur outside $R_{\rm th}$
\citep{1998ApJ...506L..19F,2006astro.ph..9071B}, since
star formation occurs
wherever the 
pressure of the
gas exceeds the critical value. 
However, while the model gives a necessary condition for star formation,
it does not give a sufficient condition: some of the galaxies in
the \citet{2001ApJ...555..301M} sample have $R_{\rm th}$ 
inside the radius at which the
gas becomes molecular, which in turn is inside the radius at which cold
atomic gas first appears
\citetext{see also \citealp{2006AJ....131..363D}, who
find evidence of a cold atomic component even in
non-star-forming regions}.
In these cases, it is possible that MRI-driven 
turbulence in the outer disk maintains 
the effective $Q$ greater than the critical value
even when some of the gas is cold \citep{PO_ApJ_2007}.

\section{MICROPHYSICS OF STAR FORMATION}

\subsection{Low-Mass Star Formation}
\label{lowmass}


Star formation is traditionally divided into two parts: Low-mass stars
form in a time short compared to the Kelvin-Helmholz time,
$t_{\rm KH}=Gm_*^2/RL$, whereas high-mass stars form in a time
$\ga t_{\rm KH}$ \citep{1974A&A....37..149K}. 
This distinction between low-mass and high-mass protostars 
is not fully satisfactory, however, since for a 
sufficiently high accretion rate any protostar
would be classified as ``low-mass."
We somewhat arbitrarily divide low and high-mass stars 
at a mass of $8\;M_\odot$.
Protostars that will form stars with masses significantly below this
value have luminosities dominated by accretion, and they form from cores that
have masses of order the thermal Jeans mass. Protostars 
above this mass have luminosities that are dominated by nuclear burning unless
the accretion rate is very high, and if they form from molecular cores, those
cores are significantly above the thermal Jeans mass.
Low-mass stars undergo extensive 
pre-main sequence evolution in the Hertzsprung-Russell diagram,
from the point on the ``birthline,'' where they cease accreting and are
revealed \citetext{\citealp{1983ApJ...274..822S}; 
see also \citealp{1972MNRAS.157..121L}}, 
to the main sequence. Here we briefly review
the current understanding of how such stars form.

\subsubsection{Theory of core collapse and protostellar infall}
\label{core_collapse}

As discussed above, low-mass stars appear to form from gravitationally bound
cores. The time scale for the collapse of these cores determines both
the time scale for the formation of a star and the accretion luminosity.
Note that the rate of infall onto the star-disk system, $\mdin$, can 
differ from the rate
of accretion onto the protostar, $\mds$, since some of the infalling gas can be
temporarily stored in the disk.
The collapse of such cores and the growth of the resulting protostars 
has been reviewed by \citet{larson03}, and we draw on this work here. 
At the outset
of theoretical studies of star formation, it was realized that
isothermal cores undergoing
gravitational collapse become very centrally concentrated, with 
a density profile that becomes approximately 
$\rho\propto r^{-2}$ \citep{1968ApJ...152..515B,1969MNRAS.145..271L}.
Prior to the formation of the protostar, there is a central,
thermally supported region of size $r\simeq \lambda_J$.
Collapse of a marginally unstable core begins near the outer
radius. The $r^{-2}$ density gradient is created as the wave of collapse
propagates inward, leaving every scale marginally unstable 
as the collapse accelerates \citetext{cf. \citealp{larson03}}. 
That is, since $\lambda_J \sim c_s/(G \rho)^{1/2}$ (eq. \ref{LJeans}), 
a sphere that is  marginally unstable at each scale, $r\sim
\lambda_J$, will have 
$\rho \sim c_s^2/(G r^2)$ when the protostar is first formed;
the corresponding infall rate is
\beq
\mdin\sim \frac{M_G}{t_G}=\frac{c_s^3}{G}~~\Rightarrow~~\mdin=\phin\frac{c_s^3}{G},
\label{eq:mds}
\eeq
where the gravitational mass and
radius are defined in equation (\ref{eq:gravscales})
and $\phin$ is a numerical factor that is typically $\ga 1$. 
(When the effect of protostellar outflows is included, the infall rate is
reduced by a factor 
$\ecore<1$.)
Although this result was first derived for an isothermal sphere,
\citet{1980ApJ...241..637S} emphasize that it should apply approximately to the
collapse of any cloud that is initially in approximate 
hydrostatic
equilibrium, with 
$c_s^2\rightarrow c_{\rm eff}^2=c_s^2+v_A^2+v_{\rm turb}^2$
including the effects of magnetic fields and turbulence as well
as thermal pressure; \citet{1987ARA&A..25...23S} suggest that 
$c_{\rm eff}\la 2 c_s$, however.
This infall rate explicitly depends only on the sound
speed, but it implicitly depends on the density of the core:
since the core was assumed to be initially in hydrostatic equilibrium,
equation (\ref{eq:mds}) is equivalent to $\mdin\sim M_{\rm core}/t_G
\propto M_{\rm core}\rho^{1/2}$.

There are two limiting cases for the gravitational collapse of an isothermal
sphere. In the first case, originally considered by \citet{1969MNRAS.145..271L}
and \citet{1969MNRAS.144..425P} and extended by \citet{1977ApJ...218..834H}, 
one begins with a static
cloud of constant density and follows the formation 
of the $r^{-2}$ density profile.
At the time when the protostar first forms (i.e., when the
central density reaches infinity in this idealized calculation),
the collapse is highly dynamic, with an infall velocity of
$3.3 c_s$. 
The infall rate onto the star is large,
rapidly increasing from $\mdin=29\cs^3/G$ at the moment of protostar
formation to
 $\mdin=47c_s^3/G$.
In the physically unrealistic case of an infinite, uniform medium,
the accretion rate would remain at this high value; in practice, the
accretion rate rapidly declines after the formation of a point mass (see below).
In the opposite case, considered
by \citet{1977ApJ...214..488S}, one assumes that the evolution to 
the $r^{-2}$ density
profile is quasi-static (most likely due to the effects of 
magnetic fields---see below), 
so that the infall velocities are negligible
at the moment of protostar formation. The resulting initial configuration
is the singular isothermal sphere (SIS), which is an unstable
hydrostatic equilibrium. The collapse is initiated at the center, and the point
at which the gas begins to fall inward
propagates outward at the sound speed (the ``expansion wave''): $R_{\rm ew}=c_st$.
This solution is therefore
termed an ``inside-out'' collapse. For $r\geq R_{\rm ew}$,
the density is that of a SIS, $\rho=c_s^2/(2\pi Gr^2)$;
for $r<R_{\rm ew}$, the gas accelerates until it reaches free fall, with
$v=-(2Gm_*/r)^{1/2}$ and $\rho_0\propto r^{-3/2}$. 
The generalized 
post-core-formation solutions of 
\citet{1977ApJ...218..834H} share the same density and
velocity scalings at small radii.
The infall rate for 
Shu's expansion wave solution
is constant in time,
\beq
\mdin=0.975 c_s^3/G=
1.54
\times 10^{-6}(T/\mbox{10 K})^{3/2}
~~~M_\odot~\mbox{yr\e}.
\label{eq:mdssis}
\eeq
The total mass inside the expansion wave at time $t$ is $2\mdin t$, so 
that about
half this mass is in the protostar (i.e., 
$f_{\rm ew}\equiv m_*/m_{\rm ew}\simeq 1/2$). 
\citet{larson03} describes the Larson-Penston-Hunter (LPH) and Shu solutions as 
``fast'' and ``slow'' collapse, respectively, and suggests that reality
is somewhere in between. 
A general discussion of the family of self-similar, isothermal
collapse solutions has been given by \citet{1985MNRAS.214....1W}.

Observations suggest that the cores that form low-mass stars initially have
density profiles that approximate those of Bonnor-Ebert spheres
(\S \ref{dynam_state}). \citet{1993ApJ...416..303F} 
used time-dependent simulations to follow
the collapse of such spheres under the assumption that support is 
by thermal pressure
alone. They found that the collapse of the innermost, nearly uniform, part of 
a critical Bonnor-Ebert
sphere (i.e., one with a center-to-edge density contrast of 14.1) 
approaches,
but does not reach,
the Larson-Penston (LP) solution prior to and at the time of core formation. 
Shortly thereafter, the infall rate begins to decline;
there is no phase of constant infall for a 
critical
Bonnor-Ebert sphere. 
On the other hand, a sphere that is initially in 
an unstable equilibrium with a larger center-to-edge density contrast
has an extended (outer) region in which the density scales as $r^{-2}$.
In this case the infall rate  
starts off as in the critical Bonnor-Ebert case and then
declines to the constant value for an SIS.
The infall rate 
eventually 
decreases below the SIS value when 
a rarefaction wave from the
boundary of the cloud reaches the origin
(see \citealp{2005MNRAS.360..675V}).
Numerical simulations of gravitational collapse in an
unmagnetized, turbulent medium
show that the initial spike and subsequent decline of the infall rate are
typical \citep{2004A&A...419..405S}.

Most of the theoretical work (except for simulations)
on low-mass star formation has neglected the role
of turbulence in the core. This is generally a valid 
approximation 
for low-mass
cores, but it becomes increasingly inaccurate as the core mass increases.
When turbulence is included, 
it is generally in the microturbulent approximation. 
\citet{1987A&A...172..293B,1992JFM...245....1B}
treated the turbulent pressure as being scale-dependent, and suggested that
turbulence could stabilize GMCs while allowing smaller scales to undergo
gravitational collapse.
\citet{1989ApJ...342..834L} introduced a phenomenological
model for turbulence, a logotropic equation of state (\S \ref{gravity}),
to treat the contraction of the core.
\citet{1992ApJ...396..631M}  and \citet{1995ApJ...446..665C} modeled
cores that are supported in part by turbulent motions 
with a density distribution that is the sum of
an $r^{-2}$ power law for a thermal core and a flatter power law for
the turbulent envelope. Turbulent cores also
can be approximately modeled as polytropes with
$\gamma_p<1$ (\S \ref{gravity}), and when cores collapse,
the adiabatic index $\gamma$ can exceed unity \citep{1995ApJ...440..686M,
1998ApJ...492..596V}.
\citet{1999PASJ...51..637O} generalized the \citet{1993ApJ...416..303F} calculation
and found that both the peak infall rate and
the rate of decline of the infall rate are increased
for $\gamma=\gamma_p>1$.
\citet{1997ApJ...476..750M} generalized the Shu solution to singular
polytropic and singular logatropic spheres. They showed that the expansion
wave accelerates in time as $R_{\rm ew}\propto t^{2-\gamma_p}$ and
the infall rate increases as $\mdin\propto t^{3(1-\gamma_p)}$
(logatropes correspond to $\gamma_p\rightarrow 0$). The ratio of the
mass in the protostar to that engulfed by the expansion wave
is $f_{\rm ew}\simeq 1/2,\, 1/6,\,1/33$ for $\gamma_p=1,\,2/3$ and a logatrope,
respectively. The infall rate for a singular polytropic sphere
can also be expressed as $\mdin=\phi_* m_*/\tff$,
where $\tff$ is the free-fall 
time
measured at the initial
density of the gas just 
accreting onto the star
and $\phi_*\simeq 1.62-0.96/(2-\gamma_p)$ \citetext{\citealp{2002Natur.416...59M};
this is valid for $0<\gamma_p<1.2$}.
Inside-out collapse solutions for clouds
that are initially contracting and that have $\gamma\neq \gamma_p$ have
been developed by \citet{2004ApJ...615..813F}.
For clouds that are supported in part by turbulence, the decay of the turbulence
can initiate the collapse of the core
\citep{1998ApJ...507L.157M}.

In the innermost regions of the collapsing core, the opacity
eventually becomes large enough that the gas switches
from approximately isothermal behavior to adiabatic behavior.
The initial calculations were carried out by \citet{1969MNRAS.145..271L}, and 
recent calculations include those by \citet{1998ApJ...495..346M},
\citet{2000ApJ...531..350M}, and \citet{2003A&A...398.1081W};
all assume spherical symmetry. The gas begins to become adiabatic at
a density $\rho\sim 10^{-13}$~g~cm\eee. The ``first core" 
forms when the gas becomes hot enough
to stop the collapse, and an accretion shock forms
at a radius $\sim 5$~AU and with an enclosed mass $\sim 0.05\;M_\odot$.
Once the gas is hot enough to dissociate the molecular hydrogen,
a second collapse ensues and the protostar is formed. 
When opacity effects are included, the maximum infall rate 
is limited to about $13c_s^3/G$ \citep{larson03}, and the average infall
rate over the time required to assemble 80\% of the final stellar mass
is about 1.5-3 times the SIS value \citep{2003A&A...398.1081W}.

\paragraph{Effects of Rotation}
\label{sec:rotcollapse}

The two classical problems of star formation are the angular momentum
problem and the magnetic flux problem: a star has far less angular momentum
and magnetic flux than an equivalent mass in the interstellar medium.
Magnetic fields effectively remove angular momentum so long as the contraction
of the core is sub-Alfv\'enic and the neutral and ionized components of  the
infalling gas are reasonably well coupled \citetext{e.g., 
\citealp{1985prpl.conf..320M,1987ppic.proc..453M}}. 
Once either of these conditions
breaks down, the gas will collapse with 
(near-)constant 
specific angular momentum, $j=\varpi v_\phi$, 
where $\varpi$ is the cylindrical radius,
provided
the transport of angular momentum by turbulence and gravitational torques is
unimportant.
For collapse at constant $j$,
a disk will form with a radius
\beq
R_d=\frac{(R_d\vkep)^2}{R_d\vkep^2}=
\frac{j^2}{R_d\vkep^2}=\frac{\Omega_0^2 \varpi_0^4}{Gm_{*d}}=3\varpi_0\brot(\varpi_0),
\label{eq:rd}
\eeq
where $\vkep=(Gm_{*d}/R_d)^{1/2}$ is the Keplerian velocity,
$m_{*d}$ is the mass of the star and disk (assumed to be equal to
the initial mass $M[\varpi_0]$),
$\brot$ is the rotational energy parameter defined in equation (\ref{eq:brot}),
and $\varpi_0$ 
and $\Omega_0$
are the initial cylindrical radius and angular velocity. 
In the collapse to a disk, the radius shrinks
by a factor $3\brot$. Note that
if the rotational velocity is proportional to the velocity dispersion, as
might be expected for a cloud supported by turbulent motions 
\citep{2000ApJ...543..822B},
then $\brot(\varpi_0)$ is constant \citep{1993ApJ...406..528G},
and the disk radius is a fixed fraction of the initial radius.
On the other hand, clouds supported primarily by thermal pressure are generally assumed to
be in uniform rotation.
Recall that \citet{1993ApJ...406..528G} found that cores typically have
$\brot(\rcore)\sim 0.02$.

As in the non-rotating case, two limits for rotating collapse have received the greatest
attention. These studies have generally assumed isothermality and have
focused on inviscid, axisymmetric flow, although
the latter conditions are likely to be violated in real disks, as discussed
in \S \ref{disks_and_winds} below.
If the core initially has constant density and is rotating slowly, then it collapses
to a disk that evolves to a configuration with a 
singular
surface density profile,
$\Sigma\propto \varpi^{-1}$
\citep{1980ApJ...239..968N,1984PThPh..72.1118N}.
The self-similar solution for the collapse of a rotating disk has
been obtained by
\citet{1998ApJ...493..342S}, who pointed out that  this solution
is the analog of the Larson-Penston-Hunter (LPH) solution for non-rotating collapse
(i.e., it includes the 
time
after the formation of the central singularity
in $\Sigma$).
A
quasi-equilibrium disk with a radius $R_d\simeq j^2/Gm_{*d}$
(eq. \ref{eq:rd}) 
grows after formation of the central singularity.
Since both $M(\varpi)$ and $j$ scale as $\varpi^2$
in the inner part of the initial
spherical cloud, it follows that $j\propto M(\varpi_0)$.
Angular momentum is conserved during disk formation, 
so when 
a
mass $M(\varpi_0)=m_{*d}$ has collapsed into the 
disk,
$R_d\propto j^2/m_{*d}\propto m_{*d}$, which grows linearly in time in the isothermal case.
Note also that since
$\Sigma\propto \varpi^{-1}$ in the disk, it follows that 
after disk formation $M$ and therefore
$j\propto \varpi$;
as a result, the rotational velocity in the disk is 
constant,
independent of $\varpi$.
The infall rate into the central disk is about $(3-11)c_s^3/G$, 
depending on the angular momentum; this is
significantly less
than that for the non-rotating LPH solution \citep{1998ApJ...493..342S}.
%
In this solution the gas outside the equilbrium disk is dynamically
contracting and is assumed to be itself in a thin disk. Numerical calculations
indicate that relaxation of the thin disk approximation increases the
accretion rate by about a factor 2 for the case they considered.

Alternatively, if the core settles into a centrally concentrated, 
spherical,
quasi-equilibrium state
prior to collapse, a slow, inside-out collapse ensues.
The density distribution of the supersonically infalling gas in the vicinity of the disk
has been determined by \citet{1976ApJ...210..377U} and
by \citet{1981Icar...48..353C} under the assumptions that
the mass is dominated by the central protostar
and that the gas is spherically symmetric far from the protostar.
The outer radius of the disk is
given by equation (\ref{eq:rd}), but the disk is far from Keplerian---there is a large inward velocity
that leads to a dynamically contracting outer disk and quasi-equilibrium inner
disk \citep{1994ApJ...431..341S}.
This solution for the inner part of the infall can be joined smoothly to
the solution for the collapse of an SIS
\citep{1984ApJ...286..529T}. 
More generally,
if the cloud initially has a power-law density profile ($\rho\propto r^{-\krho}$)
with $\krho>1$, then it is straightforward to
show that, for $\Omega_0=\,$const, the disk radius is
%
$R_d\propto m_{*d}^{(\krho+1)/(3-\krho)}.$
For the collapse of a slowly rotating SIS ($\krho=2$), 
this equation 
implies that the disk radius varies
as $R_d\propto m_{*d}^3\propto t^3$ \citep{1981Icar...48..353C}.
This rapid increase in disk radius with time is based on the assumption
that the cloud 
can evolve to a rigidly rotating SIS.
Subsequent work (described below)
shows that even when magnetic fields are included, 
this condition is difficult to realize, and
$R_d$ 
tends to increase
only linearly with time.

\paragraph{Effects of Magnetic Fields}
 \label{sec:magcollapse}

Poloidal magnetic fields prevent gravitational collapse when they are sufficiently
strong (subcritical cores), and they inhibit contraction otherwise (supercritical
cores), as discussed in \S \ref{gravity}. Magnetic tension acts to dilute
the force of gravity. In 
nonrotating
disks, this effect can be modeled approximately
by adopting an effective gravitational constant 
\citep{1997ApJ...485..240B,1997ApJ...480..701N,1997ApJ...475..251S},
$\geff=(1-\mu_\Phi^{-2})G,$
where $\mu_\Phi\equiv M/M_\Phi$ (\S \ref{gravity}) 
is assumed to be independent of $r$ and
must be greater than unity for
gravitationally bound clouds;
this equation
is exact if the disk is also
thin \citep{1997ApJ...475..251S}. As a result,
the mass of a thin, 
nonrotating
isothermal, magnetically supercritical disk in equilibrium
is $M(r)=[1/(1-\mu_\Phi^{-2})]2c_s^2r/G$ \citep{1996ApJ...472..211L}, 
which can be much larger than in the absence of magnetic support.
If such a disk is initially in static equilibrium (which is
difficult to arrange), the infall rate
resulting from an inside-out collapse 
has $\phin=1/(1-\mu_\Phi^{-2})$ in equation (\ref{eq:mds})
to within about 5\% \citep{1997ApJ...475..237L,2003ApJ...599..351A}.

Mouschovias and his students have carried out an extensive set of
calculations of the evolution of a magnetized
disk assuming that the disk is thin and axisymmetric, which reduces
the calculation to one spatial dimension.
They followed the evolution 
from a subcritical initial state to supercritical collapse under
the influence of ambipolar diffusion
\citep{1993ApJ...415..680F}, including the effects of
charged grains \citep{1994ApJ...425..142C,1996ApJ...468..749C,1998ApJ...504..280C}
and rotation \citep{1994ApJ...432..720B,1995ApJ...452..386B,1995ApJ...453..271B}.
In these calculations, the magnetic field has a characteristic hour-glass shape
in which the field is normal to the disk and flares above and below it;
observations that are consistent with this geometry have been obtained
recently at a resolution of several hundred AU \citep{G06}.
\citet{1993ApJ...417..220G,1993ApJ...417..243G}
show that even if the magnetized core began with a spherical shape, it
would collapse to a disk, which they term a ``pseudodisk" since it is
not rotationally supported. The calculations of Mouschovias and his students 
cited above
typically began in a very subcritical state,
with $\mu_\Phi\simeq 0.25$, and
stopped when the central density reached $10^{9.5}$~cm\eee,
which is about the point at which the central regions are expected
to become opaque and non-isothermal. They found that thermal pressure exerts
an outward force $\simeq 30$\% of that due to gravity, whereas
centrifugal accelerations are negligible in this phase of evolution.
The central part of the core
undergoes an extended phase of evolution until it becomes supercritical,
at which point the contraction accelerates and mass-to-flux ratio
increases more slowly. 

\citet{1997ApJ...485..240B}
has given a semi-analytic treatment of these results.
He showed that the surface density profile in
the inner core is $\Sigma(r,t)\simeq
\Sigma_c(t)/[1+(r/R)^2]^{1/2}$, where $\Sigma_c$ is the central
surface density and $R(t)=2c_s^2/G\Sigma_c(t)$ is 
the radius of the region in which thermal pressure is sufficiently strong
to maintain an approximately constant density
(\citealp{1984PThPh..72.1118N} found a similar result for rotating collapse
without a magnetic field). 
The supercritical core has a radius $R_{\rm crit}$ obtained by
setting the central surface density equal
to twice the critical value [$\Sigma_c=\mu_\Phi B_0/(2\pi G^{1/2})$ with $\mu_\phi=2$,
where $B_0$ is the initial field strength in the core].
He showed that the slow increase in the
mass-to-flux ratio,
$\mu_\Phi\propto \Sigma^{0.05}$, results in a significant reduction
in magnetic support at high densities. 
The density profile in the inner part of the disk ($r\ll R_{\rm crit}$)
has $\krho\simeq 2$, but
the increasing relative importance of magnetic forces in the outer
regions cause 
it to flatten out so that the mean value in the entire core is $\krho\simeq 1.6$.
Basu
found that the rotational velocity is independent of $r$, just as
\citet{1998ApJ...493..342S} did for the non-magnetic case.
Extending the problem to include the time after protostar formation,
\citet{1998ApJ...504..247C} obtained
a similarity solution to the non-rotating collapse problem
with ambipolar diffusion and
found an infall rate
$\mdin=5.9 c_s^3/G$ at the time
of protostar formation. Two-dimensional numerical calculations have
confirmed this result and have shown that the infall rate 
subsequently drops by somewhat less than a factor 2 
\citep{1998ApJ...504..257C}. 

\citet{2004ApJ...601..930S} 
have considered very different initial conditions,
in which a uniform field threads a singular isothermal sphere. 
Since the flux-to-mass
ratio is zero at the origin and increases outward, 
magnetic effects are negligible
at the center and become important only at a characteristic length scale
$R_{\rm ch}=\pi\cs^2/G^{1/2}B$, corresponding to the condition that
$M_G\sim M_\Phi$ (a similar length scale arises in studies of collapse
with ambipolar diffusion, as may be inferred from 
\citealp{1995ApJ...453..271B}).
They follow the collapse under the assumption 
that the flux is frozen to the gas and find that 
the infall rate declines after the expansion wave reaches $R_{\rm ch}$; this is
analogous to the result of \citet{1995ApJ...454..194C}, who 
found a similar result for the case of collapse with ambipolar diffusion when
the expansion wave reached a point at which the ambipolar diffusion was
inhibited by photoionization.
The initial conditions assumed by \citet{2004ApJ...601..930S} lead
to a poloidal field that decreases inward, whereas calculations that
start from non-singular initial conditions and include ambipolar diffusion
find that the poloidal field strongly increases inward.

A full similarity solution
for the evolution of the collapsing core after it has fallen into
a thin disk and a protostar has formed at the center, 
including rotation, magnetic fields and ambipolar diffusion,
has been obtained by \citet{2002ApJ...580..987K}.
They assumed that the gas is in a thin disk with a constant rotational velocity
(see above); how this assumption would be affected by turbulence,
which would thicken the disk 
and transport angular momentum, is unclear.
The infalling gas goes through two shocks, 
a C-shock (which has a structure dominated by ambipolar 
diffusion---e.g., \citealp{1993ARA&A..31..373D})
and a shock at the outer edge of the centrifugally supported disk. 
When the protostar
reaches $1 M_\odot$, the C-shock
is at about $10^3$~AU, 
and the centrifugal shock is at 
about $10^2$~AU, consistent with
data on T Tauri systems (see \S \ref{diskobs} below). 
Within the self-similar framework, they find that magnetic braking
can be adequate to maintain accretion onto the 
central protostar; in this case there would be  no need for 
internal disk stresses to drive accretion. 
The infall rate in
their fiducial case is $4.7c_s^3/G$; for a gas at 10~K, this corresponds to
a star formation time $t_{\rm sf}=1.3\times 10^5(m_*/M_\odot)$~yr. 
Their solution does
not include an outflow, but they show how one might be included and estimate
that this could reduce the accretion 
rate by a factor $\la 3$.
In sum, based on the theoretical work to date, it is clear that the infall
rate is proportional to $c_{\rm eff}^3/G$, where $c_{\rm eff}$ is an 
effective sound speed \citep{1980ApJ...241..637S}, but the value of
the coefficient and its time dependence have yet to be determined in
realistic cases.

	The magnetic flux problem in star formation is that stars have
very large values of the mass-to-flux
ratio ($\mu_\Phi\sim 10^4-10^5$ in magnetic stars,
$\sim 10^8$ in the Sun---\citealp{1983PASJ...35...87N}),
whereas they form from gas with $\mu_\Phi\sim 1$.
This problem does not have an adequate solution yet, but it appears
that it must be resolved in part on scales $\la 1000$~AU
and
in part on smaller ($\sim$~AU) scales.
Detailed calculations of the ionization state of the
infalling and accreting gas 
show that the ionization becomes low enough that the field decouples
from the gas at densities of order $10^{10.5}-10^{11.5}$~cm\eee\ 
\citep{1991ApJ...368..181N,2001ApJ...550..314D,2002ApJ...573..199N};
decoupling occurs at a somewhat lower density after the formation of the
central protostar, due to the stronger gravitational force \citep{1998ApJ...504..257C}.
\citet{1996ApJ...464..373L} showed that once the field decouples from
the gas, magnetic flux accumulates 
in the accretion disk as the gas flows
through the field and
onto the protostar. The pressure associated with this field is
strong enough to drive a C-shock (which has a structure dominated by
ambipolar diffusion) into the infalling gas. The radius of the shock is
predicted to be several thousand AU at the end of the infall phase
of a 1 $M_\odot$ star; inside the shock, 
the field is approximately uniform (except close to the star)
and the gas settles into an infalling,
dense disk that they identified with the ``outer disk'' observed in HL Tau
\citep{1993ApJ...418L..71H}. 

These results have been confirmed and improved upon by
\citet{1998ApJ...504..247C},
\citet{1998ApJ...504..257C}
and \citet{2002ApJ...580..987K}.
\citet{2005ApJ...618..783T} have carried out 2D axisymmetric calculations
with careful attention to the evolution of the ionization and find that
the location of the shock oscillates, leading to fluctuations in the
accretion rate; 
it is important to determine if this effect persists in 
a full 3D simulation.
\citet{2007astro.ph..2036T,2007astro.ph..2037T,2007astro.ph..2038T} find that
the magnetic field in the central region ($r\la 10$~AU) is about 0.1 G
at the end of their calculation, 
when the central star has a mass $\sim 0.01 M_\odot$;
this is at the low end of the fields inferred
in the early solar nebula from meteorites, which are in the range
$0.1-10$~G \citep{1993prpl.conf..939M}.
They show that ohmic dissipation becomes as
important as ambipolar diffusion at densities $\ga 10^{12.5}$~cm\eee,
but it does not affect the total magnetic flux.
However, even though these processes 
significantly reduce the field
within a few AU of the protostar, 
they are not sufficient to reduce the magnetic
flux in the protostar to the observed value 
\citep{1986MNRAS.221..319N,1996ApJ...464..373L,1998ApJ...497..850L,1998ApJ...504..257C,2005ApJ...618..783T}. 
It is possible that turbulent diffusion \citep{1996ApJ...464..373L} or
magnetic reconnection \citep{1967MNRAS.137...95M,1993ApJ...417..243G}
plays a role in further reducing the magnetic flux. Reconnection 
alters the topology of the field and 
can displace the region in which the flux crosses
the forming or accreting disk. 
However, reconnection cannot actually destroy flux 
(a common misconception), since at sufficient distance from
the protostar the plasma is a good conductor and the total flux inside this 
conductor must be conserved. At the present time, the solution to 
the magnetic flux problem
remains incomplete.

As remarked above (\S \ref{sec:rotcollapse}),
magnetic fields are thought to play a critical role in solving the 
the classical angular momentum problem 
by means of magnetic braking
\citep{1985prpl.conf..320M,1987ppic.proc..453M}. 
Indeed, magnetic braking 
when the field is frozen to the matter
is so effective that \citet{2003ApJ...599..363A}
and \citet{2006ApJ...647..374G}
 have 
argued that magnetic reconnection is required to reduce the field and
therefore the braking enough that a Keplerian disk can form.
The infall solution of \citet{2002ApJ...580..987K}, which includes 
ambipolar diffusion, and the
numerical simulations of \citet{2000NewA....4..601H}, 
which include both turbulence and ambipolar
diffusion, suggest that Keplerian disks can form without reconnection,
but nonetheless indicate that predicting the evolution of the specific
angular momentum of the infalling gas is a complex problem. 
However, it is not clear that any of these theoretical models 
are consistent with
the observations of \citet{1997ApJ...488..317O}, which show that the 
specific angular
momentum in gas associated with several protostars in Taurus is constant for 
$10^{-3} \,\pc<r<0.03$~pc.

	Numerical simulations, as opposed to numerical integration of
the underlying partial differential equations, are required 
to study core collapse in 2D or 3D without additional assumptions
(such as self-similarity or a thin-disk condition).
A critical review of numerical simulations of low-mass star formation is
given by \citet{2007prpl.conf...99K}. To date, such simulations 
have not included
ambipolar diffusion, nor have they simultaneously included radiative transfer
and magnetic fields; most simulations have also stopped prior 
to the formation of
the protostar. A prediction of these simulations is that a slow 
($v\sim c_s \sim 0.2\, \kms$) outflow should
occur at large radii ($\sim 10^3$ AU; 
\citealp{1998ApJ...502L.163T,2002ApJ...575..306T,2006ApJ...641..949B}). 
These authors suggest that
this outflow is related to the observed bipolar outflows, but 
\citet{2003ApJ...599..363A}
disagree. In any case, this large scale outflow could be
important in setting the outer boundary conditions for the jets and 
higher-velocity outflows that are observed 
(see \S\S \ref{windobs}, \ref{mag_winds}).

\subsubsection{Bondi-Hoyle Accretion}
\label{bondi}

Once a protostar star has
formed by gravitational collapse of a core, it can continue to grow by
gravitational accretion from the ambient medium.
Most treatments of this process do not distinguish between the
gas that accretes directly onto the star and the gas that first
falls onto the disk.
\citet{1939PCPS...34..405H} first developed
the theory of accretion by a moving point mass, and
\citet{1952MNRAS.112..195B}
extended the theory to the
case in which the star is at rest in a medium of
finite temperature. Today, gravitational accretion
by a stationary object is generally referred to
as Bondi accretion, whereas that by a moving
object is referred to as Bondi-Hoyle accretion.
If the density and sound speed far from the star are
$\rho$ and $c_s$, respectively, and the star is moving
at a velocity $v_0=\calm_0c_s$ through the ambient medium,
then the characteristic radius from which the star
accretes is
\beq
\rbh=\frac{Gm_*}{(1+\calm_0^2)c_s^2}.
\eeq
The accretion rate is
\beq
\dot M_{\rm BH}=4\pi\phibh \rbh^2\rho c_s(1+\calm_0^2)^{1/2}=\frac{4\pi\phibh\rho
G^2m_*^2}{(1+\calm_0^2)^{3/2}c_s^3},
\label{eq:mbh}
\eeq
where $\phibh$ is a number of order unity that fluctuates somewhat
due to instabilities in the flow \citep{1994ApJ...427..351R}.

There are a number of assumptions that go into this result:
(1) The mass inside $\rbh$ is dominated by
the mass of the star---i.e., the self-gravity of
the accreting gas is negligible. One can show that
\beq
\frac{M(\rbh)}{m_*}=\frac{5.85}{(1+\calm_0^2)^3}\left(\frac{m_*}{M_{\rm BE}}\right)^2,
\eeq
so that this condition is 
equivalent to requiring that the stellar mass be small compared
to the Bonnor-Ebert mass 
(eq. \ref{MBE_def}) in the ambient medium
(for $\calm_0\la 1$).
(2) The tidal gravitational field is negligible; when it is not, $\rbh$ is replaced
by the tidal radius \citep{2001MNRAS.323..785B}.
(3) The magnetic field is negligible. 
Based on dimensional scalings,
a rough approximation for the effect of a magnetic field
on the accretion rate would be
to make 
the replacement
$v_0\rightarrow (v_0^2+v_A^2)^{1/2}$.
(4) The ambient gas is moving at a uniform velocity.
In fact, gas in molecular clouds is generally supersonically
turbulent, so that an accreting star experiences large
fluctuations in both the density and velocity of
the accreting material. 
\citet{2006ApJ...638..369K} showed
that the mean accretion rate in a turbulent
medium is given by equation (\ref{eq:mbh}) 
with $\rho$ equal to the mean 
density, $\calm_0$ replaced by $\calm_{\rm turb}$, and
$\phibh\simeq 3.5\ln(0.70\calm_{\rm turb})$,
provided that the 3D Mach number $\calm_{\rm turb}$
of the turbulence is large compared to $\calm_0$ and 
compared to unity (they verified
this result for $\calm_0=0$ and $3\leq\calm_{\rm turb}\leq 10$). 
This result was
derived from simulations of isothermal gas,
but it should be approximately valid for
other equations of state also. 
The median accretion
rate is significantly less than the mean, however.

	The dominant paradigm for star formation is gravitational
collapse, but an alternative is that stars 
(or at least relatively massive stars) are formed primarily 
by the capture and subsequent
accretion 
of matter that is initially  unbound to the star
\citep{1982NYASA.395..226Z,1997MNRAS.285..201B}.
Since protostars compete for gas from a common reservoir,
this process is termed ``competitive accretion.''
The simulations of \citet{1997MNRAS.285..201B} show that a few of the fragments
gain most of the mass; these are the ones that 
reside primarily in the central regions of the clump
and have 
the highest 
accretion rates. Since gravitational
accretion scales as $m_*^2$,
initial differences in protostellar masses are amplified. 
This process has the potential
of producing the initial mass function, and
it also naturally leads to massive stars being
centrally concentrated in clusters, as observed \citep{2001MNRAS.324..573B}.
A key issue for competitive accretion is,
what is the level of turbulence in the ambient medium?
There is general agreement that competitive accretion 
is ineffective if the medium 
has turbulent energy comparable to gravitational energy, with
$\alpha_{\rm vir}$ order unity (see eq. \ref{alpha_vir_def}),
whereas it is effective if the turbulence is sufficiently weak, $\avir\ll 1$ 
(\citealp{2001MNRAS.323..785B,2005Natur.438..332K};
magnetic fields, which tend to suppress accretion, have not
been considered yet).
Bonnell and his collaborators (\citealp{2006MNRAS.370..488B}
and references therein) argue that the gas throughout
star-forming clumps has a very low turbulent velocity
so that protostars in clusters can accrete efficiently. 
On the other hand, \citet{2005Natur.438..332K}
argue that stellar feedback and the cascade of turbulence from
larger scales ensure that the star-forming clumps are sufficiently
turbulent to be approximately virialized and to therefore have
negligible competitive accretion. Analysis of data from several
star-forming clumps shows that stars in these clumps could grow by
only $(0.1-1)\%$ in a dynamical time, far too small to be significant.
The timescale for the formation of star clusters is an important discriminant
between these models: Star clusters form in about $2t_{\rm ff}$ if
turbulence is allowed to decay,
whereas it can take significantly longer if turbulence is maintained
\citep{2003MNRAS.343..413B}. The observational evidence discussed
by \citet{2006ApJ...641L.121T}  and \citet{2006astro.ph..6277K}  
favors the longer formation time.
This controversy can be resolved through more detailed observations
of gas motions in star-forming clumps and through more realistic simulations
that allow for the evolution of the turbulent density fluctuations as
the clump forms and evolves to a star-forming state, and
that incorporate stellar feedback.

\subsubsection{Observations of low-mass star formation}
\label{infallobs}


The growth of protostars can be inferred through observations
of the mass distribution surrounding the protostar,
the velocity distribution of this circumstellar gas,
and the non-stellar radiative flux.
The mass and/or temperature distribution 
on both small and large spatial scales can be inferred by
modeling the spectral energy distribution (SED) of the continuum.
Protostellar SEDs are conventionally divided into four classes,
which are believed to represent an evolutionary progression
(
\citealp{1987ApJ...319..340M} divided sources into two classes;
\citealp{1987IAUS..115....1L} introduced Classes I-III;
\citealp{1987ApJ...312..788A} discussed a similar classification; 
and \citealp{1993ApJ...406..122A}
introduced Class 0). \citet{2000prpl.conf...59A}
have summarized the classification scheme:

\begin{itemize}
\item[] {\bf Class 0}: sources with a central protostar that are
extremely faint in the optical and near IR (i.e.,
undetectable at $\lambda<10$~\micron\ with the technology
of the 1990's) and that have a significant
submillimeter luminosity, $L_{\rm smm}/L_{\rm bol}> 0.5\%$.
Sources with these properties 
have $M_{\rm envelope}\ga m_*$. Protostars
are believed to acquire a significant fraction, if not most, of their
mass in this embedded phase.

\item[] {\bf Class I}: sources with 
$\alpha_{\rm IR}>0$, where
$\alpha_{\rm IR}\equiv d\log \lambda F_{\lambda}/d\log\lambda$
is the slope of the SED
over the wavelength range between 2.2 \micron\ and 10--25 \micron.
Such sources are believed to be relatively evolved protostars with
both circumstellar disks and envelopes.

\item[] {\bf Class II}: sources with $-1.5<\alpha_{\rm IR}<0$ are believed to
be pre-main sequence stars with significant circumstellar disks
(classical T Tauri stars).

\item[] {\bf Class III}: sources with $\alpha_{\rm IR}<-1.5$ are pre-main sequence
stars that are no longer accreting significant amounts of matter 
(weak-lined T Tauri stars).

\end{itemize}

\noindent
These classes can also be defined in terms of the ``bolometric temperature,'' 
which is the 
temperature
of a blackbody with the same mean frequency as 
the SED of the 
YSO \citep{1993ApJ...413L..47M}.

Unfortunately, the geometry of the source can confound this 
classification scheme
(e.g., \citealp{2000ApJ...531..350M}).
It is well-recognized that a given source can appear as a Class II source at
small or moderate inclination angles (so that the central star is
visible) and as a Class I source at large inclination
angles (so that the central source is obscured by the disk).
A similar ambiguity can involve
Class 0 sources if
the protostellar envelope 
is flattened due to the presence of a large-scale magnetic
field or contains cavities 
created by protostellar jets. 
\citet{2007prpl.conf..117W} summarize the observational evidence that
many of the properties of Class I and Class II sources are similar, which is
consistent with inclination effects confusing the evolutionary interpretation of 
the SEDs. This ambiguity can be alleviated by radio
or sub-millimeter observations of the envelopes, 
which yield masses that are independent
of inclination; 
\citet{2001A&A...365..440M} find that about 40\% of
the sources in Taurus that are classified as Class I on the basis of 
their SEDs have envelope
masses $< 0.1 M_\odot$ and are thus  unlikely to be true protostars.
More sophisticated modeling of the SEDs
can also clarify the evolutionary sequence of young stellar objects; for example,
\citet{2006ApJS..167..256R} have calculated
$2\times10^5$ model SEDs, including the effects of outflow cavities,
that can be automatically compared with observed SEDs to infer the
properties of the source. Counts of sources at different evolutionary stages together
with an estimate for the age for one of the stages allows one to infer
the lifetimes for all the stages. Typical estimates for the ages 
are $1-2\times 10^5$~yr for Class I sources 
and $1-3 \times 10^4$~yr for Class 0 sources \citep{2000prpl.conf...59A}.

There is a significant discrepancy between the 
protostellar accretion rates that are observed and those that are expected,
resulting in the so-called 
``luminosity problem'' \citep{1990AJ.....99..869K}.
The luminosity due to accretion onto the star is
\beq
L_{\rm acc}=f_{\rm acc}\;\frac{Gm_*\dot m_*}{R_*}=
3.1f_{\rm acc} \left(\frac{m_*}{0.25 M_\odot}\right)
\left(\frac{\dot m_*}{10^{-6}\;M_\odot\;\mbox{yr\e}}\right)
\left(\frac{2.5\,R_\odot}{R_*}\right)~~~L_\odot,
\eeq
where $f_{\rm acc}$ is the fraction of the gravitational potential
energy released
by accretion, the rest being carried off in a wind or absorbed by the star
(e.g., \citealp{1995ApJ...447..813O}),
$0.25 M_\odot$ is the typical mass of a protostar (i.e, half the mass of
a typical star), and 
$2.5R_\odot$ is the corresponding radius
\citep{1988ApJ...332..804S}. 
There are 
two main
ways to estimate the expected 
average
accretion rate,
$\avg{\mds}$.
(1) Since both the mass ejected from the disk and the mass 
stored in the disk are generally
a small fraction of the stellar mass (see \S \ref{disks_and_winds}),
the average accretion rate $\avg{\mds}$ should be comparable to
the infall rate $\mdin$. Theoretically, for $T=10$~K, this is
$\mdin= 1.5
\times 10^{-6}\epsilon_{\rm core}\phin\;M_\odot$~yr\e\
(eqs. \ref{eq:mds}, \ref{eq:mdssis}).
If the fraction of the core mass that goes into the star
is $\epsilon_{\rm core}\simeq 1/3$ 
\citep{2000ApJ...545..364M,2007A&A...462L..17A},
and if the envelope infall
rate is that expected from rotating, magnetized collapse 
($\phin\simeq 5$---\citealp{2002ApJ...580..987K}),
then the infall rate is 
$\mdin
\simeq 
2.5
\times 10^{-6}\;M_\odot$~yr\e.
Observationally, 
the properties of the envelopes
around Class I objects inferred from the SEDs 
give similar infall
rates \citep{1993ApJ...414..676K},
so this estimate for $\ecore\phin$ cannot be too far off. (2)
A direct estimate of the 
average accretion rate  is that
forming an $0.5 \;M_\odot$ star in $2\times 10^5$ yr, the estimated
upper limit on the duration of the embedded stage, requires 
$\avg{\mds}=2.5\times 10^{-6}\;M_\odot$~yr\e, comparable to the estimated
infall rate.
The average luminosity corresponding to this accretion rate 
is $\avg{L}\simeq 8\;L_\odot$. 
The problem is that
the observed median luminosity of the {\it bona
fide} Class I sources (i.e., those with significant molecular envelopes)
in Taurus is about $0.5 - 1\; L_\odot$ 
(\citealp{2004ApJ...616..998W} and \citealp{2001A&A...365..440M},
respectively), almost an
order of magnitude smaller. The problem is significantly
worse than this, however, since only a small fraction of the luminosity is
due to accretion \citep{1998AJ....116.2965M};
\citet{2004ApJ...616..998W} 
find that the fraction is about 25\%. The clearest statement of
the luminosity problem is as an accretion rate problem: The observed
accretion rates in Class I protostars are 1 -- 2 orders of magnitude
smaller than those needed to form a star during the lifetime of a 
Class I object.

\citet{1990AJ.....99..869K} suggested two solutions to this problem.
One solution is that significant accretion continues into the T Tauri stage,
but this appears to be ruled out by the fact that such stars accrete 
very slowly 
($10^{-8}\;M_\odot$~yr\e; see \S \ref{diskobs}),
and there  is 
not a significant disk or envelope mass reservoir that they can draw 
on for episodic accretion. The other solution is that most of the
accretion occurs in the embedded stage, but it is episodic, so that the median
accretion rate is much smaller than the mean. They 
suggested that the high accretion-rate stage of protostellar accretion
could correspond to FU Orionis objects,
which are very luminous 
(typically $200-800\; L_\odot$ --- \citealp{1996ARA&A..34..207H})
and have have accretion rates $\sim 10^{-4}\: M_\odot$~yr\e.
Such outbursts could be due to thermal instability 
\citep{1994ApJ...427..987B,1995ApJ...444..376B},
although this would only affect the inner disk, and hence the
outburst would be limited in duration and total mass accreted;
or to
gravitational instability in the disk 
(see \S \ref{self_grav_acc} and 
\citealt{2005ApJ...633L.137V,2006ApJ...650..956V}).
However, \citet{1996ARA&A..34..207H} estimate that the observed FU Ori objects
can account for only about $5 - 20$\% of the mass of stars forming in the
solar vicinity; this discrepancy has not disappeared in the intervening decade.
Recent observational studies of the central stars in Class I sources
differ on their evolutionary status:
\citet{2004ApJ...616..998W} argue that the
Class I protostars are similar to T Tauri stars and are thus past the main
protostellar accretion phase, whereas
\citet{2005AJ....130.1145D} come to the opposite conclusion.
If protostars are close to their final mass by
the time they become Class I sources, then they must
gain most of their mass in the Class 0 stage.
In this case the luminosity problem remains, albeit in a milder form: 
The mean accretion rate required
to form an $0.5\;M_\odot$ star in $3\times 10^4$~yr (the estimated upper limit
on the lifetime of a Class 0 source---\citealp{2000prpl.conf...59A}) 
is $1.7\times 10^{-5}\,M_\odot$~yr
\e,
corresponding to $\avg{L}\simeq 50 f_{\rm acc}(2.5\,R_\odot/R_*)L_\odot$; 
by contrast, the median
luminosity of the Class 0 sources listed by \citet{2000prpl.conf...59A}  is about $10 \,L_\odot$.
This luminosity problem could be alleviated if 
a significant fraction
of the accretion energy
is carried off by the powerful protostellar outflows that accompany these sources,
so that $f_{\rm acc}\la 1/2$. 
However, accounting for 
a value of the infall rate as high as the inferred 
accretion rate onto the star
remains a theoretical challenge: The magnetized collapse
models discussed above give infall
rates of a few $\times 10^{-6}\;M_\odot$~yr\e\ 
(for $\epsilon_{\rm core}\sim 1/3$), significantly less than required.
A resolution of the luminosity problem thus remains elusive.

Spectroscopic observations using molecular transitions
can give both the mass and velocity distributions 
in collapsing cores
\citep{2000prpl.conf..217M}, but
to date the spatial resolution of these data is generally $\ga 100$~AU.
Evidence for infall in unresolved cores is provided by the
``infall asymmetry" \citep{1976A&A....46..473L,1977ApJ...214L..73L,1996ApJ...465L.133M}:
optically thick, infalling gas in which the excitation temperature rises toward the center
produces a characteristic line profile in which the blue wing is stronger than
the red wing. Observations of samples of 
starless cores \citep{1999ApJ...526..788L,2001ApJS..136..703L},
Class 0 sources \citep{1997ApJ...484..256G},
and Class I sources \citep{2000ApJ...533..440G} show a ``blue excess"
[(blue asymmetries - red asymmetries)/(number of sources)]
of about $0.25-0.35$, indicating
that many of these sources are undergoing collapse \citep{2000prpl.conf..217M}.
Unfortunately, it has proved difficult to carry out unambiguous 
observational tests of the theoretical models for protostellar
accretion. 
\citet{2006ApJ...653.1369F} mapped the
infall in a young Class 0 source and found 
reasonably
good agreement with
the LPH solution
($\phin\simeq 20$ in eq. \ref{eq:mds}); this source appears to 
be very young, since for $r>100$~AU there is no evidence for the
$\rho\propto r^{-3/2}$ density profile expected for accretion onto a protostar of significant
mass.
\citet{1998ApJ...504..900T} and \citet{2001ApJS..136..703L}
found that infall is more extended than expected in inside-out
collapse models, although this infall may reflect the formation of small clusters
rather than individual stars. They also found
that the infall velocity is
faster than expected in standard ambipolar diffusion models; however,
the velocities are consistent with the collapse of 
magnetically supercritical cores \citep{2000ApJ...529..925C}.
A potentially important result is that \citet{1997ApJ...488..317O} found that cores
in Taurus are in solid body rotation on scales $\ga 0.03$~pc
but conserve angular momentum on smaller scales. The 
physical significance of this length scale could be inferred
by determining its value in other molecular clouds.

\paragraph{Brown Dwarfs}

Since brown dwarfs
represent the low-mass extreme of star formation, they can
shed light on the earliest stages of star formation. As a result of a great deal
of observational work over the past decade, it has been established that
most brown dwarfs form by the same mechanism as most stars 
\citep{2007prpl.conf..443L,2007prpl.conf..459W}: the initial mass function,
velocity and spatial distributions at birth, multiplicity, accretion rates, circumstellar
disks, and outflows are all continuous extensions of those for hydrogen-burning
stars. This is to be expected, since stars near
the H-burning limit at $0.075\,M_\odot$ reach their final mass long before
hydrogen burning commences. Following \citet{2007prpl.conf..459W}
and \citet{2007prpl.conf..623C}, we shall assume that
brown dwarfs form by gravitational instability on a dynamical timescale,
and that their composition reflects that of the ambient interstellar medium.
By contrast, planets are 
believed
to form in circumstellar disks and to
have an elemental composition with an excess of heavy elements. 
With these definitions, the
observational distinction between giant planets with
masses $\ga M_J$ and small brown dwarfs is somewhat indistinct,
but should eventually be amenable to spectroscopic determination
\citep{2007prpl.conf..623C}.
The lower limit to the mass of a brown dwarf
is set by the condition that the star become opaque to the radiation it emits
while undergoing gravitational collapse \citep{1976MNRAS.176..367L}; including helium, this
is about $4\times 10^{-3} M_\odot\simeq 4 M_J$ \citep{2007prpl.conf..459W}.
The smallest brown dwarfs detected to date have masses $\sim (0.01-0.02)M_\odot$
\citep{2007prpl.conf..443L}.

In order for a brown dwarf to form, its mass must exceed the Bonnor-Ebert mass,
even if it forms via shock compression \citep{1978ApJ...220.1051E}; 
the pressure at the surface of the 
core that forms the
brown dwarf must therefore 
be $P/k_{\rm B}\ga 
10^9(T/10\;\mbox{K})^4(10^{-2}\;M_\odot/m_{\rm BD})^2$~K~cm\eee\ 
at the surface of the brown dwarf.
Assuming that brown dwarfs form by turbulent fragmentation,
\citet{2004ApJ...617..559P} show that such pressures can be reached in a large enough fraction of the
mass of the cluster IC348 to account for the brown dwarfs observed there.
This model is based on the asumption that the gas is isothermal and that
as a result the density PDF is a log-normal (eq. \ref{log-norm}).
The remaining mystery is, why is the IMF of brown dwarfs relatively constant
(at least to within a factor 2)
when their numbers are exponentially sensitive to the mean of the log-normal,
which depends on the Mach number at the largest scale in the cloud?
This mystery is part of the larger mystery as to why the IMF appears
to be universal, but in the brown dwarf regime
the exponential sensitivity to ambient conditions
potentially offers an opportunity
to determine how ambient conditions affect the IMF.

\citet{2007prpl.conf..459W} 
review a number of other mechanisms for brown dwarf formation that could
contribute in some cases: (1) Hierarchical fragmentation---Protostellar cores
can fragment as they collapse, and indeed this process is believed to
lead to the formation of binary 
and multiple star systems (\S \ref{sec:binary}).
Simulations are as yet inadequate to 
evaluate 
the effectiveness of
this process in producing brown dwarfs; in particular, it is 
important to include proper treatment of the
radiative transfer \citep{2000ApJ...528..325B,2006MNRAS.367...32W}.
(2) Disk fragmentation---Several analyses have shown
that fragmentation in disks around low-mass stars is suppressed for
$r\la 100$~AU
\citep{2005ApJ...628..817M,2005ApJ...621L..69R,2007prpl.conf..459W},
but again accurate treatment of radiative transfer is essential 
for quantifying this further.
\citet{2007A&A...466..943G} 
have suggested that brown dwarfs form in disks beyond 100 AU and that the resulting binaries 
are disrupted by passing stars.
(3) Premature ejection of protostellar embryos \citep{2001AJ....122..432R} ---This is a variant of
the hierarchical fragmentation scenario, in which low-mass protostars that
form via fragmentation are ejected before they can accrete enough matter to
reach the hydrogen-burning limit. \citet{2001AJ....122..432R} suggested that the ejected brown
dwarfs would have a higher velocity dispersion, a more extended
spatial distribution and smaller disks than their more massive cousins. This has
not been observed \citep{2007prpl.conf..443L}, but
SPH simulations suggest that in fact these differences
between brown dwarfs and hydrogen-burning stars are relatively small 
\citep{2005AN....326.1040G}; more accurate calculations are needed to
determine the magnitude of the differences. 
A difficulty with the ejection model is that the 
cluster simulations that support it produce
too many single stars to be consistent with observation \citep{2005A&A...439..565G}.
Observations of BD-BD binaries can provide a strong test of models
for brown dwarf formation, particularly the ejection model \citep{2007prpl.conf..427B}.
(4) Photoevaporation---Cores that are close to an O star can undergo a 
radiation-driven implosion \citep{1980SSRv...27..275K,1989ApJ...346..735B}; the subsequent
equilibrium photoevaporation produces very high pressures \citep{1990ApJ...354..529B}.
This process may produce brown dwarfs (e.g., \citealp{2004A&A...427..299W}), but 
more work is needed to determine if the number of such brown dwarfs is significant.

\paragraph{Binaries}
\label{sec:binary}

Stars are roughly evenly divided between those that are in multiple systems
(mainly binaries) and those that are single. For stars of mass $\sim 1 \,M_\odot$,
\citet{1991A&A...248..485D} 
found that the fraction of stellar systems that are multiple---i.e.,
the ratio of the total number of binaries, triples, etc. divided by the
total number of systems, including single stars---is $\fm=0.58$.
Multiplicity declines for smaller masses, and
\citet{1997AJ....113.2246R} find $\fm\simeq 0.3$ for
stars in the solar neighborhood, which are primarily M stars.
\citet{2006ApJ...640L..63L} finds a similar result when one averages over the
entire IMF: the majority of stellar systems 
(as opposed to the majority of stars)
are single. 
The fraction of singles is smaller at birth, however: Higher-order multiples
(triples, etc.) are often dynamically unstable, and in dense environments
collisions among stellar systems can disrupt 
wide
binaries \citep{1999NewA....4..495K}.
Observations summarized by \citet{2007prpl.conf..379D} show that 
the multiplicity among 
T Tauri stars in low-density associations such
as Taurus-Auriga and Ophiuchus and among Class I sources
in high-density regions such as L1641 in Orion is about twice 
as high as among field stars. 
In dense star-forming regions like Orion, however, this excess multiplicity is soon 
erased---for example, the 
multiplicity in the Orion Nebula Cluster (which is older than
L1641) is the same as in the field.
This introduces a complication in comparing the core mass function with
the IMF (\S \ref{clumps}): on average, cores produce more than one
star, and the properties of the resulting stellar systems evolve with time.
This is not a major complication, however, since each core typically
produces only 2-3 stars \citep{2005A&A...439..565G},
and the distribution of secondary masses in typical
binaries appears to follow the 
field star IMF \citep{2007prpl.conf..133G}.
It should be noted that the multiplicity of the stars 
is imprinted on their spatial
distribution: \citet{1995MNRAS.272..213L} found a clear break in the density of companions in
Taurus at about 0.04~pc, separating binaries and multiple
stars from larger scale clusters.

Binaries raise two important issues in the theory of star formation:
What is the role of binaries in 
reducing the angular momentum 
inherited by protostars?
What determines how molecular cores fragment?
Binaries do not appear to be effective in taking up the angular momentum 
of the initial core:
\citet{2004ApJ...600..769F} 
has shown that 
turbulent molecular cores 
must
lose 99\%-99.9\% of their initial angular momentum
in order to
qualitatively account for a number of features of the binary population
with periods $\ga 10^3$~day.
%
This angular momentum loss is generally assumed to
be due to magnetic braking (\S \ref{sec:magcollapse}), but 
\citet{2004A&A...423....1J} suggest that gravitational torques can also contribute.
%
%
%
Binaries can remove 
angular momentum on small scales 
by ejecting a
a companion, which
hardens the remaining binary. However, \citet{2005A&A...439..565G} 
point out two limitations on this process: (1) it tends to create equal 
mass binaries, which are
not common for typical stars, and (2) it would create a population of 
single stars 
significantly larger than observed.

There is an extensive literature on the fragmentation of protostellar cores
into 
binary and 
multiple protostars that is reviewed by 
\citet{2000prpl.conf..675B}, \citet{2007prpl.conf..379D} and 
\citet{2007prpl.conf..133G}.
Simulations of fragmentation
are very challenging because of the enormous range of scales involved,
and it does not appear that any of the simulations carried out to date have
enough resolution and enough physics (i.e., including MHD and radiative 
transfer) to adequately address the problem \citep{2007prpl.conf...99K}. In particular,
a number of simulations produce 5-10 fragments per core, whereas observations
show that most cores produce only 2-3 fragments \citep{2005A&A...439..565G}.

\subsection{Disks and Winds}
\label{disks_and_winds}

\subsubsection{OBSERVATIONS OF DISKS}
\label{diskobs}

Because protostellar cores are rotating, collapse with conservation of
angular momentum results in the formation of a centrifugally-supported
disk (\S \ref{sec:rotcollapse}).  Observed sizes and rotation
parameters for low-mass dense cores 
predict disk sizes $\simlt 1000$ AU, consistent with high-resolution
submillimeter continuum observations that
indicate average (dust) disk sizes around T Tauri stars of 
$\approx 200$AU \citep{2006astro.ph.10813A}; similar results are
obtained using mm interferometry \citep{2002ApJ...581..357K}.  
The disks around T Tauri disks extend inward to $\sim 0.04$ AU based
on modeling of observed CO vibrational emission lines \citep{N06};
these inner radii are smaller than the inner disk radii inferred for
dust disks, presumably because the dust sublimates.
The initial sizes of circumstellar 
disks are more difficult to determine, because protostellar systems in
the earliest stages (prior to the T Tauri stage) 
are still enshrouded in dusty envelopes that emit
at similar wavelengths to the disk; in a few cases where the inner
envelope emission can be spatially separated out, disk sizes appear
similar (e.g. \citealt{2005ApJ...632..973J}). 

Masses of protostellar disks are estimated using the continuum flux in
millimeter and submillimeter wavelengths.  The first disk mass
estimates for T Tauri stars were obtained from observations of the
total flux at a single millimeter wavelength under the assumption of
optically thin emission (e.g., \citealt{1990AJ.....99..924B}), but
subject to an uncertainty in the overall normalization due to the 
uncertainty in
the dust opacity coefficient (since 
$M_{\rm disk} \propto F_\nu/(\nu^2\kappa_\nu)$).
Multiwavelength (sub-mm to cm) observations suggest that the dust
opacity law $\kappa_\nu \propto \nu^\beta$ has a distinctly shallower
slope $\beta$ than holds for dust in the diffuse ISM, presumably due
to grain growth (e.g.,
\citealt{1991ApJ...381..250B,2000prpl.conf..533B}).  Interpretation of
the multiwavelength flux data as implying a change in $\beta$ is
complicated by the fact that some of the short-wavelength emission can
be optically thick (which 
would yield $F_\nu \propto \nu^2$ independent of
$\beta$ for a disk that is optically thick at all $\nu$
).  However, spatially resolved observations can be combined
with modeling to correct for optically-thick contributions
varying with $\nu$ and $R$
, with the
resulting median $\beta \simlt 1$ (e.g.
\citealt{2007prpl.conf..767N,2006astro.ph.10813A,2006astro.ph.10667L}),
suggesting that the largest ``grains'' are in fact cm-size pebbles (e.g.
\citealt{2005ApJ...626L.109W,2006A&A...446..211R}).  Total disk masses
for T Tauri systems are estimated to be in the range
$\sim 10^{-3}-10^{-1}\,\Msun$,
 with a median near $0.005\,\Msun$ from
submillimeter observations (e.g. \citealt{2005ApJ...631.1134A}), but
these may severely underestimate the true masses if a large fraction
of the grains have grown
to mm or cm sizes 
and thus emit only weakly in the submm
\citep{2006ApJ...648..484H,2007prpl.conf..767N}. 
Determining the distribution of mass within
disks is difficult because submm emission is likely optically thick in
the inner regions, while at longer wavelengths there is insufficient
resolution to probe the inner-disk regions \citep{2006astro.ph.10813A}.
Finally, we note that disk mass determinations assume a cosmic ratio
of gas to dust; at late evolutionary stages, photoevaporation may 
preferentially remove gas
and planet formation may preferentially remove dust.

The thermal structure of protostellar disks is likely quite complex.
Disks can be heated both externally via irradiation from the central
star, and internally from dissipation and thermalization 
of orbital kinetic energy as the gas
accretes \citep{1991ApJ...380..617C,1997ApJ...490..368C}.
As a consequence, the vertical temperature
distribution depends on details of the system and can have a local
minimum at intermediate altitude \citep{1998ApJ...500..411D}.
The vertical temperature distribution together with its dependence on
radius must be self-consistently calculated, since the flaring of the
disk surface affects the amount of radiation intercepted from the
central star (see discussion and references in \citealp{2007prpl.conf..555D}). 
In addition, gas and dust temperatures may differ in the upper
atmospheres where the densities are low and stellar 
X rays strongly heat the gas \citep{N06}.
Quite sophisticated radial-vertical  
radiative models (including grain growth and
settling) have been developed that agree
well with observed spectral energy distributions from $\mu$m to mm
wavelengths (see e.g. \citealp{2004A&A...421.1075D}, 
\citealp{2006ApJ...638..314D} and references therein).  
The IR emission signatures, including PAH
features at $3-13\,\mu$ and edge-on silhouette images, as well as
scattered-light/polarization observations in optical and near-IR,
indicate that although some grains have grown to large
sizes, small grains still remain in disk atmospheres (see references
and discussion in \citealp{2007prpl.conf..555D} and \citealp{2007prpl.conf..767N}).

Disk lifetimes are inferred based on stellar ages combined with IR and
mm/sub-mm emission signatures, which are sensitive to warm
dust. Multiwavelength {\it Spitzer} observations of the nearby
star-forming cluster IC 348 \citep{2006AJ....131.1574L} 
show that for $\sim 70\%$ of stars, disks
have become optically-thin in the IR (implying inner disks $R\simlt
20$AU have 
been removed) 
within the 2-3 Myr age of the system; disk
fractions are slightly higher ($\sim 50\%$) for Solar-type stars than
in those of higher or lower mass.
Observations of other clusters are consistent with these results
\citep{2006ApJ...638..897S}.  $L$-band observations of disk
frequencies in clusters spanning a range of ages  
\citep{2001ApJ...553L.153H} suggests 
that overall disk lifetimes are $\approx 6$~Myr.  
Even in the 10Myr old
cluster NCG 7160, however, a few percent of stars still show IR
signatures of disks
\citep{2006ApJ...638..897S}, and disk lifetimes appear to be inversely
correlated with the mass of the star \citep{2007astro.ph..1476H}.
Signatures (or their absence) of dusty disk emission are also well
correlated with evidence (or lack) of accretion in gaseous emission line
profiles (see below) in systems at range of ages, indicating that gas
and dust disks have similar lifetimes
\citep{2006ApJ...648.1206J,2006AJ....132.2135S}.
\citet{2005ApJ...631.1134A} found, for a large sample of YSOs in
Taurus-Auriga, that in general those systems with near-IR signatures
of inner disks also have sub-mm signatures of outer disks, and vice
versa; they conclude that inner and outer disk lifetimes agree within
$10^5$~yr.  

Accretion in YSO systems is studied using a variety of diagnostics
(see e.g. \citet{2000prpl.conf..377C}), including continuum ``veiling'' of
photospheric absorption lines and optical emission lines, which are
respectively believed to arise from hot (shocked) gas on the stellar
surface and from gas that is falling onto the star along magnetic flux
tubes.  \citet{1998ApJ...492..323G} 
measured a median accretion rate for million-year-old T Tauri stars of
$\sim 10^{-8}\Msun$~yr\e, and \citet{2001ApJ...556..265W} found similar
accretion rates for the primaries in T Tauri binary systems.
A recent compilation of observations
\citep{2003ApJ...582.1109W,2003ApJ...592..266M,2004AJ....128.1294C}
shows an approximate dependence of the accretion rate on
stellar mass 
$\dot M_{\rm disk} \propto m_*^2$, 
although with considerable
scatter \citep{2005ApJ...625..906M}. 
This  scaling of the accretion rate with stellar mass is potentially explained
by Bondi-Hoyle accretion from the ambient molecular cloud 
\citep{2005ApJ...622L..61P}. 
However, such a model accounts only for the
infall rate onto the star-disk system, not the disk accretion rate; these
need not agree.  In addition, it does not 
account for the accretion seen in T Tauri stars outside molecular
clouds \citep{2006ApJ...648..484H}.
During their embedded stages
(a few $\times 10^5$ yr), low-mass stars have typical disk accretion rates 
similar to or slightly larger than 
those of TTSs \citep{2007prpl.conf..117W}. 
As discussed in \S \ref{infallobs}, the infall rates
from protostellar envelopes typically exceed disk
accretion
rates by a factor 10-100, so it is
possible that mass is stored in the disk and released intermittently,
in brief but prodigious accretion events similar to FU Ori outbursts 
\citep{1990AJ.....99..869K,1996ARA&A..34..207H}.

For high-mass protostars, 
observations suggest that there are at least two classes of
disks \citep{2006astro.ph..3093C}.
In moderate-luminosity sources corresponding to B stars ($L\la$~few$\;
\times 10^4\;L_\odot$), the disks appear to be
Keplerian, with masses significantly less than
the stellar mass and time scales for mass transfer $\sim 10^5$~yr. In luminous
sources ($L\ga 10^5\;L_\odot$), the disks are large ($4-30\times 10^3$~AU) and
massive ($60-500\;M_\odot$). Consistent with the discussion in
\S\S \ref{sec:rotcollapse} and \ref{sec:magcollapse}, the
disks are observed to be non-Keplerian on these large scales.
To distinguish these structures from the disks observed around B stars,
\citet{2005Ap&SS.295....5C} terms them ``toroids." The inferred 
infall rates in these disks are of order $2\times 10^{-3}-2\times 10^{-2}\;M_\odot$~yr\e,
corresponding to mass transfer time scales of order $10^4$ yr \citep{2005IAUS..227..135Z}.
In view of their large size and mass, they may be
circumcluster structures rather than circumstellar ones. Indeed, one of the 
best studied luminous sources, G10.8-0.4, is inferred to have
an embedded cluster of stars with a total mass $\sim 300\;M_\odot$
\citep{2005ApJ...624L..49S}.
Simulations of the formation of an individual
massive star in a turbulent medium
give a disk size $\sim 10^3$~AU, significantly smaller than the size of the toroids
\citep{2005IAUS..227..231K}. To date, no disks have been observed
in the luminous sources on scales $\la 10^3$~AU. 
Most likely, this is because of the observational difficulties in observing
such disks; it should be borne in mind, however, that there is no
direct evidence that these sources are in fact protostellar.
Including disks around both B stars and 
the toroids around luminous sources, \citet{2005IAUS..227..135Z}
finds that the mass infall rate in the disks scales as 
$\dot M_{\rm disk}\propto m_*^{2.2}$,
although there are substantial
uncertainties in the data for the luminous sources.

\subsubsection{ACCRETION MECHANISMS}

The most fundamental theoretical question about YSO disks is what
makes them accrete; while many mechanisms have been investigated,
the problem is still open.  In large part this is because the
accretion process depends on a complicated interplay of MHD, radiative
transfer, chemistry, and even solid state physics.  The MHD is itself
non-ideal, since the medium is partially ionized, and in addition
self-gravity is important in many circumstances.  Self-gravity effects
and the level of 
electrical conductivity are very sensitive to thermal and ionization
properties, which in turn are determined by chemistry and radiative
transfer (including X-rays and cosmic rays), and the latter
are strongly affected by grain properties that evolve in time due to
sticking and fragmentation.  Compounding the difficulty imposed by the
interactions among the physical processes involved is the lack of
exact knowledge of initial and boundary conditions: how does collapse
of the rotating protostellar core shape the distribution of mass in
the disk, starting from the initial disk-building stage and
continuing (although at a reduced rate) with later infall?  Finally,
there is the difficulty imposed by the huge dynamic range in space and
time; disks themselves span a range of $\sim 10^4$ in radius and
$10^6$ in orbital period, while the small aspect ratio $H/R\ll 1$
(where $H$ is the scale height of the disk)
implies a further extension in dynamic range is required for numerical
models that
resolve the disk interior.  

Processes proposed to transport angular momentum in YSO disks
generally fall into one of three categories: purely hydrodynamic
mechanisms, MHD mechanisms, and self-gravitating mechanisms
(e.g., see the reviews of \citealt{2000prpl.conf..589S} and \citealt{2005ASPC..341..145G}).  
Within the last decade, it has become
possible to investigate mechanisms in each class using high-resolution
time-dependent
numerical simulations in two and three dimensions, in which the
stresses that produce transport are explicitly obtained as spatial
correlations of component velocities, magnetic fields, and the density
and pressure for a self-consistent flow.  
Prior to the computational revolution that made these investigations
possible, and continuing into the present for modeling in which large
radial domains and long-term evolution is required, many studies have
made use of the so-called ``alpha prescription'' for angular momentum
transport. In this approach
\citep{1973A&A....24..337S,1974MNRAS.168..603L,1981ARA&A..19..137P},
a stress tensor is defined that yields an effective viscous torque between
adjacent rings in a differentially rotating disk.  On dimensional
grounds, and using the fact that the shear stress should be
zero for solid-body rotation, this stress can be written as 
$T_{R,\phi}\equiv -\alpha P d \ln \Omega/d \ln R$; i.e. the effective 
kinematic viscosity is taken to obey $\nu\equiv \alpha
\sth^2/\Omega=\alpha \sth H$.  
This effective viscosity {\it Ansatz} makes it
possible to study disk evolution with a purely hydrodynamic,
one-dimensional model.  While the ``$\alpha$-model'' approach 
has been essential to
progress on modeling disk observables, it is limited in its ability
to capture realistic dynamics since the coefficient is arbitrary (and
usually taken as spatially constant) and the adopted functional form
for $T_{R,\phi}$, while
dimensionally correct, may not reproduce the true behavior of
nonlinear, time-dependent, three-dimensional flows 
(e.g., see 
\citealt{2003MNRAS.340..969O,2006astro.ph.12404P}). For a Keplerian disk,   
$-d \ln \Omega/d \ln R =3/2$  and in steady state the 
mass accretion rate is $\dot M_{\rm disk}=3\pi \Sigma \nu=3 \pi \Sigma \alpha
\sth^2/\Omega$; i.e. the ratio of radial inflow speed to orbital speed
is $(v_R/v_\phi)=(3/2) \alpha (\sth/v_\phi)^2=(3/2)\alpha (H/R)^2$.
Observed accretion rates of TTSs require $\alpha \sim 10^{-2}$ 
\citep{1998ApJ...495..385H}.  Since the effective viscosity is equal
to a characteristic length scale for angular momentum transport 
times a characteristic transport speed, the empirically-determined 
viscosity corresponds to a few
percent of the value that would obtain if transport occurred at sonic
speeds over distances comparable to the scale height of the disk.

Using the infall rate scaling of equation (\ref{eq:mds}), the ratio of
the disk accretion rate to the infall rate is 
\begin{equation}
\frac{\dot M_{\rm disk}}{\dot  m_{\rm in}} \sim
\frac{\alpha}{\phin} 
\left( \frac{M_{\rm disk} }{m_* }\right) 
\left( \frac{R}{H }\right)
\left( \frac{T_{\rm disk} }{T_{\rm core}}\right)^{3/2},
\end{equation}
where we
have assumed that the gravitational potential is dominated by the star.
The outer-disk temperature is not much larger than the temperature
in the core, and $R/H\sim 10$ for the outer disk, so the disk accretion
rate is much lower than the infall rate unless 
$M_{\rm disk}/m_*$
or 
$\alpha/\phin$
exceeds $\sim 0.1$.  This is not the case for TTSs, but during the
embedded stages the disk masses may be larger, and (possibly as a
consequence of larger $M_{\rm disk}$ and self-gravity; see below) the
values of $\alpha$ may be larger as well.

\paragraph{Hydrodynamic Mechanisms}

The simplest transport mechanisms would be purely hydrodynamic.
Turbulence generated either through convection (due to vertical or
radial entropy gradients), through shear-driven hydrodynamic instabilities, or
through external agents (such as time-dependent, clumpy infall) could
in principle develop velocity field correlations $\langle \rho \delta v_R
\delta v_\phi\rangle$ of the correct sign ($>0$) to transport angular momentum
outward.  \citet{1992ApJ...388..438R} showed, however, that convective modes
tend to transport angular momentum inward, rather than outward, and 
\citet{1996ApJ...464..364S} confirmed from three dimensional
 numerical simulations with
turbulence driven by convection that angular momentum transport is inward.
Convection driven by radial entropy gradients also transports angular
momentum inward, and is generally stabilized by differential rotation
\citep{2006ApJ...636...63J}.

Several analytic studies have shown that purely hydrodynamic
disturbances in Keplerian-shear disks are able to experience large
transient growth
\citep{2003A&A...402..401C,2004ApJ...606.1070K,2004A&A...427..855U,
2005ApJ...626..978J,2005ApJ...629..373A},
especially for the case of two-dimensional (i.e. $z$-independent)
columnar structures.  Conceivably, transient growth of sheared waves 
could lead to self-sustained 
turbulence with outward transport of angular momentum,
if new leading wavelets could be continually reseeded in the flow
via nonlinear interactions
\citep{2007astro.ph..2046L}.
While transient growth is indeed seen in
two-dimensional ($R-\phi$)
numerical simulations, it is subject to secondary Kelvin-Helmholtz
instability that limits the growth when $|k_R \delta v_\phi|/\Omega
\simgt 1$ \citep{2006astro.ph..9131S}.  The turbulence that results
also appears to decay without creating leading wavelets to complete 
the feedback loop, but this may be due to limited numerical
resolution. Other numerical
evidence, together with analytic arguments, suggest that nonlinear 
shear-driven 
hydrodynamic instabilities are unable to maintain turbulence for
Rayleigh-stable rotational profiles (in which angular momentum
increases outward, i.e. $\kappa^2/\Omega^2=2d\ln (\Omega R^2)/d\ln R >0$)
\citep{1996ApJ...467...76B,1999ApJ...518..394H}.  Since simulations
using the same numerical methods show that analogous Cartesian
shear flows do exhibit nonlinear instability, rotating systems are
presumably stabilized by Coriolis forces and the epicyclic motion
that results.  One potential concern is that the effective Reynolds
numbers of numerical experiments are too low to realize nonlinear
shear-driven instabilities and self-sustained
turbulence.  Very recently, however, \citet{2006Natur.444..343J} reported
from laboratory experiments at Reynolds numbers up to millions
that hydrodynamic flows with Keplerian-like rotation profiles in fact show
extremely low levels of angular momentum transport, corresponding to
$\alpha <10^{-6}$.

Although it may be difficult to grow perturbations from instabilities
in uniform Keplerian disks, it is still possible that disks are born
with large internal perturbations, and that ongoing infall at all
radii can continually resupply them.  Simulations have shown that
two-dimensional disks with non-uniform vorticity tend to develop
large-scale, persistent vortices that are able to transport angular
momentum outwards \citep{2004A&A...427..855U,2005ApJ...635..149J}.
Three-dimensional simulations, however, show that vortex columns tend
to be destroyed \citep{2005ApJ...623.1157B,2006astro.ph..9131S}.
While off-midplane vortices can be long-lived
\citep{2005ApJ...623.1157B}, the angular momentum transport in
three-dimensional simulations is an order of magnitude lower than for
the two-dimensional case \citep{2006astro.ph..9131S}, and secularly
decays.  Further investigation of this process is needed, and it is
particularly important to assess whether vorticity can be injected at
a high enough rate to maintain the effective levels of $\alpha \sim
10^{-2}$  needed to explain observed TTS accretion.

\paragraph{MHD Mechanisms}

The introduction of magnetic fields considerably alters the dynamics
of circumstellar disks.  The realization by
\citet{1991ApJ...376..214B} that 
weakly-or-moderately magnetized, 
differentially
rotating disks are subject to a powerful local instability -- now
generically referred to as the magnetorotational instability (MRI) --
revolutionized the theory of accreting systems.  Early axisymmetric
numerical simulations showed robust growth and development of the
so-called channel solution \citep{1991ApJ...376..223H}, while
three-dimensional numerical simimulations showed emergence of
quasi-steady state saturated turbulence
\citep{1995ApJ...440..742H,1995ApJ...445..767M,1995ApJ...446..741B} 
in which the angular
momentum transport is outward, and is dominated by the magnetic stresses,
$\langle -B_R B_\phi/(4\pi) \rangle$.  Much effort has been devoted to
exploring the MRI as a basic mechanism driving accretion in a variety
of systems; \citet{1998RvMP...70....1B} and
\citet{2003ARA&A..41..555B} summarize many of the these developments.
The effective value of $\alpha$ depends on the mean vertical magnetic
flux, which presumably evolves over long timescales, and can easily
exceed 0.1
(e.g., \citealt{1996ApJ...464..690H,1996ApJ...463..656S,2004ApJ...605..321S}).

While MRI almost certainly plays an important role in driving
accretion in YSO systems, it is not a magic bullet.  The difficulty is
that substantial portions of these disks may have ionization too low
for MRI to be effective
\citep{1996ApJ...457..798J,1996ApJ...457..355G,1997ApJ...480..344G,1998ApJ...500..411D,1999ApJ...518..848I},
creating a ``dead zone''.  
A critical review of the
requirements for MRI to develop in
partially-ionized disks is given in 
\citet{2005ASPC..341..145G}; 
Ohmic diffusion appears to be the main
limiting effect, with the saturated-state value of $\alpha$ dropping
when $v_{A,z}^2/\eta \Omega\simlt 1$,
where $\eta$ is the resistivity
\citep{2002ApJ...577..534S,2006astro.ph.12552T}.  In the very
innermost parts of YSO disks ($R\simlt 0.1$ AU), alkali metals are
collisionally ionized where the stellar irradiation maintains the
temperature above $\sim 2000$K, so MRI can operate.  In the outer disk
(beyond several AU), and in the mid-disk's surface layers, column
densities are low  enough 
($\Sigma < \Sigma_a \sim 100 \g\, \cm^{-2}$)
that X rays or cosmic rays can penetrate the
disk to ionize it.  (For comparison, the surface density in the 
minimum Solar nebula is
$\Sigma=1700 (R/{\rm AU})^{-1.5}
\g\, \cm^{-2}$ -- \citealp{1985prpl.conf.1100H}.)
Unfortunately, the extent of the MRI-active region
in the outer disk is very sensitive to the presence and size
distribution of dust particles; if small grains are present and
well-mixed, the active region is quite limited, while it can become
very large if all the dust is incorporated in large particles or
settles to the midplane
\citep{2000ApJ...543..486S,2002MNRAS.329...18F,2004ApJ...608..509D,2005MNRAS.361...45S}.
Even if ionizing radiation is limited to the surface layers 
by high total disk
columns, if small grains are absent (an extreme assumption) 
the gas-phase recombination rate
is low enough such that turbulence with rapid vertical mixing can
maintain non-negligible ionization in the interior.
\citet{2006astro.ph.12552T} have shown, using direct numerical
simulations, that  
the dead zone can be effectively eliminated in this (optimistic) scenario; 
while the very center of the disk at 1 AU is not unstable to MRI,
the interior is still conductive enough that magnetic fields generated
nearer the surface can induce accretion in the midplane.

One of the possible consequences of spatially-varying conductivity in
disks is that the accretion rate will, in general, vary with radius.
If only a surface layer $\Sigma_a$ is ``active'', 
in the sense of
being sufficiently conductive to support MRI with effective viscosity
coefficient $\alpha_a$, then the accretion rate in that layer will be
$\dot M_a = 3 \pi \Sigma_a \alpha_a \sth^2/\Omega$.  Since $\Sigma_a$
varies slowly with radius (for the case of external ionization) while
the combination $\sth^2/\Omega$ tends to decrease inward, dropout from
the accretion flow can accumulate within the ``dead zone'' that is
sandwiched between active layers \citep{1996ApJ...457..355G}.  If the
``dead zone'' remains completely inactive, then matter will build up
until it becomes dynamically unstable and begins to transport angular
momentum by gravitational stresses (see below), potentially leading to
transient bursts of accretion
\citep{1996ApJ...457..355G,2001MNRAS.324..705A}.

Finally, we note that MHD winds (see \S\ref{mag_winds}) 
may remove angular momentum from
disks, driving the matter remaining in the disk to accrete in order to
maintain centrifugal balance.  The angular momentum deficit is 
tranferred to the disk by magnetic stresses, so that only the matter
that is well-coupled to magnetic fields will be affected.  Thus, the
above considerations regarding ionization also apply to wind-driven accretion.

\paragraph{Self-Gravitational Mechanisms}
\label{self_grav_acc}

Accretion disks that have sufficiently small values of the Toomre
parameter 
$Q=\kappa \sth/(\pi G \Sigma)\sim (H/R) (m_*/M_{\rm disk})$ 
are subject to nonlinear
growth of density perturbations via the swing amplifier (see \S\S 
\ref{gravity}, \ref{GMC_formation}).  Then, in addition to 
hydrodynamic Reynolds stresses $\langle \rho v_R v_\phi \rangle$ and
MHD Maxwell stresses $-\langle B_R B_\phi/(4\pi) \rangle$,
gravitational ``Newton stresses'' 
$\langle g_R g_\phi /\linebreak[1] (4 \pi G)\rangle$  
(where ${\bf g}=-\nabla \Phi$)
also contribute to the radial transport of angular momentum. 
\citet{2001ApJ...553..174G} showed that if the disk is in equilibrium
such that cooling removes the energy dissipated by mass accretion at a
rate per unit area $\Sigma \sth^2/[(\gamma-1)t_{\rm cool}]$, then 
$\alpha^{-1} = (9/4)\gamma (\gamma-1) \Omega t_{\rm cool}$, where
$\gamma$ is the effective (two-dimensional) adiabatic index
(which takes into account vertical degrees of freedom, and 
depends on the three-dimensional index and degree of self-gravity).
Numerical simulations with simple cooling prescriptions (constant 
$t_{\rm cool} \Omega$) show that the disk can settle into a
self-regulated state with $Q$ near unity \citep{2001ApJ...553..174G,
2004MNRAS.351..630L,2005MNRAS.364L..56R,2005ApJ...619.1098M}, 
provided that $t_{\rm
cool}\Omega$ is not too small (in which case the disk fragments).  For
disks that are not externally illuminated, \citet{2003ApJ...597..131J}
performed
 two-dimensional simulations with realistic
opacities (and a ``one-zone'' vertical radiative transfer
approximation for cooling), and 
found 
that the transition between fragmentation and non-fragmentation
lies in the range $t_{\rm cool} \Omega = 1-10$.  The corresponding
$\Sigma$ at the transition point increases with $\Omega$, such that
outer disks are the most active regions gravitationally.
Values of $\alpha$ up
to $0.5$ are possible, with the equilibrium condition prediction satisfied
down to $t_{\rm cool} \Omega \approx 3$ and $\alpha \approx 0.1$.
Using a three-dimensional model of an $0.07\Msun$ disk
with realistic cooling, \citet{2006ApJ...651..517B}
find a value of $\alpha \sim 0.01$ over a large range of radii $>20$ AU.

In view of the limitations on $\alpha$, \citet{2006astro.ph..3093C} 
argue that accretion rates are limited
to values substantially smaller than inferred for the formation
of high-mass stars (\S \ref{hiinfall}).
On the other hand, \citet{2007ApJ...656..959K} find that disks around
high-mass protostars can transfer mass inward at the same rate that
it falls in. They carried
out simulations of high-mass star formation in
a turbulent medium and included radiative transfer rather
than prescribing the heating and cooling rates. They found that large amplitude
$m=1$ modes develop that give effective values of $\alpha$ of order
unity, in qualitative agreement with the isothermal disk results of
\citet{1994ApJ...436..335L}.

Disks that are illuminated sufficiently strongly will have
the temperature set by the external radiation field, rather than
internal dissipation of energy.  In that case, whether self-gravity is
important or not depends essentially on the amount of matter present
in a given region.  Where the surface density is high enough so that
$Q$ is near but not below the critical value $\approx 1.4$, self-gravitational
stresses will be appreciable but not so large as to cause
fragmentation.  
Analytic estimates assuming steady state and accretion heating as well
as irradiation 
\citep{2005ApJ...628..817M,2005ApJ...621L..69R} indicate that
fragmentation is only possible in the outer portions of disks,
although more massive disks, around more massive stars, are more
subject to fragmentation \citep{2006MNRAS.373.1563K}.
At temperatures comparable to those in
observed systems, disks with masses $\simgt 0.1 \Msun$ are candidates
for having significant mass transport due to self-gravitating torques 
\citep{2004ApJ...609.1045M}.  Thus, self-gravity is likely to be
particularly important during the embedded stage of disk evolution,
when disk masses are the largest.
\citet{2005ApJ...633L.137V,2006ApJ...650..956V} propose, based on
results of two-dimensional simulations, that recurrent 
``bursts'' of accretion due to self-gravity are likely to develop
during the early stages of protostellar evolution.
A number of other results from models of self-gravitating disk
evolution (with an emphasis on criteria for planet formation through
fragmentation) are presented in the review of
\citet{2007prpl.conf..607D}.

\subsubsection{DISK CLEARING}

While a large proportion of the mass in the disk ultimately accretes
onto the star, conservation of angular momentum requires that some of
the matter be left behind.  MHD winds during the main lifetime of the
disk remove some of this material (see \S\ref{mag_winds}).  What remains
is either incorporated into planets, or removed by photoevaporation.  
Although planet formation is inextricably coupled to disk 
evolution, recent developments in this exciting -- and rapidly expanding --
field are too extensive to summarize here.  A number of excellent
recent reviews appear in {\it Protostars and Planets V}. 

Disks can be irradiated by UV and X ray photons originating either in
their own central stars, or in other nearby, luminous stars (see e.g.
reviews of \citealt{2000prpl.conf..401H} and \citealt{2007prpl.conf..555D}).
EUV radiation penetrates only the surface layer of the disk, where it heats
the gas to $\sim 10^4\,\K$ (the ionization and heating depth is
determined by the Str\"omgren condition); FUV penetrates deeper into
the disk (where densities are higher), but heats gas to only a few
$100$~K \citep{1994ApJ...428..654H,1998ApJ...499..758J}.  The characteristic
radial scale in the disk 
for a thermally-driven wind is the gravitational radius 
$r_g = G m_* \mu /(kT)$, where $T$ is the temperature at the base of
the flow.  Pressure gradients enable flows to emerge down to $(0.1-0.2)
r_g$ \citep{1983ApJ...271...70B,2004ApJ...607..890F,2004ApJ...611..360A}.
EUV-driven winds
are most important in the inner disk, since the gravitational
potential there is too deep for FUV-heated regions at modest
temperatures to escape.

Observations discussed above (see also
\citealt{1995ApJ...450..824S} and \citealt{1996AJ....111.2066W}) indicate that the
inner and outer disks surrounding YSOs disperse nearly simultaneously
and on a very
short ($\sim 10^5$~yr) 
timescale, based on the small number of transition
objects between classical and weak T Tauri systems and the typical CTT
lifetimes of a few to several Myr.  Since the accretion time of the
outer disk itself determines the system lifetime, rapid removal of the
outer disk must be accomplished by other means; photoevaporation is
the most natural candidate.   Models of photoevaporation that also
include viscous disk evolution (which allow spreading both inward and
outward) have very recently shown that rapid and near-simultaneous
removal of the whole disk indeed occurs 
\citep{2001MNRAS.328..485C,2006MNRAS.369..216A,2006MNRAS.369..229A}. 
In this process, the accretion rate declines slowly over time until
the photoevaporative mass loss rate at some location 
in the inner disk exceeds the rate at which mass is supplied from
larger radii.  The inner disk, which is no longer resupplied from
outside, then drains rapidly into the star.  At the same time, the
radiative flux onto the outer disk grows as it is no longer attenuated
by the inner disk's atmosphere; the photoevaporation rate in the outer
disk climbs dramatically, and it is removed as well.

\subsubsection{OBSERVATIONS OF YSO JETS AND OUTFLOWS}
\label{windobs}

Young stellar systems drive very powerful winds.  The clearest
observable manifestations of YSO winds are the central ``Herbig-Haro''
jets consisting of knots of ionized gas 
($v\simgt 100\,\kms$), 
and the
larger-scale bipolar outflows consisting of expanding lobes of
molecular gas ($v\sim 10\,\kms$).  ``Jet-like'' outflows
(i.e. high-$v$, narrow molecular structures) are also observed in some
circumstances (see below).  The high velocities of jets indicate that
they represent (a part of) the primary wind from the inner part of the
star-disk system, while the low velocities and large masses of (broad)
molecular outflows indicate that they are made of gas from the star's
environment that has been accelerated by an interaction with the wind.
In addition to these observed signatures, there may be significant gas
in a large-scale primary wind surrounding the jet, which remains
undectected due to lower excitation conditions (low density,
temperature, and/or ionization fraction).

Outflows are ubiquitous in high-mass star formation as well as in low-mass
star formation \citep{1996ApJ...472..225S}. 
Outflows from 
high-mass protostellar objects
with $L<10^5\;L_\odot$
(corresponding to $m_*< 25\;M_\odot$---\citealp{1996snih.book.....A})
are collimated \citep{2002A&A...383..892B}, but somewhat less so than 
those in low-mass protostars \citep{2004A&A...426..503W}.
In some cases, jets are observed with the outflows, and in these cases the
momentum of the jet is generally large enough to drive the observed
outflow \citep{2005IAUS..227..237S}.
No well-collimated flow has been observed in a source
with $L>10^5\;L_\odot$; as remarked above, disks that are clearly
circumstellar have not been observed in such sources either.
\citet{2005ccsf.conf..105B} have proposed an evolutionary 
sequence that is consistent with
much of these data: A protostar that eventually will become an 
O star first passes through the HMPO stage with no H~II region and
with a well-collimated jet. When the star becomes sufficiently massive 
and close to the main sequence that it produces
an H~II region, the outflow becomes less collimated. The collimation
systematically decreases as the star grows in mass and the H~II region
evolves from hypercompact to ultracompact (see \S \ref{photo}).  
The remainder of this section focuses on winds and outflows from
low-mass stars, which have been observed in much greater detail than
their high-mass counterparts.

Recent reviews focusing on the observational properties of jets
include those of
\citet{2000prpl.conf..815E}, \citet{2001ARA&A..39..403R}, and
\citet{2007prpl.conf..231R}.  Jets are most commonly observed at high
resolution in optical forbidden lines of O, S, and N, as well as
H$\alpha$, but recent
observations have also included work in the near-IR and near-UV.  For
CTTs, which are YSOs that are themselves optically revealed, observed
optical jets are strongly collimated (aspect ratio at least 10:1, and
sometimes 100:1), and
in several cases extend up to distances more than a parsec from the
central source \citep{2007prpl.conf..215B}.  The jets contain both
individual bright knots with bow-shock morphology, and more diffuse
emission between these knots.

The emission diagnostics from bright knots are generally consistent
with heating by shocks of a few tens of $\kms$
\citep{1987ApJ...316..323H,1994ApJ...436..125H}, producing
post-shock temperatures $T_e\approx 10^4\, \K$.  
The electron density $n_e$, ionization
fraction $x_e=n_e/n_{\rm H}$, and temperature $T_e$ can be
estimated using line ratios \citep{1999A&A...342..717B}.  
Analyses of spectra from a
number of jets yields a range of parameters $n_e=(50-
3\times 10^3)\, \cmt$ and $x_e=0.03-0.6$ so that $n=(10^3-10^5)\, \cmt$
\citep{2006A&A...456..189P}.  
The total mass loss rate in jets $\dot M_{\rm jet}$, and hence the
total jet momentum flux, $\dot M_{\rm jet} v_{\rm jet}$, 
can be estimated using jet densities and velocities
together with an emission filling factor, yielding
$\dot M_{\rm jet} = (10^{-8} -10^{-7}) \Msun\, \yr^{-1}$
for CTTs \citep{2006A&A...456..189P}.  For  
Class 0 sources, which are much more luminous 
and have much higher accretion rates, estimated 
mass-loss rates in jets based on O I emission (from shocked gas)
extend up to $\dot M_{\rm jet} \sim 10^{-6} \Msun\,\yr^{-1}$
\citep{1997ApJ...476..771C}.
Inferred
values of $\dot M_{\rm jet}$ are generally correlated with estimates
of $\dot M_{\rm disk}$ from veiling \citep{1995ApJ...452..736H}, with
the ratio in the range $0.05-0.1$ \citep{2007prpl.conf..231R}.
%

The densities and temperatures obtained from jet diagnostics indicate
internal pressures 
$P/k_{\rm B}=(10^7-10^9)\, \K\; \cmt$ in the jet, exceeding the
ambient pressure in the surrounding core and GMC by a factor
$10^2-10^4$.  In principle, infalling envelope gas could provide a
``nozzle'' to collimate an emerging wind, 
but simulations indicate that only relatively weak winds can be so
confined as to produce a narrow jet \citep{2000ApJ...530..923D}.
This implies that observed jets must be contained within
a broader wind, with collimation likely produced by magnetic hoop
stresses (see below).  
Emission line analyses in fact indicate that a
lower-velocity [$\sim (10-50)\, \kms$] wind component is present near
the source, surrounding the high-velocity flow of a few $100 \,
\kms$ that emerges as the large-scale jet
\citep{1995ApJ...452..736H,1997A&AS..126..437H,2000ApJ...537L..49B,2005JKAS...38..249P}.  
Since velocities of
MHD winds scale with the Keplerian rotation speed of the footpoint
(see below), the presence of both high- and low-velocity components 
suggests that winds are driven from a range of radii
in the disk.  Recent high-resolution observations have detected
signatures of differential rotation in jets, using near-UV, optical,
and near-IR lines \citep{2002ApJ...576..222B,2007prpl.conf..231R};
these also indicate a range of wind launch points.  

Recent reviews of the observational properties of molecular outflows
include those of \citet{1999osps.conf..227B}, \citet{2000prpl.conf..867R}, and
\citet{2007prpl.conf..245A}. Like jets, classical molecular 
outflows can extend to distances
$0.1-1\pc$ 
from the central star, but they have much lower velocities
(up to a few tens of $\kms$) 
and collimation (aspect ratio $\sim 3-10$).  In a few very young, embedded
sources, molecular jets with much higher velocities and aspect ratios
have been observed in H$_2$, CO, and SiO lines (e.g., 
\citealt{1999A&A...343..571G,2002A&A...387..931B,2007astro.ph..1284L}).  
The total momentum
flux carried in CO outflows is correlated with the bolometric 
luminosity of the source and 
is discussed in \S \ref{GMC_evol}.
For embedded sources with $L_{\rm bol} = 1-10^{5} L_\odot$, the
momentum flux is $10^{-4}-10^{-1}\Msun \,\kms\, \yr^{-1}$
\citep{2000prpl.conf..867R}; 
in optically-revealed sources, 
this declines considerably (e.g. \citealt{1996A&A...311..858B}).

Detailed spectroscopic and morphological analysis of outflows enable
intercomparisons with theoretical models.  
Mapping of
outflows reveals both simple expanding shells, and more complex
features such as multiple cavities and bow shock structures that are
suggestive of episodic ejection events \citep{2002ApJ...576..294L};
outflow lobes become broader and more irregular over time
\citep{2006ApJ...646.1070A}.  Channel maps and position-velocity
diagrams in some sources show parabolic structures that are consistent
with driving by wide-angle winds, and in other sources show spur
structures that are consistent with jet driving 
\citep{2000ApJ...542..925L,2001ApJ...557..429L}.  The strongly curved
morphology of internal bow shocks (as seen in both molecular and atomic
tracers) indicates that the
wind must have velocities that decrease away from the poles, 
since a time-variable
wind with latitudinally-constant velocity produces nearly flat
internal shocks \citep{2001ApJ...557..429L}.  
This implies, in turn, that the wind is driven from a range
of radii in the disk, rather than arising from only a narrow region.

\subsubsection{DRIVING MHD WINDS AND JETS}
\label{mag_winds}

It was recognized very early on that jets and outflows contain more
momentum than could possibly be driven by radiation pressure 
\citep{1985ARA&A..23..267L}, whereas the high efficiencies and
velocities found by \citet{1982MNRAS.199..883B} for MHD
winds driven from accretion disks in near-Keplerian rotation 
suggested that the same magnetocentrifugal 
mechanism could drive winds in YSO systems 
\citep{1983ApJ...274..677P}. The main requirement for these winds to
develop is for the disk to be threaded by magnetic fields of
sufficient strength.  The mathematical theory of MHD winds and jets is 
presented in e.g. \citet{1996epbs.conf..249S} and \citet{2004adjh.conf..187P}.

Over the years, two main types of MHD wind models for YSO systems have
been explored.  One, the ``x-wind'' model 
(see \citealt{1994ApJ...429..781S} and references in 
\citealt{2000prpl.conf..789S} and \citealt{2007prpl.conf..261S}),
focuses on the interaction region between the stellar magnetosphere
and the inner accretion disk as the source of the wind.  In this
model, a large portion of the stellar dipole flux is taken to be
concentrated into a small range of radii near the point
where the magnetosphere and disk corotate.  Since YSOs are rapid
rotators,
 the corotation point is close to the star, and the wind that would
be launched could have terminal speed of a few $100\,\kms$, as is
observed in jets.  The second class of MHD wind models assumes that a
much larger region of the disk is threaded by open field lines, such
that there would be a range of terminal wind speeds, reflecting 
the range of rotation speeds at the magnetic field's footpoints in the
disk (see references in
\citealt{2000prpl.conf..759K} and \citealt{2007prpl.conf..277P}).  For disk winds, 
the poloidal magnetic flux could in part be generated locally
(e.g. by an MRI dynamo), in part be advected inward with the 
collapse of the prestellar core, and in part originate in the star and
diffuse outward into the disk.  Since one type of wind would not
exclude the other, it is likely that both x-winds and disk winds 
are present at some level.
This might help, for example, explain particular features of jets such
as their strong central density concentration as well as the apparent
decrease in velocity from inside to outside.

The observed rotation velocities in jets can be used to infer the
launch point in the disk
\citep{2003ApJ...590L.107A}.  From the Bernoulli equation for a 
cold flow along a streamline which rotates with angular velocity
$\Omega_0$, 
the quantity 
$
{\cal E}=\frac{1}{2}|{\bf v}|^2 + \Phi_g - v_\phi \Omega_0 R
$
is 
constant,
where 
$\Phi_g=-G m_*/r$ is the
gravitational potential
and in this section $R$ denotes the cylindrical radius.
At $R_{\rm obs}$, where the wind is observed 
(sufficiently beyond the Alfv\'en transition),
the dominant terms in 
the $\cal E$ equation 
are the first and the last terms on the RHS. For the
cases of interest,
$v_{\phi, \rm obs}\ll
v_{p, \rm obs}$,
where $v_p$ is the poloidal velocity,
 and 
$|{\cal E}|^{1/2}=(3/2)^{1/2}\Omega_0 R_0\ll v_{p, \rm obs}$, 
so that $\Omega_0 \approx v_{p, \rm obs}^2/(2 v_{\phi, \rm obs}
R_{\rm obs})$.  
The specific angular momentum $j=R(v_\phi- B_\phi B_p/[4\pi \rho
v_p])$ is also conserved along streamlines. 
One can show that
this is equal to 
$\Omega_0 R_A^2$,
where $R_A$ is the Alfv\'en radius of the wind. Since $j$
is dominated by the kinetic term at large distance
(where the wind is super-fast-magnetosonic), 
observations can be used to infer the ratio $R_A/R_{\rm obs}\approx \sqrt{2} 
v_{\phi, \rm obs}/v_{p,\rm obs}$.
For the low velocity component of 
DG Tau, \citet{2003ApJ...590L.107A} find from 
calculating $\Omega_0$ as above 
that the wind launch point
radii are $\sim 0.3-4$AU, implying a disk wind.  
The high velocity
component could originate as either an x-wind or a disk wind from
smaller radii. For DG Tau, the inferred ratio
$R_A/R_0\approx 2-3$ is also consistent with numerical solutions that
have been obtained for disk winds (see \citealt{2007prpl.conf..277P}
for a summary).  
This implies that the angular
momentum carried by the wind,
$\dot M_{\rm wind}\Omega_0 R_A^2$, which equals the angular
momentum lost by the disk, $\dot M_{\rm disk}\Omega_0 R_0^2$,
can drive accretion at a rate 
$\dot M_{\rm disk}/\dot M_{\rm wind}=(R_A/R_0)^2 \sim 4-9$.

The acceleration of MHD winds is provided by a combination of
the centrifugal ``flinging'' effect produced by rigid 
poloidal fields, and gradients in the toroidal magnetic pressure in
the poloidal direction (e.g. \citet{1996epbs.conf..249S}).  Beyond the
Alfv\'en surface, magnetic hoop stresses will tend to bend streamlines
toward the poles.  Full cylindrical streamline collimation, in the sense of 
${\bf v_p} \parallel \hat z$ asymptotically, can only occur if $B_\phi
R$ is finite for $R\rightarrow \infty$ \citep{1989ApJ...347.1055H}.
Using solutions in which all velocities scale as $v, v_A \propto r^{-1/2}$ and
the density and magnetic field respectively scale as $\rho \propto
r^{-q}$ and $B\propto r^{-(1+q)/2}$, \citet{1997ApJ...486..291O}
showed, however, that cylindrically-collimated disk winds are slow, in the
sense that the asymptotic value of $v_p/\Omega_0 R_{0}$ is at
most a few tenths.  Since observed jets are fast, they must either
have their streamlines collimated by a slower external wind, or else
be collimated primarily in density rather than velocity.
Time-dependent 
simulations have also shown that the degree of collimation in the flow
depends on the distribution of magnetic flux in the disk; cases with
steeper distributions of $B$ 
with $R$ tend to be less collimated in terms of streamline shapes 
\citep{2006ApJ...651..272F,2006MNRAS.365.1131P}.

The idea that nearly 
radially-flowing wide-angle MHD winds may produce a ``jetlike''
core, with density stratified on cylinders, was first introduced by 
\citet{1995ApJ...455L.155S} in the context of x-winds.  This effect
holds more generally, however, as can be seen both 
analytically \citep{1999ApJ...526L.109M} and
in simulations (see below).  
Asymptotically, the density approaches 
$\rho \rightarrow |B_\phi| R k/(\Omega_0 R^2)$ where $k$ is the (conserved)
mass flux-to-magnetic flux ratio
(also termed the mass-loading parameter).
Since nearly 
radially-flowing winds must 
be nearly force-free,
$|B_\phi| R$ varies weakly with $R$, such that if the
range of $k/\Omega_0$ over footpoints is smaller than the range of $R$
over which the solution applies 
(which is generally very large),
the wind density will vary as $R^{-2}$.  
The $R^{-2}$ dependence cannot continue to the origin;
\citet{1999ApJ...526L.109M} suggested that precession, internal shocks due to
fluctuating wind velocity, or magnetic instabilities would result
in a flattening of the density close to the axis
so that the momentum
flux in the wind $\rho v_w\propto (1+\theta_0^2-\cos^2\theta)^{-1}$, where
$\theta$ is the angle of the flow relative to the axis and
$\theta_0\ll 1$ measures the size of the flattened region.
This distribution gives approximately equal amounts of momentum in each logarithmic
interval of angle for $\theta>\theta_0$.
Several time-dependent numerical MHD simulations have demonstrated
this density collimation effect for wide-angle winds 
\citep{2003ApJ...582..269G,2003ApJ...595..631K,2005ApJ...630..945A}.

Magnetized winds are subject to a variety of instabilities 
(e.g. \citealt{2000ApJ...540..372K,2004Ap&SS.293..117H}), which may
contribute to 
enhancing the confinement of the jet,
structuring the jet column (yielding wanders, twists, and clumps), 
and mixing with
the ambient medium at interfaces.  Since jets
are likely surrounded by wider winds, they are to some extent
protected from the development of Kelvin-Helmholtz and helical modes that 
that disrupt jets propagating through ambient gas, although
development of axisymmetric pinch modes may still contribute to the
formation of HH knots \citep{2002ApJ...576..204H}.  
In addition,
lightly-loaded poloidal flux within the central core of the wind/jet may
help suppress the growth of large-scale pinch and kink instabilities
\citep{1995ApJ...447..813O,2006ApJ...653L..33A}.
Time-dependent 
simulations focusing on the portion of the wind flow above the disk
show that while steady winds are possible in certain ranges of the
mass-loading parameter for a given distribution of magnetic flux, in
other ranges no steady solution is possible 
\citep{1999MNRAS.309..233O,2005ApJ...630..945A}.  Since the spectral
diagnostics of HH objects indicate shock speeds of a few tens of
$\kms$, it is plausible that they form due to nonlinear steepening and
shocking of wind instabilities.

\subsubsection{ORIGINS AND EFFECTS OF OUTFLOWS}
\label{section_outflows}

Overall, the structure and kinematics of molecular outflows suggest
that they are driven by winds that originate from a range of radii in
the disk, with a dense central core (seen as a jet) surrounded by a
lower-density, lower-velocity wide-angle wind.  Jet driving and wind
driving of outflows have traditionally been explored separately,
although in practice they would operate in tandem.

Jets drive outflows as bow shocks, with ambient material swept into a
thin shell and carried away from the body of the jet as the shock
overtakes and entrains it
\citep{1993A&A...278..267R,1993ApJ...414..230M}, mixing newly shocked material
with material that is already flowing outward
\citep{1997A&A...323..223S}.  
The leading bow shock is itself created 
due to pressure forces 
at the circumferential boundary of the working surface
at the head of the jet, 
which drive transverse flows.
Jets with internal shocks can create
analogous bow shocks, with the difference that internal bow shocks
would propagate into the wind that surrounds the jet, whereas the
leading bow shock would propagate into the ambient medium.
Leading bow shocks tend to be fairly narrow, because the cooling of
shocked gas in the working surface limits the
transverse thrust that can be applied to the shell 
\citep{1999A&A...345..977D}.  As a consequence,
the width of the shell increases only as the cube root of the distance
from the head of the jet
\citep{1993ApJ...414..230M,2001ApJ...557..443O}.  Thus, bow shocks
have difficulty explaining broad outflows.  On the other hand, the
 ``convex spur'' velocity features seen in some systems agree
well with the predictions of bow shock models
\citep{2001ApJ...557..429L,2001ApJ...557..443O}.

For wide-angle winds, the momentum flux contained in the transverse
bulk motion of the wind is large compared to the thrust that could be provided
by pressure forces in the shell of shocked (strongly cooling) gas,
so that a momentum-conserving ``snowplow'' flow is a good approximation.
\citet{1991ApJ...370L..31S} developed the ``wind-swept shell'' model
of outflows based on this concept, which was able to explain the large
opening angles seen in most outflows.
\citet{1996ApJ...472..211L} and \citet{1999ApJ...526L.109M} extended
the wind-swept shell analytic model to incorporate the characteristic
$R^{-2}$ density stratification and logarithmic collimation of
streamlines of asymptotic wide-angle MHD winds, also allowing for 
latitudinal density stratification in the surrounding core.  The
mass-velocity and position-velocity relations for these analytic models agree
well with those in observed outflows.  Numerical simulations of
outflows swept up by wide-angle
winds \citep{2001ApJ...557..429L,2006ApJ...649..845S} are in good
agreement with the results of analytic models.

Outflows affect both the immediate environment of the forming star
(removing mass from the core before it can collapse into a disk), the
clump in which the core forms (also removing mass, and injecting
energy), and the larger-scale cloud (injecting energy).  The effects
of energy injection on clumps forming clusters of stars, and on GMCs as a
whole, are discussed in \S\ref{clusters} and \S \ref{GMC_evol},
respectively. Mass removal by winds is related to the star formation
efficiency, as we next discuss. 

The star-formation efficiency $\epsilon$ can be defined for individual
cores, for star-forming clumps, or for GMCs. The correspondence between
the Core Mass Function and the IMF has been discussed in \S \ref{clumps};
they are related by the core star-formation efficiency, $\ecore\equiv
m_*/M_{\rm core}$. (The individual-star IMF must also take into account
the multiplicity of the stars formed in a given core.) 
\citet{1995ApJ...450..183N}  
showed that outflows from protostars could reverse the infall and
determine $\ecore$; they assumed spherical winds and found
$\ecore\sim$~a few percent. \citet{2000ApJ...545..364M} 
calculated the dynamics of the outflows including collimation and
obtained $\ecore\sim 0.25-0.75$, depending on the
degree of flattening of the core due to magnetic support.
Subsequent observations suggest $\ecore\simeq 1/5 - 1/3$
(\S \ref{obs_imf_cmf}), at the low end of this range.
They also evaluated the star-formation efficiency for a star-forming clump,
and found typical values somewhat less than 0.5.
The predicted values of 
$\ecore$
 are inversely
proportional to the momentum per unit mass in the outflow, $p_w/m_*$;
they are consistent with observation for $p_w/m_*\sim 40$~km~s\e\ as
assumed, but not if $p_w/m_*$ is much smaller (see \S \ref{GMC_evol}
for a discussion of the values of $p_w$ inferred from observation).
Both \citet{1995ApJ...450..183N} and \citet{2000ApJ...545..364M} 
found that $\ecore$ is only weakly dependent on the
core mass, so that the CMF and the IMF should be similar in shape,
as observed (\S \ref{clumps}).

\subsection{High-Mass Star Formation}
\label{highmass}



High-mass protostars are characterized by Kelvin-Helmholtz times less than the
accretion time, so that they undergo nuclear burning while still accreting
(\S \ref{lowmass}). This leads to two powerful feedback effects that do not
apply to low-mass protostars, radiation pressure and photoionization
\citep{1971A&A....13..190L}. 
Furthermore, high-mass
protostars tend to form in dense clusters, so that interaction with other
protostars and newly formed stars may be important in their evolution.
Drawing on the review of \citet{2007prpl.conf..165B}
 we first summarize work on infall onto
high-mass protostars and then discuss the feedback effects.

\subsubsection{Protostellar Infall}
\label{hiinfall}

High-mass star
formation is generally taken to be a scaled up version of low-mass star formation:
The accretion rate is $\dot m_*\sim c_{\rm eff}^3/G$, 
where
the effective sound speed $c_{\rm eff}$ includes the effects of
thermal gas pressure, magnetic pressure, and turbulence
(Stahler et al. 1980, although they did
not address the issue of high-mass star
formation). As discussed in \S \ref{lowmass}, there may be a numerical
factor of a few in front of the $c_{\rm eff}^3/G$. Wolfire \& Cassinelli
(1987) found that accretion rates of order
$10^{-3}\;M_\odot$~yr$^{-1}$ are needed to overcome the effects of
radiation pressure for the highest
stellar masses, and attributed this to the high values of
$c_{\rm eff}$ in high-mass star forming regions. 
\citet{1992ApJ...396..631M} used their ``TNT" model (\S \ref{gravity}) to infer formation times
for $(10-30)\; M_\odot$ stars of $(6-10)\times 10^5$~yr;
the turbulent envelopes allow equilibrium cores to have greater densities
and shorter collapse times than those supported by thermal pressure alone.
\citet{1995ApJ...446..665C} extended this to more massive stars and found
formation times $>10^6$~yr for stars of $100\, M_\odot$, 
a significant fraction of the main sequence
lifetime. On the other hand, by modeling the spectral energy distributions (SEDs) of
high-mass protostars, \citet{1999ApJ...525..808O} inferred
that high-mass stars form in somewhat less than $10^5$ yr,
and favored a logatropic model for the density distribution of
the core. \citet{2000ApJ...534..976N}  inferred an accretion rate of $10^{-2}\;M_\odot$~yr\e\
(corresponding to a formation time of a few thousand years)
for the source IRc2 in Orion based on the assumption that
the accretion 
rate is $\sim 10 c_{\rm eff}^3/G$, with the effective sound
speed $c_{\rm eff}$ determined from the observed line width.

The turbulent core model for high-mass star formation
\citep{2002Natur.416...59M,2003ApJ...585..850M}  
follows from the assumption that such stars form in turbulent,
gravitationally bound cores (virial parameter
$\avir\sim 1$). The turbulence is
self-similar on all scales above the Bonnor-Ebert 
scale, where thermal pressure dominates. The star-forming clump and
the protostellar cores within it are assumed to be centrally
concentrated so that the 
pressure and density have a power-law
dependence on radius, $P\propto r^{-k_P}$,
$\rho\propto r^{-\krho}$. It follows that
the cores are polytropes (\S \ref{gravity}), and since
the Bonnor-Ebert scale is small, the cores
are approximately singular.
The protostellar infall rate is determined by the surface density of
protostellar core, which in turn is comparable to that of the clump in 
which it is embedded.
The regions of high-mass star formation studied by \citet{1997ApJ...476..730P}
have surface densities
$\Sigma_{\rm cl}\sim 1$~g~cm$^{-2}$, 
corresponding to visual extinctions
$A_V \sim 200$~mag; these values are similar to those for 
observed star clusters in the Galaxy 
(e.g., $\sim0.2$~g~cm$^{-2}$ in the Orion Nebula Cluster,
0.8~g~cm\ee\ for the median globular cluster and
$\sim 4$~g~cm$^{-2}$ in the Arches Cluster).
By contrast, regions of 
low-mass star formation have $\Sigma\sim 0.03$ g~cm\ee,
corresponding to $A_V\sim 7$~mag \citep{1996ApJ...465..815O}.
The radius of a protostellar core is 
\beq
R_{\rm core}=\left(\frac{M_{\rm core}}{\pi \Sigma_{\rm core}}\right)^{1/2}
\simeq 0.06\left(\frac{m_{*f}}{60\ecore\; M_\odot}\right)^{1/2}
\frac{1}{\Sigma_{\rm cl}^{1/2}}~~~\mbox{pc},
\eeq
where $m_{*f}$ is the final stellar mass. The second expression
is based on the result that the surface density of a typical core is
comparable to that of the clump in which it is embedded;
cores near the center of a clump have higher surface densities, 
and the sizes are correspondingly smaller.
Using the results of \citet{1997ApJ...476..750M} for the inside-out collapse of
a singular polytrope and adopting $\krho=\frac 32$, a typical density power law
from \citet{1997ApJ...476..730P}, \citet{2003ApJ...585..850M} found
that the typical 
infall
rate and the corresponding
time to form a star of mass $m_{*f}$ 
are
\setlength{\arraycolsep}{0.5mm}
\begin{eqnarray}
\dot m_*&\simeq& 0.5\times 10^{-3}\left(\frac{m_{*f}}{60\ecore\; M_\odot}
\right)^{3/4}
\Sigma_{\rm cl}^{3/4}\left(\frac{m_*}{m_{*f}}\right)
^{0.5}~~M_\odot\;\mbox{yr\e},
\label{eq:mds_hi}\\
t_{*f}&\simeq &1.3\times 10^5 \left(\frac{m_{*f}}{60\ecore\;M_\odot}\right)
^{1/4}\Sigma_{\rm cl}^{-3/4}~~~\mbox{yr},
\end{eqnarray}
where $\Sigma_{\rm cl}$ is the surface density 
(in g~cm\ee) of the several thousand $M_\odot$ clump
in which the star is forming.
For typical values of $\Sigma_{\rm cl}\sim 1$~g~cm\ee, the
star formation time is of order $10^5$~yr and the infall
rate is of order $10^{-3}\,M_\odot$~yr\e. This
infall rate is large enough to overcome the effects
of radiation pressure at the dust destruction front,
thereby addressing one of the key theoretical difficulties
for models of high-mass star formation (see below).
The mean infall rate could be somewhat larger than
given in equation (\ref{eq:mds_hi}) if the core was initially overdense or
contracting, and turbulence in the core could generate
large fluctuations in the infall rate.
However, the infall rate given above is only a few times greater
than the free-fall value and is unlikely to be much larger.

The key assumptions in this model are that
stars form from pre-assembled cores (although
since the cores are turbulent, there will be significant
mass exchange with the ambient medium); that
the cores and the clumps in which they are embedded
are in approximate virial equilibrium; and that they
are magnetically supercritical, so that the magnetic
field does not significantly limit the rate of accretion.
Evidence in support of the first assumption
has been obtained by \citet{2005ApJ...634L.185B}
and 
\citet{
2005ApJ...634L..57S};
the remaining assumptions are also subject to observational test.
The model is necessarily approximate, since it treats the turbulence as
a local pressure (the microturbulent approximation),
and since it 
incorporates all the feedback effects due to radiation pressure and photoevaporation
in the core star formation efficiency, $\ecore$, which 
was assumed to be
of order 1/2. Some of the large density fluctuations in the supersonically
turbulent cores will form low-mass stars, but most of the mass
of the core is assumed to go into one or two massive stars.
\citet{2005MNRAS.360....2D} have criticized the model on the ground that
the massive cores would fragment and form many low-mass
stars rather than a single massive star, but radiative heating by the rapidly accreting 
high-mass protostar
strongly suppresses fragmentation
\citep{2006ApJ...641L..45K,2007ApJ...656..959K}.
The turbulent core model is consistent with 
the correspondence between the core mass function and
the IMF (\S \ref{clumps}), and it naturally allows for the
disks and winds associated with high-mass stars 
(see \S \ref{disks_and_winds}) since it is an extrapolation
of low-mass star formation theory. The cores are predicted to
be denser than the clump in which they are embedded by
about $(M_{\rm clump}/M_{\rm core})^{1/2}$, 
which is much greater than unity for 
stellar mass cores embedded in clumps with $M>10^3\;M_\odot$;
this naturally overcomes the crowding problem. 

An alternative class of gravitational collapse models involves
rapidly accelerating accretion ($\mds\propto m_*^q$ with $q>1$, so that $m_*\rightarrow
\infty$ in a finite time in the absence of other effects).
Building on  the work of \citet{2000A&A...359.1025N},
\citet{2001A&A...373..190B} 
assumed that the accretion rates are proportional
to the mass outflow rates observed in protostellar outflows;
since the outflows are swept-up material,
the 
justification
 for this assumption is unclear. They found
$t_{*f}\sim 3\times 10^5$ yr for massive stars, with most
of the growth occurring in the last 10\% of this time.
In the competitive accretion model (\citealp{1997MNRAS.285..201B};
see \S \ref{bondi}), massive stars form via Bondi-Hoyle-Lyttleton
accretion ($\mds\propto m_*^2$). 
\citet{2002ApJ...580..980K,2003ApJ...599.1196K} has
studied this model further, focusing on the associated H~II regions.
For a $10\;M_\odot$ star in a typical high-mass star-forming
clump observed by \citet{1997ApJ...476..730P}, which has a mass $\sim 4000 M_\odot$
and a virial parameter of order unity,
the 
Bondi-Hoyle 
accretion rate is much smaller than that expected
in the turbulent core model \citep{2003ApJ...585..850M}, even after allowing
for the turbulent enhancement factor $\phibh$ (eq. \ref{eq:mbh}).
The rate of
Bondi-Hoyle accretion 
increases
if the virial parameter is small, if the infall occurs onto
a cluster of stars that is much more massive than a single star 
(as Keto comments), or if the infall occurs from
a significantly less massive clump. In the latter two cases
the assumptions underlying 
Bondi-Hoyle 
accretion
begin to break down,
and further study is needed to determine the infall rate.
\citet{2004MNRAS.349..678E} have shown that radiation pressure
halts Bondi-Hoyle accretion when the star is moving supersonically relative
to the gas for $m_*>10\;M_\odot$, since the luminosity is large
enough that
radiation pressure deflects gas 
away from the star.

In view of the challenges facing conventional theories of high-mass star
formation, \citet{1998MNRAS.298...93B} made the radical suggestion that high-mass stars form
via stellar collisions. This model requires stellar densities $\sim 10^8$ stars
pc\eee\ during the brief period in which the stars grow by merging.
This coalescence model produces an IMF that is in qualitative
agreement with observations, although no feedback effects were
included in the calculations \citep{2001MNRAS.324..573B}. This model faces a number of
challenges: (1) The required stellar density 
is far greater than has been observed in any Galactic star cluster.
For example, W3 IRS5 is one of the densest clusters observed
to date, with 5 proto OB stars in a sphere of radius 0.015 pc \citep{2005ApJ...622L.141M};
the corresponding stellar density $\sim 4\times 10^5$~pc\eee\ is 
lower than required by the coalescence model 
by more than 2 orders of magnitude, although it must be borne in mind
that the number of lower mass stars in that volume is currently unknown.
(2) For large OB protoclusters, 
the hypothesized ultra-dense state would produce a very luminous,
compact source, yet this has never been observed.
(3) The mass loss that is hypothesized to reduce the cluster 
density to observed values must be finely tuned in order to leave
the cluster marginally bound. (4) Finally, it is difficult to see how
the model could account for observed associated disks and outflows
(\S\S \ref{diskobs}, \ref{windobs}) 
above.
 \citet{2005AJ....129.2281B} discuss a number of observational tests of the coalescence
model, and suggest that the wide-angle outflow from OMC-1 
in the Orion molecular cloud could be due to the merger of two protostars
that released $10^{48}-10^{49}$~erg. Two variants of the coalescence
model have been suggested: \citet{2000prpl.conf..327S} proposed that gas bound to
the protostars could increase the cross section for collisions,
although they did not explain why this would result in stellar coalescence
rather than the formation of a binary. \citet{2005MNRAS.362..915B}
have proposed an explanation for this: assuming that the
gas has negligible angular momentum (which is plausible if
the turbulence is weak, as assumed in the competitive accretion model),
then accretion drives the stars in the binary to closer separations
and ultimately to a merger. 
The stellar density required for the binary
coalescence model is $\sim 3\times 10^6$ stars~pc\eee, substantially
smaller than in the direct coalescence model but higher than observed nonetheless.
On the other hand,
\citet{2006astro.ph.11822K} argue that pre-main sequence evolution
of tight, high-mass protostellar binaries can lead to equal mass
binaries, as often observed, rather than to mergers.

\subsubsection{Observations of high-mass protostars}
\label{himassobs}

\citet{2007prpl.conf..165B} have summarized the current state
of observations of high-mass star formation.
They divide the formation of individual high-mass stars into four stages:
\begin{itemize}
\item[1.] High-Mass Starless Cores (HMSCs)
\item[2.] High-Mass Cores harboring accreting Low/Intermediate-Mass Protostar(s)
destined to become high-mass star(s)
\item[3.]High-Mass Protostellar Objects (HMPOs), with $m_*\ga 8 M_\odot$
\item[4.] Final Stars
\end{itemize}
\noindent
The earliest stages of high-mass star formation may occur in the Infrared Dark Clouds
(IRDCs--\citealp{1998ApJ...494L.199E}), which have properties consistent
with being the dense clumps out of which clusters eventually form
\citep{2006ApJ...653.1325S}.
To date, few true HMSCs have been detected---high-mass cores often
appear to have some signatures of star formation. In the turbulent core model,
this could be because the central densities in the cores are much greater
than the mean densities (in contrast to the case
for low-mass cores), and the time scale for gravitational collapse is
correspondingly shorter. The lack of true HMSCs is also
consistent with the competitive accretion model
or the coalescence model, since in these models
HMSCs do not exist. 
Evidence for High-Mass Cores harboring low/intermediate mass
protostars, or possibly relatively low-mass HMPOs, has been obtained only recently
\citep{2005ApJ...634L.185B,2005ApJ...634L..57S}.
HMPOs are 
often (but not always)
associated with Hot Molecular Cores (HMCs), which have a rich chemistry
\citep{2005IAUS..227...70V}.
HMPOs are often associated with H$_2$O and Class II CH$_3$OH maser emission,
although the interpretation of this emission remains ambiguous.
HMPOs are also associated with H~II regions (see \S \ref{photo}); many should have
hypercompact H~II regions and some should be associated with ultra-compact
H~II regions, but most ultra-compact HII regions are associated with the final stars.

Observational tests of infall models for high-mass protostars are difficult
due to their large distances (typically $\ga 2$~kpc), crowding,
large extinctions, and confounding effects of H~II regions.
Several tests are possible: if confirmed, the correspondence between the
core mass function and the IMF (\S \ref{clumps}) would be
consistent with the turbulent core model; the properties
of disks and winds associated with HMPOs can provide important
clues (\S \ref{diskobs}); the spectral energy distributions
(SEDs) of embedded sources 
provide
information
on the distribution of circumstellar matter on scales smaller
than can be directly resolved 
\citep{1999ApJ...525..808O,2005ApJ...631..792C,2005IAUS..227..206W}; 
and chemical clocks can provide direct
measures of the time scale for the growth of HMPOs.
To this end, \citet{2006A&A...454L...5D} have developed the first model
for the chemical evolution of an HMPO, including the evolution
of the central source, infall, and adsorption and desorption of ices from grains.
They find that the time scale for the warm chemistry is set by
the time it takes for matter to flow through the warm region,
and that the total age of the HMPO they study (AFGL 2591) is
$(0.3-1)\times 10^5$~yr.

\subsubsection{Forming stars in the presence of radiation pressure}
\label{rad}

One measure of the importance of radiation pressure is to compare
the stellar luminosity with the luminosity at which
the force due to radiation pressure balances gravity.
Since dust provides the dominant opacity to non-ionizing radiation in
the interstellar medium, this generalized Eddington luminosity is
\beq
L_{E,\, d}=\frac{4\pi cGm_*}{\kappa_d},
\eeq
where $\kappa_d$ is the dust opacity per unit mass, and $c$ is the
speed of 
light. 
The dust in the infalling gas sublimates
when it reaches the dust
destruction front at $r=\rdd\simeq 1.2\times 10^{15}(L/10^5\,
L_\odot)^{1/2}$~cm \citep{1987ApJ...319..850W}. We approximate the radiation field
outside $\rdd$ as a blackbody with a temperature $T$ that declines
with radius. As an example, 
consider the \citet{1994ApJ...421..615P} dust model: $\kappa_d(T)$ first rises with
temperature as the average frequency increases, but then declines
for $T\ga 600$~K as some of the grain species sublimate. The
maximum opacity is $\kappa\simeq 8$~cm$^2$~g\e, which
leads to $L_{E,\, d}\simeq 1600(m_*/M_\odot)\;L_\odot$.
Since main sequence stars have
luminosities $L\simeq 10(m_*/M_\odot)^3\;L_\odot$
for $7\,M_\odot\la m_*\la 20\, M_\odot$ 
(inferred from \citealp
{1996snih.book.....A}),
the infalling gas and dust pass through a region in
which the force due to the infrared radiation exceeds that due to gravity if
$m_*\ga 13\, M_\odot$. For somewhat larger masses, the net force is outward over
a sufficiently large region that the infall is stopped. At the
dust destruction front, the gas and dust are exposed to the stellar UV radiation,
for which $\kappa\sim 200~$cm$^2$~g\e. However, this
radiation interacts with the matter only once with this opacity, since it is emitted in
the infrared after absorption; as a result, the condition for the
infall to persist is that its momentum exceed that of the radiation,
$\dot M_{\rm in}v_{\rm in}> L/c$ 
\citep{1971A&A....13..190L,1974A&A....37..149K,1987ApJ...319..850W}.
High infall rates $\sim 10^{-3}\;M_\odot$~yr\e\ can overcome the UV
radiation problem, but not the IR one.

Several mechanisms have been proposed to permit the formation of massive stars
in the face of radiation pressure:

\begin{enumerate}
\item {\sl Reduced dust opacity}: Based on 1D, multifluid calculations of
steady flows with both graphite and silicate grains with a range of
sizes, \citet{1987ApJ...319..850W}  
found that a reduction in the dust-to-gas ratio
of at least a factor 4 is needed in order for accretion to proceed for stars with $m_*\geq 60 
M_\odot$.

\item {\sl Rotation}: \citet{1989ApJ...345..464N} showed that the higher ram pressure
associated with disk accretion helps overcome the UV radiation pressure problem,
and escape of the infrared radiation from the disk alleviates the infrared radiation
pressure problem. He found that accretion could continue onto a $100\, M_\odot$ star
with an accretion rate as small as $10^{-4}\;M_\odot$~yr\e, 50 times smaller
than for spherical accretion. \citet{1996ApJ...462..874J} showed that there
are a range of conditions for which infrared radiation pressure cannot 
halt infall prior to the formation of a disk
in the context of the \citet{1984ApJ...286..529T} 
model for rotating collapse, even for quite massive stars.
\citet{2002ApJ...569..846Y}  have carried out the most detailed 
axisymmetric numerical simulations
to date. Using frequency-dependent radiative transfer, they found that radiation
pressure limited the maximum stellar mass that could be formed from a
$120\;M_\odot$ core to $43\;M_\odot$; they point out that this is an upper
limit, since it did not allow for fragmentation or for the effects of outflows.

\item {\sl Rapid infall}: \citet{2003MNRAS.338..962E} have shown that 
radiation pressure becomes moot if the
protostellar core is sufficiently dense that the protostellar mass is inside
the dust destruction front. The models they considered to achieve this
condition were far from virial equilibrium and had constant density,
which led to very large accretion
rates $\sim 10^{-2}\; M_\odot$~yr\e. However, models with
centrally concentrated initial density profiles ($\rho\propto r^{-1}$) and
normal dust 
could not produce
stars with 
$m_*>16\;M_\odot$.

\item {\sl Beaming}: \citet{1989ApJ...345..464N} pointed out that disks redirect
the infrared radiation toward the poles, reducing the radiative force in the plane
(\citealp{1999ApJ...525..330Y} termed this the ``flashlight effect" 
and emphasized its observational
importance). \citet{2005ApJ...618L..33K} showed that the cavities produced by outflows from
massive stars would allow the infrared radiation to escape, reducing the
the radiation pressure in the infalling gas and permitting infall over a substantial range of
solid angle.

\item {\sl 3D effects}: 3D simulations with flux-limited, gray radiative transfer
show that 
the accreting gas is subject to radiation-driven Rayleigh-Taylor instabilities,
which 
facilitate
 the escape of the
radiation in low column regions and the accretion of the gas
in high column regions 
\citep{2005IAUS..227..231K}.
There is no evidence that radiation
pressure halts the accretion up to $m_*=35\;M_\odot$, a substantially
higher mass than was found in axisymmetric simulations with gray
transfer \citep{2002ApJ...569..846Y}. \citet{2007astro.ph..1800T}
have shown that the dusty envelopes of HMPOs are subject to
the photon bubble instability, which further promotes infall.

\end{enumerate}

\subsubsection{Photoionization feedback: H~II regions}
\label{photo}

The H~II regions associated with HMPOs provide strong feedback on
infall and accretion, and may play a role in defining the maximum
stellar mass. They are classified into two types: {\it Ultra-compact} H~II
(UCHII) regions have diameters $(0.01-0.1)$~pc, densities
$\geq 10^4$~cm\eee, and emission measures $\int n_e^2 dl\geq
10^7$~pc~cm$^{-6}$ \citep{1989ApJS...69..831W}. {\it Hypercompact} H~II (HCHII) regions
have diameters $<0.01$~pc with emission measures $\geq 10^8$~pc~cm$^{-6}$
(\citealp{2007prpl.conf..165B}; 
for a slightly different definition and a review of both
types of H~II region, see \citealp{2007prpl.conf..181H}).
HCHII regions often appear in tight groups in high-mass star-forming
regions, and they often have broad radio recombination lines with
widths that can exceed 100 km s\e.

The high accretion rates characteristic of HMPOs delay the point at
which the stars reach the main sequence \citep{2003ApJ...585..850M,2006astro.ph.11822K}, thereby delaying the time at which the photosphere is hot enough to produce
an H~II region. High accretion rates also quench the 
emission of ionizing photons once the star has reached the main
sequence \citep{1995RMxAC...1..137W}. Close to the star---i.e., inside the gravitational
radius $r_g=Gm_*/c_i^2=3.2\times 10^{15}(m_*/30\,M_\odot)~\mbox{cm}$,
where $c_i\simeq 10$~km~s\e\ is the isothermal
sound speed of the ionized gas---spherically accreting gas is
in free fall, with $\rho\propto r^{-3/2}$.
For an ionizing photon luminosity
$S$, the radius of the HCHII region is
\beq
R_{\rm HCHII}=R_*\exp(S/S_{\rm cr}),
\eeq
where
\beq
S_{\rm cr}=\frac{\alpha^{(2)}\mds^2}{8\pi\muh^2 Gm_*}
=5.6\times 10^{50}\left(\frac{\mds}{10^{-3}\,M_\odot\;\mbox{yr\e}}\right)^2
\left(\frac{100\,M_\odot}{m_*}\right)
~~~\mbox{s\e},
\eeq
\citep{2002MNRAS.332...59O},
where $\alpha^{(2)}$ is the recombination rate to excited states of hydrogen and where
we have replaced the proton mass in their expression with $\muh=2.34\times 10^{-24}$~g,
the mass per hydrogen nucleus. Provided the accretion is spherical,
the H~II region is quenched for $S\la S_{\rm cr}$.
If $S/S_{\rm cr}$ is not too large ($\la 7$),
$R_{\rm HCHII}$ is less than  $r_g/2$ and the infall velocity 
at the Stromgren radius exceeds $2c_i$,
the minimum velocity of an R-critical ionization front;
as a result there is no shock in the accretion flow and the H~II region
cannot undergo the classical pressure-driven expansion \citep{2002ApJ...580..980K}. 
If the accretion is via
a disk, as is generally expected,
then the ionizing photons can escape out of the plane of the disk,
and the H~II region will not be trapped 
\citep{2006ApJ...637..850K,2006astro.ph..3856K}.
Disk accretion is often associated with the production of winds, and
\citet{2003IAUS..221P.274T}  have suggested that 
such winds confine HCHII regions: the winds clear the gas 
along the axis,
and the ionizing radiation then illuminates the inner surfaces of the winds.
If correct, this offers the possibility of a powerful diagnostic for
determining the nature of disk winds associated with 
massive stars. \citet{2005A&A...437..947V} found very compact radio emission aligned with
the outflows in two high-mass
protostellar sources, consistent with this picture.
When the ionizing luminosity becomes large
enough, however, the wind will become ionized and the H~II region will
evolve to a UCHII state. 
The ionizing photons will photoevaporate the surface of the disk
at a rate of order $\dot m_{\rm evap}\sim$~few~$\times 10^{-5} 
(S/10^{49}~\mbox{s\e})^{1/2}\;M_\odot$~yr\e;
absorption of ionizing photons by dust can significantly
affect this \citep{1994ApJ...428..654H,1997A&A...327..317R}.
This mass-loss rate is too small to be important in setting the maximum mass of
the star (although it can be important in primordial star formation---McKee \& Tan, in preparation).
Absorption of ionizing photons by dust
must also be taken into account when inferring the ionizing luminosity of the 
central star from the properties of the H~II region (\citealt{2006ApJ...639..788D} and
references therein).

\subsubsection{Star Formation in Clusters}
\label{clusters}

Most stars are born in clusters (e.g., \citealp{2003ARA&A..41...57L,
2007prpl.conf..361A}),
and this is particularly true of high-mass stars.
The mass distribution of clusters appears to obey a universal 
power law, $d\ncl/d\ln M\propto M^{-\alpha}$,
with $\alpha\simeq 1$. With this distribution, $Md\ncl/d\ln M=$ const: 
taken together, clusters in each decade
of mass have the same total number of stars. 
\citet{2003ARA&A..41...57L} find that 
very young clusters within 2 kpc of the Sun 
that are still embedded in their natal molecular clouds 
obey this power law for
$M\ga 50\;M_\odot$; the upper limit of the observed distribution
is set by the largest cluster expected in the area they surveyed. The mass distribution
of OB associations in the Galaxy also has a power-law distribution with
$\alpha\simeq 1$ \citep{1997ApJ...476..144M}; 
they inferred that the distribution extended from $\sim 50 \;M_\odot$ to
$2\times 10^5\;M_\odot$ and could account
for all the stars formed in the Galaxy. \citet{1989ApJ...337..761K} 
found that the luminosity distribution
of H~II regions in disk galaxies obeys $d\caln /d\ln L\propto L^{-1\pm 0.5}$,
which is consistent with an $M^{-1}$ distribution since the luminosity is
proportional to mass for associations that are large enough to fully
sample the IMF. The distribution of OB associations in the SMC 
has $\alpha=1$ from the largest associations down to associations
with a single OB star \citep{2004AJ....127.1632O}.
The star clusters in the ``Antennae" galaxies show $\alpha=1$ over
the mass range $10^4\;M_\odot<M<10^6\;M_\odot$ \citep{1999ApJ...527L..81Z};
this is one of the best determined cluster mass functions, and has an error, including
systematic errors, estimated as $\pm 0.1$.
The mass distributions of open clusters and globular
clusters are also consistent 
with an $M^{-1}$ distribution at birth \citep{1997ApJ...480..235E}.
\citet{2006astro.ph.11586D} found $\alpha\simeq 0.9$ for
clusters in irregular galaxies and $\alpha\simeq 0.75$ in
disk galaxies, but comment that this 
result could be affected by the low 
spatial resolution of the data.
The $M^{-1}$ mass distribution of clusters is intermediate between
the high-mass part of the IMF ({$M^{-1.35}$) on the one hand, and
the observed mass distribution of GMCs ($M^{-0.6}$) and the clumps within
them ($M^{-0.3}$ to $M^{-0.7}$; \S \ref{dynam_state}) on the other.
It is important to understand the origin of the 
differences among these power laws, which appear to be real.

The structure of star clusters contains clues to their formation.
High-mass stars in Galactic clusters that are massive enough to contain a number of 
such stars are observed or inferred to be segregated toward the center of the cluster.
The large fraction of O stars that
are runaways can be naturally explained if they originate
in dense, mass-segregated 
clusters and undergo dynamical interactions \citep{1992MNRAS.255..423C}.
\citet{1998ApJ...492..540H} analyzed the spatial distribution of stars in the Orion Nebula Cluster
(ONC) and concluded that the high-mass stars were born preferentially
near the center. Using $N$-body simulations,
\citet{1998MNRAS.295..691B} showed that it takes a relaxation time,
$t_{\rm relax}\simeq 0.1(\caln_*/\ln\caln_*)t_{\rm cross} \;(\simeq 14 t_{\rm cross}$
for $\caln_*=1000$),
for high-mass stars to collect near the center of a cluster due to
dynamical interactions. In the case of the ONC, they argued that
significant dynamical mass segregation has occurred, but not enough to
account for the observed central concentration of OB stars;
they concluded that therefore the observed mass segregation is primordial.
\citet{2006ApJ...641L.121T} suggested both a greater age and a longer crossing time for
the ONC, but the basic conclusion does not change. 
In NGC 3603, the most luminous
Galactic star cluster that is not heavily obscured, \citet{2006AJ....132..253S} found that the 
maximum mass of the stars decreases away from the center
of the cluster, with all the most massive stars being quite close to the center,
and again concluded that the segregation is primordial.
These arguments for primordial mass segregation have been weakened
by the realization that sub-clustering in the initial cluster
significantly accelerates the rate of dynamical mass segregation
\citep{2007ApJ...655L..45M}. 
In addition, most estimates do not account for dynamical friction
between the stars and the surrounding gas, which can considerably
reduce the mass segregation timescale \citep{1999ApJ...513..252O}.
However, recent observations 
have reinforced the argument for primordial mass segregation:
\citet{2005ApJ...622L.141M}  have found a Trapezium-like cluster
in W3~IRS5 that is deeply embedded in molecular gas and is only half
the radius of the ONC. In $\rho$~Oph, \citet{2006A&A...447..609S} found direct
evidence for primordial mass segregation by showing that the
the core mass function exhibits mass segregation as well; this also
supports the correspondence between the core mass function and the IMF
discussed in \S \ref{clumps}. The cluster R136 in the LMC appears to
be an exception to the rule that high-mass stars are centrally concentrated,
since half the massive stars are located outside the central core
\citep{2006AJ....132..253S}.

Only a small fraction of clusters survive as bound clusters to an age of
$10^8$~yr; \citet{2003ARA&A..41...57L}  estimate this fraction as
$4-7\%$. 
In order for a cluster
to remain bound, its natal clump must have a high star formation efficiency.
Analytic estimates suggest that
if the gas in the clump is removed suddenly, such as by an H~II region,
one requires $\eclump> 0.5$ in order for the cluster to remain bound, whereas
if the gas is removed gradually, the cluster will expand adiabatically and
lower values suffice \citep{1980ApJ...235..986H,1983ApJ...267L..97M}.
Numerical calculations show that a fraction of the cluster survives
even if the mass ejection is abrupt \citep{1984ApJ...285..141L}.
\citet{2001MNRAS.321..699K} modeled the evolution of the ONC with $\eclump=0.3$ and a 
sudden mass ejection; they concluded that 30\% of the mass of
the ONC 
would
remain bound, and that it 
would
evolve into a cluster
like the Pleiades.
Star formation efficiencies in the embedded 
clusters in the solar neighborhood are 
observed to be $\simeq 0.1-0.3$ \citep{2003ARA&A..41...57L}; since
star formation is ongoing in these clusters, the final value
of the star formation efficiency, $\eclump$, is near the upper limit
of this range.
\citet{2000ApJ...545..364M}  calculated the star formation
efficiency for clumps in which the mass loss
is dominated by protostellar outflows ($M\la 1-3\times 10^3\;M_\odot$).
(Note that the clump star formation efficiency, $\eclump$, which
is the fraction of the mass of a clump that goes into a cluster of stars, 
is distinct from the core star formation efficiency, $\ecore$, which
is the fraction of a core mass that goes into a single or binary star.)
They estimated $\eclump\simeq 0.4$ for clumps with escape velocities $v_{\rm esc}\simeq
2$~km~s\e, comparable to the
observed value. The star formation efficiency is predicted to rise with
$v_{\rm esc}$---i.e., with increasing mass and/or density of the clump.

The star formation efficiency for larger clusters, ranging up to globular clusters
and super star clusters (SSCs; e.g., \citealp{1996ApJ...466L..83H}) is
most likely determined by the H~II regions
that form in the clusters. Since globular clusters are much more centrally
concentrated than open clusters that form in the disk of the Galaxy today,
it is likely that their star formation efficiency, $\eclump$, was higher.
(Note that a high star-formation efficiency over the life of the clump, $\eclump$, 
is consistent with a low
value of the star-formation efficiency per free-fall time, 
$\epsilon_{\rm ff,\, clump}$,
only if clusters form over a number of free-fall times, and conversely,
cluster formation in $1-2$ dynamical times requires a relatively high
value of $\epsilon_{\rm ff,\, clump}$, -- see \S \ref{large-scale_SFR}.)
Using a simple phenomenological model,
\citet{1997ApJ...480..235E} showed how $\eclump$ should increase with both the mass of
the natal clump and its pressure, $P\propto \Sigma_{\rm cl}^2$.
They point out that high pressures are naturally produced in merging galaxies, accounting for the
large number of super star clusters seen in such systems. The high
surface densities of globular clusters implies that they necessarily formed
in high pressure environments (see also \citealp{2003ApJ...585..850M}).
High star formation efficiencies are possible in a clump with embedded H~II
regions since their destructive effect is significantly reduced
in a clump composed of dense cores \citep{2001sgnf.conf..188T,2005MNRAS.358..291D}.
For sufficiently massive and concentrated clusters, the escape velocity
exceeds the sound speed of ionized gas, and this can further increase the
star formation efficiency \citep{2002MNRAS.336.1188K,
2002ApJ...566..302M,
2004ASPC..322..263T}.

How do stars form in clusters? 
The natal clumps of embedded clusters in the solar neighborhood
have densities $\sim 10^{4-5}\;\cmt$ and masses $\sim 10^{2-3}\;M_\odot$
\citep{2003ARA&A..41...57L}, and more broadly distributed high-mass
star-forming clumps in the Galaxy have densities $\sim 10^5\;\cmt$ and
masses $\sim 10^{3.5}\;M_\odot$ \citep{1997ApJ...476..730P}.
These extreme conditions have 
led to suggestions that the process of high-mass star formation is
qualitatively different from that observed in regions of low-mass star formation
(\S \ref{hiinfall}) or that it is triggered by an external effect.
In a review of triggered star formation, \citet{1992sfss.conf..381E}
pointed out that triggering generally does not affect the star formation efficiency
by more than a factor 2. In any case, 
triggered star formation loses much of its meaning in a theory of star formation 
based on turbulence, 
since in most cases the
triggering event is just a manifestation of the intermittency of the turbulence.
The observed correspondence between the core mass function and
the IMF (\S \ref{clumps}) and the constancy of the star formation
efficiency per free-fall time (\S \ref{large-scale_SFR})
suggest a more unified picture in
which stars form via gravitational collapse in a turbulent medium
over most, if not all, the range of observed clustering.

\section{OVERVIEW OF THE STAR FORMATION PROCESS}

Key goals of a theory of star formation are
 to predict the rate of star formation
and the distribution of stellar masses on the macroscopic scale, and to predict
the properties of individual stars from the initial conditions on the 
microscopic scale.
In the past decade, there has been a paradigm shift in the theory 
from star formation
in a quasistatic medium to star formation
 that occurs in a supersonically turbulent
one, and this has led to significant progress on both fronts. 
Based on our current understanding,
the narrative of star formation contains the following elements:

\begin{itemize}

\item The road to star formation in a disk galaxy like the Milky Way
begins when massive ($\sim
  10^7\Msun$) bound structures condense out of the diffuse ISM as a 
result of gravitational instabilities,
frequently initiated within spiral arms.

\item The most massive structures (GMAs or HI superclouds) inherit
  high levels of internal turbulence from the diffuse ISM, and this
  combines with self-gravity to cause fragmentation into GMCs of a
  range of masses, as well as clumps within the GMCs.

\item The turbulence within GMCs is highly supersonic
  and approximately Alfv\'enic. It imposes a log-normal distribution of
  densities, and creates a spectrum of gas condensations over a wide
  range of spatial scales and masses.  This structure is hierarchical. 
  
  \item This turbulence damps in about one crossing time, and as yet 
  it is not understood exactly how, and for how long, the highly intermittent
  sources of energy in the interstellar medium 
  (including within GMCs themselves)
  can maintain the observed universal level of turbulence in GMCs.

\item Spatially-defined structures within GMCs tend to have internal
  velocity dispersions that increase with size as $\sigma \propto
  \ell^{0.5}$, which is understood to reflect the underlying power
  spectrum scaling expected for supersonic turbulence.

\item Some of the densest regions created by turbulence become
  self-gravitating cores with masses that are typically
  of order the Bonnor-Ebert mass. The distribution of core masses
  appears to be similar to the initial mass function (IMF) for stars, 
  and turbulence appears to be important in defining this distribution.

\item   These cores are frequently clustered, due
  to the dominance of large scales in the turbulent flow.  Forming cores 
  sample from the local vorticity of the turbulence to determine their spins.
 The rate of core formation can be estimated based on the
 turbulent properties of a GMC.

\item Dense cores that begin or become magnetically supercritical
  undergo collapse, first becoming strongly stratified internally.
  Observations show that magnetic fields in cores are roughly
  critical, and this is consistent with inferred core lifetimes.

\item Continued accretion after the  collapse of a core
can occur if the surrounding ambient medium has a sufficiently
low level of turbulence, but it is not yet known how much this can
increase the masses of stars.

\item The collapse of a core leads to the formation of a rotating
  disk interior to an accretion shock; significant magnetic flux is
  lost in this collapse process, although 
based on current results this is
not enough to account
  for the small fluxes observed in stars.

\item Disks accrete due to a combination of processes that transport
  angular momentum outward; these transport mechanisms include
  gravitational stresses when the surface density is high enough, and
  magnetic stresses when the ionization is high enough.

\item Powerful winds are magnetocentrifugally driven from the surface
  of circumstellar disks at a range of radii.  The inner portion of
  the wind, which arises nearest the central star, becomes collimated
  into a jet-like flow due to magnetic hoop stresses.

\item The impact of a wide-angle, stratified disk wind on the
  protostellar core sweeps up much of the ambient gas into a massive
  molecular outflow.  This reduces the net efficiency of star
  formation to $\sim 1/3$.  The combined action of many outflows also
  helps to energize dense, star-cluster-forming clumps.

\item Massive stars form from cores that are considerably more massive
than a Bonnor-Ebert mass, and are most likely highly turbulent.
Radiation pressure strongly affects the dynamics of massive star
formation, but can be overcome by the combined action of
disk formation, protostellar outflows and radiation-hydrodynamic instabilities
in the accreting gas. It is not clear whether protostellar feedback
determines the maximum mass of the stars that form.

\item Massive, luminous stars ionize their surroundings into HII
  regions.  The expansion of these regions into ambient gas at $\sim
  10\;\kms$ energizes GMCs, contributing to the large-scale turbulent
  power.  However, this process is difficult to regulate, and can
  unbind GMCs within a few dynamical crossing times.  By the time they
  are finally destroyed, GMCs may have lost much of their original mass by
  photoevaporation.

\item The destruction of GMCs returns almost all of the gas they
  contain to the diffuse phase of the ISM, with a mean star formation
  efficiency over the cloud lifetime of $\sim 5\%$.  This low
  efficiency can be understood as a consequence of the small fraction
  of mass that is compressed into clumps dense enough that turbulence
  does not destroy them before they collapse.

\item  The return of GMC gas to the diffuse ISM completes the cycle of
  star formation, which then begins anew.

\end{itemize}
\noindent
The coming decade will test and revise this narrative of
star formation, particularly with the advent of ALMA and JWST and
the continued advances in numerical simulation. 
Turning this narrative into a quantitative, predictive theory
will provide a foundation for addressing 
many of the outstanding questions in astrophysics today, ranging from
the formation of planets to the evolution of 
galaxies and the origin of the elements.

\paragraph{Acknowledgements}
We are grateful to our expert readers, 
J. Bally,
 G. Basri, S. Basu, 
E. Bergin,
L. Blitz,
R. Crutcher, B. Elmegreen, C. Gammie, L. Hartmann, M. Heyer,
R. Kennicutt, S. Kenyon, M. Krumholz, C. Lada,
 Z.-Y. Li, C. Matzner, 
S. Offner, P. Padoan, J. Tan, E. Vazquez-Semadeni, and E. Zweibel, for their
insightful comments on draft sections of the manuscript,
and to our editor, E. van Dishoeck, for her comments on the entire manuscript.
We are also
grateful to C.-F. Lee for his help producing the figure of HH111
and to Nathan Smith for his help with the figure of the Carina Nebula.
The work of
CFM and ECO was supported by the National Science Foundation under
grants AST 0606831 and AST 0507315, respectively.  In preparing this
review, we have relied upon the search and archive facilities provided
by NASA's Astrophysics Data System Bibliographic Services.


\newcommand\araa{ARAA}
\newcommand\apjl{ApJL}
\newcommand\aap{A\&A}
\newcommand\aaps{A\&A Supp}
\newcommand\apj{ApJ}
\newcommand\pre{Phys Rev E}
\newcommand\mnras{MNRAS}
\newcommand\apss{Ap\&SS}
\newcommand\jgr{JGR}
\newcommand\aj{AJ}
\newcommand\pasp{PASP}
\newcommand\apjs{ApJS}
\newcommand\pasj{PASJ}
\newcommand\nat{Nature}

\end{document}